\documentclass[a4paper,single column,12pt]{article}
\usepackage{epsfig,amsmath,amssymb}
\usepackage[colorlinks=true,linktocpage=true,linkcolor=blue,citecolor=blue]{hyperref}
\usepackage{graphicx}
\usepackage{dcolumn}
\usepackage{bm}
\usepackage{subfig}
\usepackage{amsmath}
\usepackage{float}
\usepackage{color,soul}
\usepackage{mathtools,slashed}
\usepackage{tabularx}
\usepackage{hhline}
\usepackage{color}
\newcommand{\old}[1]{}

\newcommand{\be}{\begin{eqnarray}}
\newcommand{\ee}{\end{eqnarray}}
\newcommand{\bi}{\begin{itemize}}
\newcommand{\ei}{\end{itemize}}

\newcommand\com{(\gamma e^- \longrightarrow \gamma e^-)}
\usepackage{enumitem}
\usepackage{cite}
\usepackage{hyperref}
\hypersetup{
	colorlinks=true,
	linkcolor=blue,
	filecolor=magenta,      
	urlcolor=cyan,
}
\usepackage{caption}

\newcommand{\roverline}[1]{\mathpalette\doroverline{#1}}
\newcommand{\doroverline}[2]{\overline{#1#2}}

\begin{document}
\begin{flushright}
{\normalsize
}
\end{flushright}
\vskip 0.1in
\begin{center}
{\large {\bf
S-Matrix approach to Compton scattering at the tree level in a strong magnetic field}}
\end{center}
\vskip 0.1in
\begin{center}
Jitendra Pal$^\dag$\footnote{jeetupal007@gmail.com} and Binoy Krishna 
Patra$^\dag$ \footnote{binoyfph@iitr.ac.in}
\vskip 0.02in
{\small {\it $^\dag$ Department of Physics, Indian Institute of
Technology Roorkee, Roorkee 247 667, India\\
} }
\end{center}
\vskip 0.01in
\addtolength{\baselineskip}{0.4\baselineskip} 

\section*{Abstract}
We have studied the Compton scattering ($\gamma   e^- \longrightarrow 
\gamma e^-$) at the tree level in a homogeneous background of
strong magnetic field ($|eB| \gg { m }^{ 2 },~\text{m is the mass of 
electron}$) through the S-matrix approach. For that purpose, using the 
Schwinger propagator for the electron, we have first calculated the 
square of the S-matrix element in the Landau gauge by summing over
the final states of electron and photon and averaging over the initial 
states of the same. In the strong magnetic field, only the lowest Landau 
level for electron is considered. Finally we have computed the crosssection 
for Compton scattering as a function of initial photon energy for the different 
strengths of strong magnetic fields, where we have found that the
crosssection in vacuum gets decreased due to the presence of strong magnetic
field. However, for a fixed initial photon energy, the crosssection increases 
linearly with the magnetic field.

\section{Introduction}
The study of the Compton scattering in a strong magnetic field is
of astrophysical importance because the strong magnetic
field tremendously affects this matter-radiation
scattering. This type of magnetic field is generally found in magnetized 
neutron stars, where the Compton scattering plays a
key role in the formation of neutron star spectra. For neutron star 
applications, it thus becomes necessary to modify the scattering properties 
{\em like} the spinors and its orthogonality and completeness condition 
etc. in a strong magnetic field.
A strong magnetic field is also  expected to be produced in recent heavy-ion
program in RHIC at BNL and in LHC at Geneva at very early stages of
collisions for the non-central 
events \cite{Shovkovy:LNP871'2013, DElia:2012ems, 
Fukushima:2012vr,Mueller:2014tea,Miransky:2015ava}.
Depending on the centralities, the strength of the magnetic field may
reach between $m_{\pi}^2$ ($\simeq 10^{18}$ Gauss) at
RHIC \cite{Kharzeev:2007jp} to 15 $m_{\pi}^2$ at
LHC \cite{Skokov:2009qp9}. At extreme cases, it may reach values of 
50 $m_{\pi}^2$ at LHC. A very strong magnetic field ($\sim 10^{23}$~Gauss)
may have existed in the early universe during the electroweak
phase transition due to the gradients in Higgs field
\cite{2c66851a101a4255af732b66c3fe121d}.

The nonrelativistic treatment of Compton scattering in strong magnetic 
field is done by \cite{Canuto:PRD3'1971, Blanford:MNRAS174'1976}.
However, the relativistic effect becomes dominant as the strength of 
magnetic field exceeds $ 10^{12} $ Gauss. The strength of the magnetic field 
decides the transition between the Landau levels, so that for the weak 
magnetic field, a number of resonances corresponding to the transition between 
the Landau levels appear. The relativistic quantum electrodynamics (QED) 
treatment allows the scattering at resonances with higher harmonics and 
can consider the  transition for electrons to their corresponding Landau 
levels.  Such results from QED for Compton scattering have
 been offered in various papers \cite{PhysRevD.19.2868,Dass:1974, 
Daugherty:1986,Bussard:1986wd} at various levels of analytic and 
numerical development. However, they consider the infinitely long lived 
intermediate states, so it will be far from resonance. In order to consider 
the resonant Compton scattering, one needs to consider the finite life time 
decay width for the electrons to do the cyclonic transitions to their 
Landau levels. In particular, the Ref.\cite{Gonthier:2014wja, 
Mushtukov:2015qul} has considered the spin-dependent influences at the 
cyclotron resonance and assumes the soft photon emitted due to transition 
of electrons from their corresponding Landau levels with very low energy 
and taken the finite decay width of landau levels.

In the presence of magnetic field, the momentum of electron ($\bf p$) is
factorized into components with respect to the direction of the magnetic
field, where the transverse component gets quantized and longitudinal
component ($p_Z$) remains unaffected. As a result, the dispersion 
relation for the electron is modified quantum mechanically into\cite{Bhattacharya:2007vz}
	\begin{align}
	E_n(p_Z)=\sqrt { p_Z ^2+m^2+2neB}~,\label{landau_quantization}
	\end{align}
	where $ n = 0,1,2,$  . .
 are the quantum numbers specifying
 the Landau levels. In strong magnetic field,
 the electrons are rarely excited to the higher Landau levels,
 only the lowest Landau levels (LLL) $(n = 0)$
 are populated. Thus the dynamics of electrons
 are effectively restricted to (1 + 1) dimensions.
	
We have studied the Compton scattering process through the S-matrix, which
is based on the second-order 
QED perturbation theory and provides a ground for detailed investigation
in a field of radiation transfer in case of strong external magnetic field.
This paper is divided into following sections.
Using the notations of four momenta of electron and photon suitable
 for the description in a strong magnetic field, we 
 revisit the solution of Dirac equation in a strong and homogeneous magnetic 
field in sections 2.1, 2.2 and 2.3 and then considered the Schwinger 
propagator in section 2.4 for calculating the S-matrix. Using those notations, 
we calculate the S-matrix
 element for the $s$-and $u$-channel in sections 3.1 and 3.2, respectively.
Then we wish to calculate the S-matrix element squared, which involves the 
M-matrix element squared for $s$-, $u$-channel and the interference 
between them. Thus we calculate the M-matrix element squared for
the $s$-channel, $u$-channel 
and the interference term, after summing over the final states and
averaging over the initial states in sections 4.1. 4.2 and 4.3,
respectively. In section 5, we first revisit the formula for 
calculating the crosssection by constructing 
the Lorentz invariant phase space, flux factor, energy-momentum conserving 
Dirac-Delta function etc. in the presence 
of strong magnetic field and finally, we evaluate the crosssections for the 
$s$ and $u$ channels in sections 5.1, 5.2, respectively because the 
matrix element for the interferene term vanishes. 
Finally we  conclude in section 6.
	\section{Fermions in strong magnetic field}
	\subsection{Notation}
As mentioned earlier, the momentum of an electron in a magnetic field
gets separated into the longitudinal and transverse components with 
respect to the direction of the magnetic field. Therefore, using the 
following convention of the metric tensor
	\begin{eqnarray}
	g^{\mu\nu}=(1,-1,-1,-1), g^{\mu_{\perp}\nu_{\perp}}
	=(0,-1,-1,0)~ {\rm and} ~ 
	g^{\mu_{\parallel}\nu_{\parallel}}=(1,0,0,-1),
	\end{eqnarray}
it will be useful to define the four momentum for the electron as
	\begin{eqnarray}
	x^\mu&=&(X^0,X,Y,Z), \nonumber\\
	p_\perp^\mu&=&(0,p^1,p^
	2,0)=(0,p_X,p_Y,0),\\
	p_\parallel^\mu
	&=&(p^0,0,0,p^3),=(E,0,0,p_Z),\\
	\widetilde{p_\parallel}^\mu
	&=&(\widetilde{p}^0,0,0,\widetilde{p}^3) =(p_Z,0,0,E).
		\end{eqnarray}
However, the usual definition for the photon four momentum is 
	\begin{eqnarray}
	k^\mu&=&=(k^0,k^1,k^2,k^3)
	=(\omega,k_X,k_Y,k_Z).		
		\end{eqnarray}

		\subsection{Spinors}
The methods of Ritus eigenfunction\cite{osti_4691541} along with Schwinger
 Proper-time formalism~\cite{PhysRev.82.664} are commonly used to 
solve the Dirac equation of charged fermions in the presence of a 
constant magnetic field. There are different approaches which have been 
adopted in the literature~\cite{herold_ruder_wunner,sokolov_ternov,
1992herm.book.....M} to
obtain the spinor in a magnetic field. However, we have employed
the procedure to solve the 
Dirac equation in a constant external field from Ref.~\cite{Bhattacharya:2007vz}. 
	
For the sake of simplicity, we assume a static and homogeneous magnetic field 
along the $Z$-direction, $\vec{B} =B \hat{Z}$. Such a magnetic field can be obtained
from a vector potential $A^\mu = (0,0,BX,0)$. The choice of vector potential is 
not unique as the same magnetic field can also be obtained in a symmetric gauge,
$A^\mu = (0,\frac{-BY}{2}, \frac{BX}{2},0)$. Thus the positive energy 
Dirac spinors with the gauge $A^\mu=(0,-BY,0,0)$ are given by the shifted
coordinate, $\xi$ (=$\sqrt{eB}\left(Y+\frac{p_Y}{eB}\right)$) ~\cite{PhysRev.81.115,Bhattacharya:2007vz}
\begin{align*}
e^{ -ip\cdot x_{\slashed Y } } U_{ s }(Y,n,{ p }_{\slashed Y }),  
\end{align*}
where $U_s$'s are
			
\begin{eqnarray}
U_+ (Y,n,\vec p _\slashed Y) = 
				\left( \begin{array}{c} 
					I_{n-1}(\xi) \\[2ex] 0 \\[2ex] 
					{\strut\textstyle p_Z \over
						 \strut\textstyle E_n+m}
					 I_{n-1}(\xi) \\[2ex]
					-\, {\strut\textstyle
						 \sqrt{2ne{ B}} 
						 \over \strut\textstyle 
						E_n+m} I_n (\xi) 
				\end{array} \right)
				 \,, \qquad 
				U_- (Y,n,\vec p _\slashed Y) 
				= \left( \begin{array}{c} 
					0 \\[2ex] I_n (\xi) 
					\\[2ex]
					-\, {\strut\textstyle
						 \sqrt{2ne{ B}} \over 
						 \strut\textstyle E_n+m}
					I_{n-1}(\xi) \\[2ex] 
					-\,{\strut\textstyle p_Z
						 \over \strut\textstyle
						  E_n+m} I_n(\xi) 
				\end{array} \right) \,. 
				\label{Usoln}
				\hspace{5mm} 
			\end{eqnarray} 
			Similarly the negative 
			 energy Dirac spinors with
			  $\widetilde \xi$ 
			(=$\sqrt{eB}\left(Y-\frac
			{p_X}{eB}\right)$)
			 are given 
			by~\cite{PhysRev.81.115,Bhattacharya:2007vz}
			\begin{eqnarray}
			V_- (Y,n,\vec p_{\slashed{Y}})
			 = N\left( \begin{array}{c} 
			{\strut\textstyle p_Z \over
				 \strut\textstyle E_n+m}
			I_{n-1}(\widetilde\xi) \\[2ex] 
			{\strut\textstyle \sqrt{2neB}
				 \over \strut\textstyle E_n+m} 
			I_n (\widetilde\xi)  \\[2ex] 
			I_{n-1}(\widetilde\xi) \\[2ex] 0
			\end{array} \right);\hspace{2mm}
			V_+ (Y,n,\vec p_{\slashed{Y}})
			 =N \left( \begin{array}{c} 
			{\strut\textstyle \sqrt{2neB} 
				\over \strut\textstyle E_n+m}
			I_{n-1}(\widetilde\xi) \\[2ex] 
			-\,{\strut\textstyle p_Z \over
				 \strut\textstyle E_n+m}
			I_n(\widetilde\xi)  \\[2ex] 
			0 \\[2ex] I_n (\widetilde\xi)
			\end{array} \right) ,
			\end{eqnarray}
where the normalization constant ($N$) is $N=\sqrt{E_n+m}$ and the symbol, $p_{\slashed Y}$
denotes the absence of the $Y$-component of momentum in the spinors. The energy
eigenvalues are given by the above Landau quantization
\eqref{landau_quantization} and the energy eigenfunctions, $I_n (\xi)$'s
 are expressed in terms of Hermite polynomials, $H_n (\xi)$ as
			\begin{eqnarray}
			I_n(\xi)&=&
			\frac{\sqrt{eB}}
			{n!2^n\sqrt{\pi}}
			e^{\frac{-\xi^2}{2}}H_n(\xi),
			\end{eqnarray}
			having their completeness relation
			\begin{eqnarray}
\sum_\nu I_\nu(\xi) I_\nu(\xi^{'}) = \sqrt{|eB|} \;
				\delta(\xi-\xi^{'}) = 
\delta (Y-Y').
				\label{completeness}
			\end{eqnarray}

As mentioned earlier, in a strong magnetic field, only the lowest Landau level (n=0) is
 populated so only $	U_- (Y,n,\vec p _\slashed Y)$ will be non-zero and the $Y$-dependence
 can also be extracted from the spinors for $n=0$ case. This will made our task easy to handle 
 the $Y$-dependence contained in $I_n(\xi)$ while solving the S-matrix. Thus, for positive energy 
 solutions, the nonvanishing spinor is
			\begin{align}
		  U_ { - }(Y,n,\vec { p } _\slashed Y)=N
		 I_{ 0 }(\xi )\left( \begin{matrix}
		 0 \\ 1 \\ 0 \\
		  \frac { -{ p }_{ Z } }
		  { E_{ 0 }+m }  \end{matrix}
		   \right). \label{factor}
			\end{align}
Similarly, for negative energy solutions, the nonvanishing spinor is
		\begin{align}
		V_ { + }(Y,n,\vec { p } _\slashed Y)=N
		I_{ 0 }(\widetilde{\xi} )\left( \begin{matrix}
		0 \\ 	\frac { -{ p }_{ Z } }
		{ E_{ 0 }+m } \\ 0 \\
	1  \end{matrix}
		\right) .
		\end{align}
	\subsection{Completeness Condition}
We can now calculate the spin sums for the particles ($P_U$) and anti-particles ($P_V$)
in the presence of external magnetic field  as~\cite{PhysRev.81.115,Bhattacharya:2007vz}
	\begin{eqnarray}
P_U (Y,Y' ,n,\vec p_\slashed Y)& =& \sum_s U_s (Y,n,\vec p_\slashed Y)		 \overline U_s (Y' ,n,
		 \vec p_\slashed Y)\label{PU}\nonumber\\
			 &=& 
			{1\over 2}
			\bigg[\left\{ m(1+\Sigma_z) +
			\rlap/p_\parallel - 
			\widetilde{\rlap/p}_\parallel
			 \gamma_5 \right\} I_{n-1}(\xi)
			I_{n-1}(\xi')\nonumber\\  
			&+& \left\{ m(1-\Sigma_z) 
			+ \rlap/p_\parallel +
			\widetilde{\rlap/p}_\parallel 
			\gamma_5 \right\} I_n(\xi)
			I_n (\xi')\nonumber\\ 
			&-& \sqrt{2ne{B}}
			 (\gamma_1 - i\gamma_2) 
			I_n(\xi) I_{n-1}(\xi') 
			- \sqrt{2ne{ B}} (\gamma_1+ i\gamma_2)
			 I_{n-1}(\xi) I_n(\xi') 
			\bigg] . 
			\label{PUs}
		\end{eqnarray}
In the ultra-relativistic ($p^2>>m^2$) and strong magnetic field ($n=0$) limits, the spin sum reduces to
		\begin{align}
		P_U (Y,Y' ,n,\vec
		 p_\slashed Y)=
		 \frac{I_0(\widetilde\xi)
			I_0(\widetilde\xi' )}{2}\big[\slashed{p
			}_\parallel
			+\widetilde{\slashed{p}}
			_\parallel\gamma_5].
		\end{align}
Similarly, the spin-sum for the negative energy spinors can also be calculated as
		\begin{eqnarray}
			P_V (Y,Y' ,n,\vec p_\slashed Y) &\equiv&
			\sum_s V_s (Y,n,\vec p_\slashed Y)
			 \overline V_s (Y',n,\vec p_\slashed Y) 
			\label{PV}\nonumber\\
		 &=& 
			{1\over 2}  
			\Bigg[ \left\{ -m(1+\Sigma_z) +
			\rlap/p_\parallel - 
			\widetilde{\rlap/p}_\parallel
			 \gamma_5 \right\}
			  I_{n-1}(\widetilde\xi)
			I_{n-1} 
			(\widetilde\xi')
			\nonumber\\ 
			&+& \left\{ -m(1-\Sigma_z)
			 + \rlap/p_\parallel +
			\widetilde{\rlap/p}_\parallel \gamma_5 \right\}
			 I_n(\widetilde\xi)
			I_n(\widetilde\xi')\nonumber\\ 
			&+&\sqrt{2ne{ B}} (\gamma_1 -
			 i\gamma_2)
			 I_n(\widetilde\xi)
			I_{n-1}(\widetilde\xi ') 
			+ \sqrt{2ne{ B}} (\gamma_1 +
			 i\gamma_2) I_{n-1}(\widetilde\xi)
			I_n(\widetilde\xi') \Bigg] \,.
			\label{PVs}
		\end{eqnarray}
In ultra-relativistic and strong magnetic field limits, the spin sum reduces 
to
		\be
		P_V (Y,Y' ,n=0,\vec
		 p_\slashed Y)=\frac{I_0(\widetilde\xi)
			I_0(\widetilde\xi')}
		{2}\big[\slashed{p}_\parallel
		+\widetilde{\slashed{p}}_\parallel\gamma_5].
		\ee
However, the $Y$-dependence will be later absorbed in the S-matrix element and
results in the spin sum in the strong magnetic field as
\begin{align}
	P_U (n=0, \vec p_\slashed Y) =\frac{1}{2}\big[\slashed{p}_\parallel
	+\widetilde{\slashed{p}} _\parallel\gamma_5].\label{spinsum}
	\end{align}
\subsection{ Electron propagator: Schwinger proper-time method}
Usually the electron propagator is obtained by the vacuum expectation value of the 
time-ordered product of the Dirac field operators
\begin{align}
S (x-x') & = i \langle 0|T\psi(x) \overline{\psi}(x')| 0\rangle \nonumber\\
&= \theta(t -t') \langle 0|\psi(x) \overline{\psi}(x') |0\rangle - \theta(t'- t) 
\langle 0|\overline{\psi}
(x') \psi(x)|
0\rangle\ ~,
\end{align}
where, $\psi(x)$ and $ \overline{\psi}(x')$ are the solutions obtained
from the Dirac equation in the strong magnetic field from \eqref{field} 
and \eqref{adjoint}, respectively. However, the electron propagator satisfies 
the Dirac-equation in the Green's function approach as
\begin{equation}
i\gamma^{ \mu  }({ \partial  }_{ \mu  }+
iq{ A }_{ \mu  }-m)S(x,x^{ ' })=-\delta^4 (x-x^{ ' })~.\label{dirac}
\end{equation}

The solution of eq.\eqref{dirac} can be obtained from the Schwinger proper-time 
method\cite{PhysRev.82.664}. However, the gauge transformation introduces 
the phase factor into the solution. Since the magnetic field breaks
the translational invariance of space so we can not take its
Fourier transform directly. To take its Fourier transform, 
a phase factor ($\phi$) is introduced in the solution, which is responsible
for the breaking of the translational invariance, thus the propagator
becomes
\begin{align}
S(x,{ x }^{ ' })=\phi (x,{ x }^{ ' })
\int { \frac { { d }^{ 4 }q }{ { (2\pi ) }^{ 4 } }
 {\ e }^{ -iq.(x-{ x }^{ ' }) } } iS(q).
\end{align}
The above phase factor is not gauge invariant and its
form is given in path representation~\cite{PhysRevD.10.2699}
\begin{equation}
\phi (x,{ x }^{ ' })
=\exp\bigg(ie\int _{ x }^{ x^{ ' } }
{ { A_{ \mu  }(x) }{ dx }_{ \mu  } } \bigg),
\end{equation}
which becomes unity for single fermionic line. Since $S(q)$ does not 
depend on the spatial coordinate due to gauge invariance, so  $S(q)$ is 
translationally invariant and it can be written in a discrete form 
as\cite{PhysRevD.91.016007,PhysRevD.10.2699,PhysRevD.42.2881, 
	Sadooghi:2008yf}
\begin{equation}
S(q)=i \exp\bigg(\frac {-q_\bot^2}{\mid eB\mid}\bigg) (-1)^{ n }\sum _{ n=0 }^{ \infty  }
{ \frac { { D }_{ n } }{ q_{ \parallel }^{ 2 } -m^{ 2 }-2neB }  }, 
\end{equation}
where
\begin{equation}
{ D }_{ n }=(\gamma^{ \parallel  }.q
_{ \parallel  }+m)\bigg[(1-i\gamma^{ 1 }
\gamma^{ 2 })L_{ n }\left(\frac { 2q_{ \bot  }
	^{ 2 } }{ eB } \right)-(1+i\gamma^{ 1 }\gamma^{ 2 }
)L_{ n-1 }\left(\frac { 2q_{ \bot  }^{ 2 } }{ eB } \right)
\bigg]+4\gamma^{ _{ \bot  } }.q_{ _{ \bot  } }
L_{ n-1 }\left(\frac { 2q_{ \bot  }^{ 2 } }{ eB } \right).
\end{equation}
In strong magnetic field limit ($n=0$), the terms containing Laguerre 
polynomial, {\em like} $L_{ -1 } (\frac {2q_{\bot}^{2}}{eB})$ will become 
zero and $L_{ 0 } (\frac { 2q_{ \bot  }^{ 2 } }{ eB } ) $ will
   	 become unity, so $D_0 $ reduces to 
\begin{align}
{ D }_{ 0 }=(\gamma ^{ \parallel  }
\cdot q_{ \parallel  }+m)
(1-i\gamma ^{ 1 }\gamma ^{ 2 }),
\end{align}
which yields the electron propagator in the momentum space in a
strong magnetic field\cite{PhysRevD.62.105014,Hasan:2017fmf}
in the extreme relativistic limit
\begin{align}
iS(q)=\left[\frac { (1+
	{ \gamma  }^{ 0 }{ \gamma  }^{ 3 }{ \gamma  }^{ 5 })
	({ \gamma  }^{ 0 }{ q }_{ 0 }
	-{ \gamma  }^{ 3 }{ q }_{ 3 }) }{ { q }_{ \parallel  }^{ 2 }+
	i\epsilon  } \right]
\exp\left(\frac { {-q}_{ \bot  }^{ 2 } }{ eB } \right)
\end{align}
and is factorizable into the longitudinal and the transverse component as 
     \begin{align}
     iS(q)=S_\parallel (q_\parallel) S_\bot (q_\bot),
\label{factorisation}
     \end{align}
where,
     \be
     S_\parallel (q_\parallel)&=&\left[\frac{ (1+{\gamma}^{0}{\gamma}^{3}
     	{ \gamma  }^{ 5 })({ \gamma  }^{ 0 }
     	{ q }_{ 0 }-{ \gamma  }^{ 3 }{ q }_
     	{ 3 }) }{ { q }_{ \parallel  }^{ 2 }
     	+i\epsilon  } \right],\quad\\
    S_\bot (q_\bot)&=&\exp\left(\frac{{-q}_{\bot}^{ 2 } }{ eB } \right).		
     \ee
\section{S-Matrix Element for Compton Scattering ($\gamma   e^-
\longrightarrow \gamma e^-$)}
In Compton scattering, both $s$- and $u$-channel
diagrams contribute, which are represented in Figure(s) 1 and 2, 
respectively. Thus
the S-matrix element for Compton scattering due to both $s$ and $u$-channel
diagrams is
\begin{align}
S  _{fi}= \langle  f \mid \left({ S }^{ s }+{ S }^{ u }\right) 
\mid i  \rangle,
\end{align}
where the initial and final states are denoted by
\be
\mid i \rangle &=&\mid { e }^{ - }({ p_\slashed Y }),\gamma ({ k }) \rangle ,\\
\mid f \rangle &=&\mid { e }^{ - }({ P }_{\slashed Y }),\gamma ({ K }) \rangle,
\ee
respectively. The operators, $S^s$ and $S^u$
denote the transition operators for the $s$ and $u$-channel, respectively 
and we will now derive their forms from the corresponding Feynman 
diagrams.

\subsection{S-Matrix element in $s$-channel}
\begin{figure}[H]
		\begin{center}
			\includegraphics[width=1.0\linewidth]{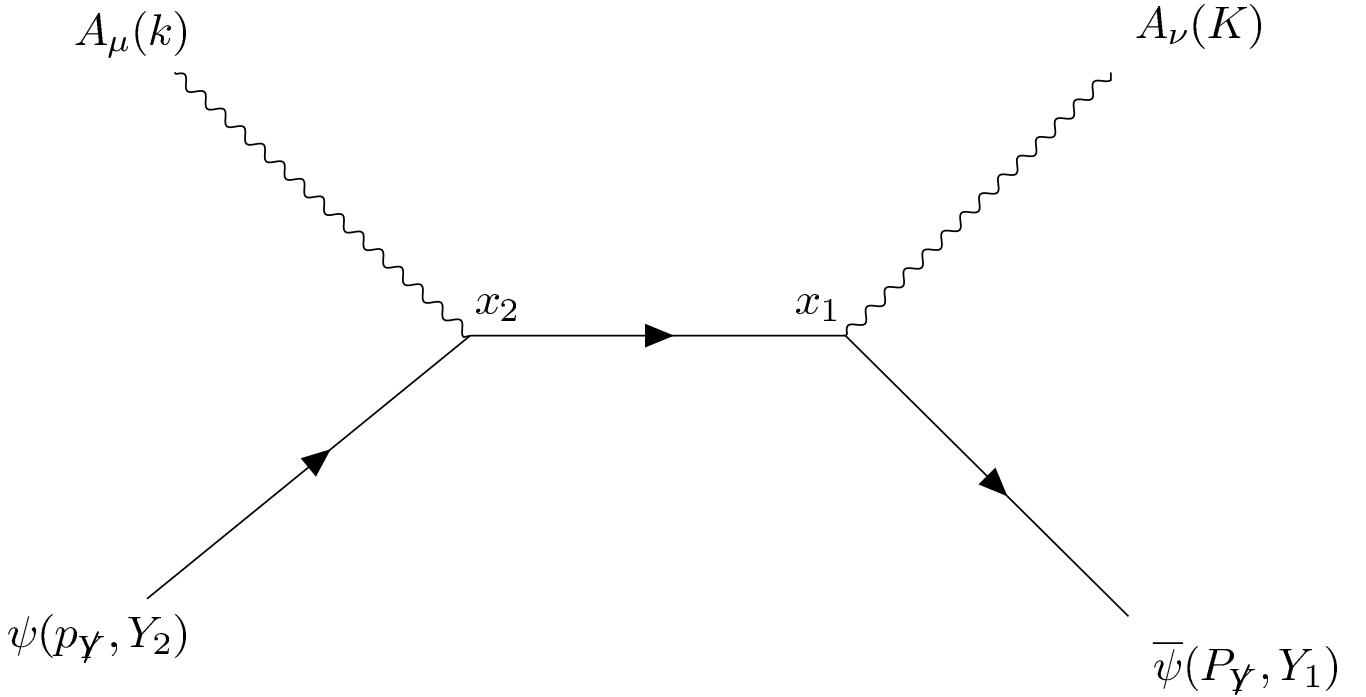}
			\caption{$s$-channel diagram}
		\end{center}
	\end{figure}
The $s$-channel diagram in Compton scattering at the lowest-order
is depicted in Figure 1, where $\psi$ and $\overline \psi$ denote the 
incoming and outgoing electron, respectively whereas $A_\mu(k)$ and $A_\nu(K)$
represent the incoming and outgoing photon, respectively. 
The incoming electron field operator is given by 
\begin{eqnarray}
\psi(x) &=&N_1 \sum_{s} \sum_{n=0}^\infty 
\int\int {dp_X \, dp_Z \over (2\pi)^2} \left[ \widehat a_s 
\ e^{-ip\cdot X_\slashed Y} U_s(Y,p_\slashed Y) +
\widehat b_s^\dagger\ e^{ip\cdot X_\slashed Y} V_s(Y,p_\slashed Y)
\right],\label{field} \nonumber\\
&\equiv& { \psi  }^{ + }(x)+{ \psi  }^{ - }(x),
\label{psi}
\end{eqnarray} 
where $\sum_s$ and $\sum_n$ denote the summation over spins and the Landau levels
and $\widehat a_s$ and $\widehat b_s^\dagger$  are the annihilation and  creation operators 
for fermions and anti fermions, respectively. $U_s(Y,p_\slashed Y)$ and $ 
V_s(Y,p_\slashed Y)$ represent the positive and negative energy spinors, respectively
and $ N_1$ is the normalization constant.

Similarly the adjoint field operator is given by  	
\be
\overline { \psi  } (x)&=&N_{ 1 } \sum_{s}
\sum _{ n=0 }^{ \infty  }\int \int \frac{ dp_{ X }\, dp_{ Z }}{ (2\pi )^{ 2 } }
\left[ \widehat a_{ s }^{ \dagger  } 
\ e^{ -ip\cdot X_{\slashed Y }}
\overline { U } _{ s }(Y,p_\slashed Y)
+\widehat { b } _{ s }\ e^{ ip\cdot X_\slashed Y }
\overline { V } _{ s } (Y,p_\slashed Y) \right]\label{adjoint}, \nonumber\\
&\equiv& \roverline{\psi}^+(x)+ \roverline{\psi}^-(x).
\ee
For the real photon, the field operator expression can be written as
\begin{eqnarray} 
A_{ \mu  }(x) & = & N_{ 2 }\sum _{ r =1 }^{ 2 } 
\int  \frac { d^{ 3 }k }{ (2\pi )
	^{ 3 }{ 2k^{ 0 } } }
\left[ \widehat { b }
_{ r  }(k)\ e^{ -ik\cdot x }
\epsilon ^{r  }_{ \mu  }(k)
+\widehat { b } _{ r }^
{ \dagger  }(k) \ e^{ ik\cdot x }
\epsilon ^{ r* }_{ \mu  }
(k) \right] \\
&\equiv & {A_\mu  }^{ + }(x)+{ A_\mu  }^{ - }(x),  
\end{eqnarray}
where the index, $r$ labels the sum over the transverse polarization states 
and ${A_\mu  }^{ + }(x)$ (${ A_\mu  }^{ - }(x)$) represents the positive 
(negative) energy  part, respectively.

Therefore, the S-matrix element for the $s$-channel in Compton scattering process is given by
\begin{align}
	S^s _{fi}=-{ e }^{ 2 }\left\langle f\left|\int { { d }^{ 4 }
	{ x }_{ 1 } } { d }^{ 4 }{ x }_{ 2 }
\roverline{\Psi}^-
({ x }_{ 1 }){ A }_{ \alpha  }^{ - }
({ x }_{ 1 }){ \gamma  }^{ \alpha  }
i{ S }_{ F }({ x }_{ 1 }-{ x }_{ 2 })
{ \gamma  }^{ \beta  }{ A }_{ \beta  }^
{ + }({ x }_{ 2 })\Psi ^{ + }{ (x }_{ 2 })\right|i \right\rangle~.
\end{align}
Since the electron and photon states  are independent of each other, we can 
factorize the initial and final states as
\be
|i\rangle &=& |{ e }^{ - }({ p_\slashed Y })\rangle \otimes |\gamma ({ k })
\rangle,\\
|f\rangle&=&|{ e }^{ - }({ P }_{_\slashed Y  })\rangle \otimes |\gamma ({ K })\rangle~.
\ee
So the S-matrix element for the $s$-channel becomes
	\begin{align}
	S^s_{fi}=\int { { d }^{ 4 }
		{ x }_{ 1 } } { d }^{ 4 }{ x }
	_{ 2 }\left\langle { e }^{ - }({ P_{_\slashed Y } })\left|
	{\roverline{\Psi}^-} ({ x }_{ 1 })
	{ \gamma  }^{ \alpha  }i{ S }_{ F }
	({ x }_{ 1 }-{ x }_{ 2 }){ \gamma  }^
	{ \beta  }\Psi ^{ + }{ (x }_{ 2 })
	\right|{ e }^{ - }({ p_\slashed Y }) \right> ~
	\left< \gamma ({ K })|{ A }
	_{ \alpha  }^{ - }({ x }_{ 1 })
	{ A }_{ \beta  }^{ + }({ x }_{ 2 })
	|\gamma ({ k }) \right\rangle .
	\end{align}
By substituting the eigenvalue equations for the electron and photon field operators 
	 	\be
	 	{ \psi  }^{ + }\big|{\ e }^{ - }({ p_\slashed Y })\big\rangle
	 	&=&U_s(p_\slashed Y,Y) \ e^{-ip\cdot x_\slashed Y}\big|0\big\rangle, \nonumber\\
	 	\big\langle {\ e }^{ - }({ p_\slashed Y })\big| 
\roverline{\Psi}^-&=&\big\langle 0 \big| {\overline  U_{ s } 
(p_\slashed Y,Y) } \ e^{ -ip\cdot x_\slashed Y }, \nonumber\\
	 	{ A }_{ \mu }^{ + }\big| \gamma(K)  \big> &=&\ e^{ -iK\cdot x }
\epsilon_\mu\big|0\big\rangle, \nonumber\\
{ A }_{ \mu }^{ - }\big| \gamma (K) \big> &=& \ e^{ iK\cdot x } 
\epsilon^*_\mu\big|0\big\rangle~,
\ee
the above matrix element becomes simplified as
	 	\begin{align}
	 S^s_{fi}=-{ e }^{ 2 }
	 	\int { { d }^{ 4 }{ x }_{ 1 } }
	 	{ d }^{ 4 }{ x }_{ 2 }{\ e }^
	 	{ i({ P_{\slashed Y}}
	 		+{ K }).{ x }_{ 1 } }{\ e }^
	 	{ -i({ p_{\slashed Y} }+{ k })
	 		.{ x }_{ 2 } } \overline { U }
	 	({ P _{\slashed Y},Y_1})
	 	\slashed\epsilon ({ K })i{ S }
	 	_{ F }({ x }_{ 1 }-{ x }_{ 2 })
	 	U(p_\slashed Y,Y_2)\slashed\epsilon (k).
	 	\end{align}
Using the Fourier transform, the electron propagator in the coordinate space is 
converted into the momentum space, thus ${ S^s_{fi}}$ becomes
	 			\be
	 			S^s_{fi}&=&-{ e }^{ 2 }\int
	 			{ { d }^{ 4 }{ x }_{ 1 } } { d }^
	 			{ 4 }{ x }_{ 2 }\int\frac { { d }^{ 4 }q}
	 			{ { (2\pi ) }^{ 4 } } {\ e }^{ i({ P_{
	 						\slashed Y } }+{ K_{  } }).
	 				{ x }_{ 1 } }{\ e }^{ -i({ p_{ \slashed Y  } }+
	 				{ k }).{ x }_{ 2 } }\overline { U } 
	 			({ P_{ \slashed Y },Y_1 }) \nonumber\\
&&\times\slashed\epsilon ({ K }){\ e }^{ -iq.(x_{ 1 }-{ x_{ 2 } })
	 			}iS(q)U(p_{\slashed Y },Y_2)~
	 			\slashed\epsilon (k),\label{bhabha-eq}
	 			\ee
where $q$ (=$p+k$) is the momentum of electron propagator in the $s$-channel diagram. 

Rearranging the exponential terms and using the factorization of the electron propagator 
into longitudinal and transverse components from (\refeq{factorisation}), the S-matrix
element takes the form
	\be
	S^s_{fi} &=& -{ e }^{ 2 }\int
	{ { d }^{ 4 }{ x }_{ 1 } } { d }
	^{ 4 }{ x }_{ 2 }\int\frac { { d }^{ 4 }
		q }{ { (2\pi ) }^{ 4 } } { \ e }^
	{ i({ P_{ \slashed Y } }+{ K }-
		q).{ x }_{ 1 } }{\ e }^{ -i({ p_
			{\slashed Y } }+{ k }-q) \cdot { x }
		_{ 2 } } \nonumber\\
&& \times \left[\overline { U } ({ P_{ \slashed Y } ,Y_1 })\slashed\epsilon ({ K }) 
S_\parallel (q_\parallel) U(p_{ \slashed Y },Y_2) \slashed\epsilon (k)\right] 
S_\bot (q_\bot).
	\ee
Now, integrating over all the components except the $X$ and $Y$ components,
we will get two-dimensional Dirac-delta function, so
	\begin{align}
	S^s_{fi}=-{ e }^{ 2 }
	{ \left( 2\pi  \right)  }^{ 2 }
	{ \delta  }_{ \slashed X,\slashed Y }
	^{ 2 }({ P+ }{ K }-p-k)
	\int { { d }X_{ 1 } } { d }{ X }_{ 2 }
	{ d }{ Y }_{ 1 }{ d }Y_{ 2 }{ { d }{ q }
		_{ X }d }{ q }_{Y }{\ e }^{ i({ K_
			{ Y} }-q_{ Y }){ Y }_{ 1 } }
	{\ e }^{ -i({ k_{ Y } }-q_{ Y })
		{ Y }_{ 2 } 
	}\nonumber\\\times{\ e }^{ i({ P_{ X } }
	+{ K_{ X } }-q_{ X }){ X }_{ 1 } }
{\ e }^{ -i({ p_{ X } }+{ k_{ X } }-q_
	{ X }){ X }_{ 2 } }[\overline { U }
({ P_{  \slashed Y },{ Y }_{ 1 } })
\slashed\epsilon {( K )}S_\parallel (q) U(p_{\slashed Y} ,{ Y }_{ 2 })
\slashed\epsilon (k)] S_\bot (q_\bot).
\end{align}	 	
	
Using the integration below over the X-variable 
	\be
&&\int { { { d }{ q }_{ X } }\delta
		({ P_{ X } }+{ K_{ X } }-q_{ X })
		\delta ({ p_{ X } }+{ k_{ X } }-q
		_{ X })\exp\left( \frac { { -q }
			_{ X }^{ 2 } }{ eB }  \right)  } \nonumber\\
	&&=\quad { \left( 2\pi  \right)  }^
	{ 2 }\delta ({ P_{ X } }+{ K_{ X } 
	}-{ p_{ X } }-{ k_{ X } })\exp\left
	( \frac { -{ ({ P_{ X } }+{ K_{ X }
			}) }^{ 2 } }{ eB }  \right)~, 
	\ee
the matrix element, $S^s_{fi}$ becomes	 	
	\be
	 S^s_{fi}&=&-{ \left( 2\pi 
		\right)  }^{ 4 }{ e }^{ 2 }
	{ \delta  }_{\slashed Y }^{ 3 }
	({ P+ }{ K}-p-k) \exp
	\left( \frac { -{ ({ P_{ X } }
			+{ K_{ X } }) }^{ 2 } }
	{ eB }  \right) \nonumber\\
&& \times \int { d }{ Y }_{ 1 }{ d }Y_{ 2 }
	{ d }{ q }_{ Y }{\ e }^{ i({ K_{ Y } }-q_{ Y }){ Y }_{ 1 } }
{\ e }^{ -i({ k_{ Y } } -q_{ Y }){ Y }_{ 2 } } \nonumber\\
&& \times \Big[\overline { U} ({ P_{ \slashed Y }, { Y }_{ 1 } })\slashed\epsilon
	({ K })S_\parallel (q_\parallel)U(p_{\slashed Y }
	,{ Y }_{ 2 })
	\slashed\epsilon (k)\Big] 
	\exp\left( \frac { { -q }_{ Y }^{ 2 }
		 }{ eB }  \right).	 	
	\ee
Our next task is to solve the spatial ($Y$) integration. So factorizing the spinors 
into the spatial and momentum components (as done in eq. \eqref{factor}) and defining
the new variables by $\overline {P_Y}= K_{Y}-q_{Y}$ and $\overline {P'}_Y={ k_{ Y } } -q_{ Y }$, 
the matrix element $S^s_{fi}$ is written as
 \be 
S^s_{fi} & =&- 
 { \left( 2\pi  \right)  }^{ 4 }{ e }^{ 2 }
 { \delta  }_{\slashed Y }^{ 3 }({ P+
 }{ K }-p-k) \exp\left( \frac {
 -{ ({ P_{ X } }+{ K_{ X } }) }
 ^{ 2 } }{ eB }  \right) \nonumber\\
&&\times \int { d } { Y }_{ 1 }{ d }Y_{ 2 }{ d }
{ q }_{ Y }{ I_{ 0 }(\xi_1 )I_{ 0 }
	(\xi_2 )\ e }^{ i(\overline P_Y){ Y }_{ 1 } } {\ e }^{ i(\overline P'_Y){ Y }_{ 2 } } \nonumber\\
&& \times \Big[\overline { U } ({ P_{ \slashed Y }
})\slashed\epsilon ({ K })S_\parallel (q_\parallel)
U(p_{\slashed Y })
\slashed\epsilon (k)\Big] \exp\left( \frac { { -q }_{ Y }^{ 2 } }{ eB }  
\right)~.	 		
\ee
	
In the strong magnetic field along the $Z$-direction, $p_\bot$ ({\em i.e.} 
$p_X$) is much smaller than $|eB|$, so $\xi$ approximately becomes
	\be
	\xi =\sqrt { e{ B } } \left( Y+{
		\frac { { p }_{ X } }{ eB } 
	} \right) \, \approx \sqrt { e{ B } }~Y,
	\ee
so the integrations over $Y_1$ and  $Y_2$  becomes decoupled in the form of standard integrals:
\be
\int { d } { Y }_{ 1 }I_{ 0 }
(\sqrt { e{ B } } Y_{ 1 }){\ e }
^{ i\overline P_Y{ Y }_{ 1 } } &=&
\sqrt{\frac{2\pi}{eB}} I_{ 0 }\left( \frac { \overline P_Y }{ \sqrt { e{ B } }  }  \right),\nonumber\\  
\int { d }
{ Y }_{ 2 }I_{ 0 }(\sqrt { e{ B } }
Y_{ 2 }){\ e }^{ i\overline P'_Y
	{ Y }_{ 2 } } &= &\sqrt{\frac{2\pi}{eB}} I_{ 0 }\left( \frac { \overline P'_Y }{ \sqrt { e{ B } }  }  \right)~. 
\ee
Thus, after performing the spatial integration, $S^s_{\rm fi }$ finally becomes	 	
\be
S^s_{\rm fi}&=&-{ \left( 2\pi
	\right)  }^{ 4 }{ e }^{ 2 }
{ \delta  }_{\slashed Y }^{ 3 }
({ P+ }{ K }-p-k) \exp
\left( \frac { -{ ({ P_{ X } }+
		{ K_{ X } }) }^{ 2 } }{ eB } 
\right)\nonumber\\
&&\times \Big[\overline { U } ({ P_{ \slashed Y } })\slashed\epsilon (
{ K })S_\parallel (q_\parallel) U(p_{\slashed Y})\slashed\epsilon (k)\Big]
\int { { d }{q }_{ Y } } \exp\left( 
\frac { { -q }_{ Y }^{ 2 } }{ eB }
\right)  I_{  0}\left( \frac { \overline P'_Y }{ \sqrt { e{ B } }  } 
\right) I_{ 0 }\left( \frac {\overline P_Y }{ \sqrt { e{ B } }  }  
\right). 	 			 
\ee

Finally after the momentum integration over $q_{Y}$ through the relation
	\be
\int { { d }{ q }_{ Y } } \exp \left( \frac { { -q }_{ Y }^{ 2 }
	}{ eB }  \right)  I_{ 0}\left ( \frac { \overline P'_Y }{
		\sqrt { e{ B } }  }  \right) I_{ 0 }\left( \frac { 
\overline P_Y	 	}{ \sqrt { e{ B } }  }  \right
	)= \sqrt { 2 } \pi~\exp\left[ -\frac { 3\left( K_Y^2+k_Y^2 \right) } { 8eB } +\frac {K_Y k_Y}{ 4eB } \right],	 
		  \ee
the S-matrix element for $s$-channel diagram results in a form
\begin{eqnarray}
&&S^s_{ fi }=-16\sqrt{2}\pi^5{ e }^{ 2 }  
\exp\left[ -\frac { 3\left( K_Y^2+k_Y^2 \right) } { 8eB } +\frac {K_Y k_Y}{ 4eB } \right] \exp\left( \frac
{ -{ ({ p_{ X } }+{ K_{ X } }) }^{ 2 } }{
        eB }  \right){ \delta  }_{\slashed Y }^{ 3 }({ P+ }{ K }-p-k)\nonumber\\&&~~~~~~~\times \Big[\overline { U} ({
P_{ \slashed Y } })\slashed\epsilon
({ K })S_\parallel (q_\parallel)
U(p_{ \slashed Y })\slashed\epsilon (k)\Big]\label{S-s matrix}. \label{s cha}                               
\end{eqnarray}
\subsection {S-Matrix element in $u$-channel}
 The S-matrix element for the $u$-channel diagram in Figure 2  is given by
\begin{figure}[H]
 	\begin{center}
 		\includegraphics[width=1.0\linewidth]{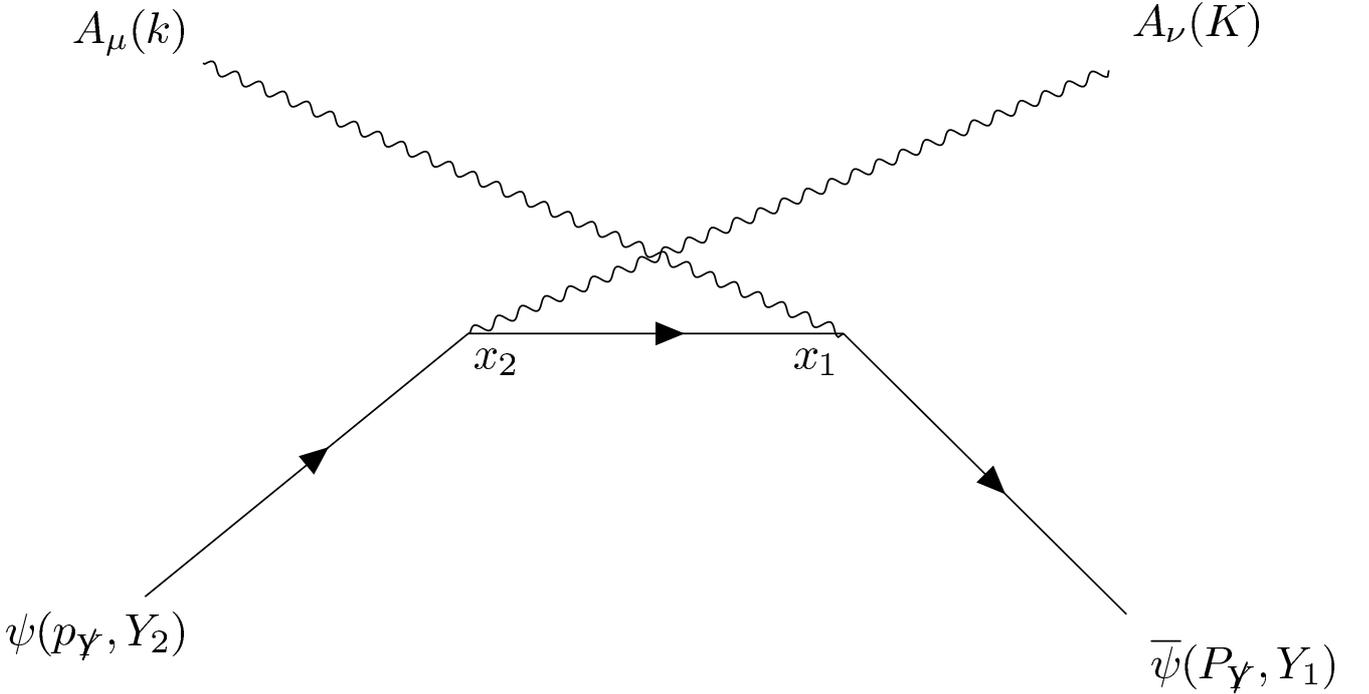}
 		\caption{$u$-channel diagram}
 	\end{center}
 \end{figure}
 \begin{align} S^u_{fi}=-{ e }^{ 2 } 
\left\langle f\left|\int { { d }^{ 4 }{ x }_{ 1 } } { d }^{ 4 }{ x }_{ 2 }
{\roverline{\Psi}^-}({ x }_{ 1 }){ A }_{ \alpha  }^{ + }
({ x }_{ 1 }){ \gamma  }^{ \alpha  }i{ S }_{ F }({ x }_{ 1 }-{ x }_{ 2 })
{ \gamma  }^{ \beta  }{ A }_{ \beta  }^{ - }({ x }_{ 2 })\Psi ^{ + }
{ (x }_{ 2 })\right|i \right\rangle ~.
\end{align}
The further simplification needs the same procedure as done for the
$s$-channel but the crossing-symmetry, {\em i.e.} replacing $K $ by
$-k$ in the $s$-channel expression \eqref{s cha}, solves it easily
as 
	\begin{eqnarray}
	& S^u_{fi}=-16\sqrt{2}\pi^5{ e }^{ 2 }
	{  \exp\left[ -\frac { 3\left( K_Y^2+k_Y^2 \right) } { 8eB } +\frac {K_Y k_Y}{ 4eB } \right] \exp\left( \frac { -{ ({ P_{ X } }-{ K_{ X } }) }^{ 2 } }{ eB }  \right
){ \delta  }_{\slashed Y }^{ 3 } }({ P+ }{ 
K }-p-k)\nonumber\\&~~~~~~~~~~~~~~~~~~~ \times
\Big[\overline { U } ({ P_
	{ \slashed Y } 
})\slashed\epsilon
({ k })S_\parallel (q'_\parallel)
U(p_{ \slashed Y })
\slashed\epsilon (K)\Big],
\label{u cha}	 			
\end{eqnarray}
where $q'$ (=$p-K$) is the momentum in the electron propagator for the
$u$-channel diagram. Finally, the total S-matrix element for the Compton scattering 
is obtained from the $s$- (\refeq{s cha}) and $u$-channel (\refeq{u cha} ) contributions
and is expressed in terms of matrix elements for $s$- and $u$-channel, {\em
viz.} $\mathfrak{M}^s$ and $\mathfrak{M}^u$, respectively 
\be
S_{fi}&=& S^s_{fi}+S^u_{fi} \nonumber\\
&=&-16\sqrt{2}\pi^5   { \delta  }
_{\slashed Y}^{ 3 }({ P+ }{ K }
-p-k)\times \left[ c\mathfrak{M}^s+
d\mathfrak{M}^u \right] ~,
\ee
where $c$ and $d$ are given by
\be
 c&=& { \exp\left[ -\frac { 3\left( K_Y^2+k_Y^2 \right) } { 8eB } +\frac {K_Y k_Y}{ 4eB } \right]  }
\exp\left( \frac { -{ ({ P_{ X } }+{ K_{ X } })
	}^{ 2 } }{ eB }  \right), \\
d&=& { \exp\left[ -\frac { 3\left( K_Y^2+k_Y^2 \right) } { 8eB } +\frac {K_Y k_Y}{ 4eB } \right]  
}\exp\left( \frac { -{ ({ P_{ X } } -{ k_{ X } }) }^{ 2 } }{ eB }  \right),
\ee
respectively.

Now we need to square $S_{fi}$, which involves the square of the three-dimensional
 Dirac delta function. So we first evaluate the square of the Dirac-delta functions, by taking the limits
$X \rightarrow \infty$, $Y \rightarrow \infty$, and 
$T \rightarrow \infty$,
\be
\left| \delta ({ { P_{ X }+ }K_{ X } }-k_{ X }-{ p_{ X } }) \right| ^{ 2 }&
=& \frac { { L }_{ X } }{ 2\pi  } \delta  ({ { P_{ X }+ }K_{ X } }-k_{ X}-
{ p_{ X } })~, \nonumber\\
\left| \delta({ { P_Z } }+{ K_Z }-{ p_Z }-k_Z)\right| ^{ 2 }
&=&\frac { { L }_Z }{ 2\pi  }\delta ({ { P_Z } }+{ K_Z }-
{ p_Z }-k_Z)~, \nonumber\\
\left| \delta({ { E' } }+\omega' - E-\omega ) \right| ^{ 2 }
&=&\frac { T }{ 2\pi  } \delta ({ { E' } }+\omega' - E-\omega ),
\ee
respectively. The values $L_X $ and $L_Z $ cancel out due to the normalization 
constant. Thus, the square of the S-matrix element in large time T for Compton 
scattering takes the form 
\begin{align} 
\frac { |S_{fi}|^{ 2 } }{ T } ={ { { \left( 2\pi  \right)  }^{ 6 }\pi  }
	\delta^{ 3 }_\slashed Y} ({ P}+{ K } -p-k)|{ c }\mathfrak{M}^s+
{ d}\mathfrak{M}^u|^{ 2 }, \label{totalmatrix}
\end{align}
where the matrix elements for the $s$- and $u$-channel are given by
\be
\mathfrak{M}^s &=& -e^2\Big[\overline { U} ({ P_ { \slashed Y } })
\slashed\epsilon ({ K }) S_\parallel (q_\parallel) U(p_{ \slashed Y })
\slashed\epsilon (k)\Big],\label{s-matrix}	\\
\mathfrak{M}^u&=& -e^2\Big[\overline { U } ({ P_ { \slashed Y } 
})\slashed\epsilon ({ k})S_\parallel (q'_\parallel) U(p_{ \slashed Y }) \slashed\epsilon (K)\Big],\label{u-matrix}
\ee
respectively.
\section{M-Matrix Element squared for Compton Scattering}
Our aim is to calculate the square of the matrix element for $s$- and $u$-channel diagrams and the interference term.

\subsection{M-Matrix Element squared for $s$-channel}
 The matrix element given by \eqref{s-matrix} can be rewritten in a compact 
form 
\begin{equation}
\mathfrak{M}^s=A\overline { U }^{s'}
( P_{\slashed {Y}}){ \Gamma  }_{ 1 } 
{ U }^{s}( p_{\slashed {Y}}),\label{matrixs}
\end{equation}
where $s$ and $s'$ denote the spins for the initial and final electrons,
respectively and $A$ and $\Gamma_1$ are given by
\be
A &=& -{ e }^{ 2 }\bigg[\frac {
	1 }{ { q }_{ \parallel  }
	^{ 2 }-m^2+i\epsilon }\bigg] ,\\
{ \Gamma  }_{ 1 }&=&{ \varepsilon  }_{ \mu}^{r' \ast  }(K){ \gamma  }^{ \mu }{ (1
	+{ \gamma  }^{ 0 }{ \gamma  }
	^{ 3 }{ \gamma  }^{ 5 })
	({ \gamma  }^{ 0 }
	{ q }_{ 0 }-
	{ \gamma  }^{ 3 }{ q }_{ 3 } }
){ \gamma  }^{ \nu }{ \varepsilon  }_{ \nu}^{r}(k),\label{gamma}
\ee
whereas $r$ and $r'$ represent the polarization for the initial and final 
photons, respectively. Thus the square of the matrix element becomes
\begin{align}
|\mathfrak{M}^s|^2=A^{ 2 }
\left[ \overline { U }^{s'}( P_{\slashed {Y}}){ \Gamma  }_
{ 1 } { U }^{s}( p_{\slashed {Y}}) \right] \left[ \overline { U }^{s'}( P_{\slashed {Y}}){ \Gamma  }_{ 1 } { U }^{s}( p_{\slashed {Y}})
\right] ^{ * }.
\end{align}

To calculate the unpolarized crosssection, we need to average over the 
quantum states of incoming particles and sum over the final states, therefore 
we replace the above matrix element squared
\begin{equation}
	|\mathfrak{M}|^2 \rightarrow \frac{1}{(2s+1)(2r+1)}
\sum_{\rm all ~states}|\mathfrak{M}|^2
	\equiv \roverline{|\mathfrak{M}|^2}~.
	\end{equation}
Now we need to do the spin sum for both electrons and photons. First we will do the spin sum 
for electrons, therefore
\begin{eqnarray}
\roverline{| \mathfrak{M}^s| ^{ 2 }}=\frac{A^{ 2 }}{6}
\sum _{ r,r'} \sum _{ s,s' }\left[\overline { U }^{s'}
( P_{\slashed {Y}}){ \Gamma  }_{ 1 }
{ U }^{s}( p_{\slashed {Y}}) \right]
\left[ \overline  { U }^{s}( p_{\slashed {Y}})
\overline { { \Gamma  }_{ 1 } } 
{ U }^{s'}( P_{\slashed {Y}}) \right],
\end{eqnarray}
where
\begin{align}
{\overline  { \Gamma  }_{ 1 } } =
{ { \gamma  }^{ 0 } }
{ \Gamma^\dagger_{ 1 }  }
{ \gamma  }^{ 0 }.
\end{align}
It can be  further rewritten in a short hand notation as 
\begin{equation}
\roverline{|\mathfrak{M}^s|^2}=\frac{A^{ 2 }}{6}\sum _{ r,r'}
{\rm Tr} \left[ Q\sum _{s' }{ { U }^{s'}(P_
	{ {\slashed Y } }) } \overline
{ U }^{s'}(P_{ {\slashed Y } }) \right], \label{ssquared}
\end{equation}
where 
\begin{equation}
Q={ \Gamma  }_{ 1 }\sum_{ s } 
{ U }^{s}( p_{\slashed {Y}})\overline 
{ U }^{s}( p_{\slashed {Y}})
{ \overline{ \Gamma  }_{ 1 } }.
\end{equation}
Now summing over the photon polarization states, $\roverline{|\mathfrak{M}^s|^2}$ becomes\footnote{Detailed calculation is given in Appendix A.1}
\begin{eqnarray}
\roverline{|\mathfrak{M}^s|^2}=\frac{A^{ 2 }}{24}
\sum _{r,r' } 
{ \varepsilon  }_{ \mu }^{ r'\ast  }(K)
{ \varepsilon  }
_{ \lambda  }^{r'}(K){ \varepsilon  }
_{ k }^{{r} \ast  }(k)
{ \varepsilon  }_{ \nu}^{r}(k) Tr\bigg[\left\{
{ \gamma  }^{ \mu }
{ (1+{ \gamma  }^{ 0 }{ \gamma  }
	^{ 3 }{ \gamma  }
	^{ 5 })({ \gamma  }^{ 0 }
	{ q }_{ 0 }-{ \gamma  }
	^{ 3 }{ q }_{ 3 }) }
({ \gamma  }^{ \nu })(\slashed{ p }_
{ \parallel  }
+\nonumber\widetilde {\slashed
	p_{ \parallel }
} { \gamma  }^{ 5 })\right\}\nonumber\\
\quad\quad \times 
\left\{({ \gamma  }^{ k })({ \gamma  }^{ 0 }
{ q }_{ 0 }-{ \gamma  }^{ 3 }
{ q }_{ 3 }){ (1-{ \gamma  }^{ 5 }
	{ \gamma  }^{ 3 }{ \gamma  }^{ 0}) }
({ \gamma  }^{ \lambda  })({\slashed P_
	{ \parallel  } }+
\widetilde {\slashed P_{ \parallel } }
{ \gamma  }^{ 5 })\right\}\bigg].\label{averaged M_s}
\end{eqnarray}
Using the spin-sum over the photon polarization states
\begin{equation}
\sum _{ r}^{ \quad  }
{ { \varepsilon  }_{ \mu }^{r\ast  } }
{ \varepsilon  }
_{ \nu }^{r}=-{ g }_{ \mu\nu },
\end{equation}
$\roverline{|\mathfrak{M}^s|^2}$ becomes
\begin{equation}
\roverline{|\mathfrak{M}^s|^2}
=\frac{{ A }^{ 2 }}{24} {\rm Tr} \bigg[\left\{{ \gamma  }
^{ \mu }{ (1+{ \gamma  }^{ 0 }{ \gamma  }^{ 3 }
	{ \gamma  }^{ 5 })
	{\slashed q_\parallel } }
({ \gamma  }^{ \nu })(\slashed
{ p }_{ \parallel  }
+\widetilde {\slashed p_{ \parallel  } } 
{ \gamma  }^{ 5 })\right\}
 \times\left\{({ \gamma _{ \nu } })
{\slashed q_\parallel }{ (1-{ \gamma  }^{ 5 }
	{ \gamma  }^{ 3 }{ \gamma  }^{ 0})
	{ \gamma _{ \mu }(\slashed{ P }_{ \parallel 
		} }+\widetilde {\slashed
		P_{ \parallel  } }
	{ \gamma  }^{ 5 })}\right\}\bigg].\label{s-squared}
\end{equation}
Applying the cyclic properties of traces and the following properties of the gamma matrices
\begin{align}
{ \gamma  }_{ \mu }  \slashed a{ \gamma  }
^{ \mu }=-2\slashed a
\quad {\rm and} \quad { \gamma  }^{ 5 }
{ \gamma  }^{ \nu }+{ \gamma  }^{ \nu }
{ \gamma  }^{ 5 }=0,
\end{align}
and multiplying all the terms carefully, $\roverline{|\mathfrak{M}^s|^2}$ will have sixteen terms and after evaluating the sixteen 
terms \footnote{Detailed calculation is given in Appendix A.1.1}, the  square of the matrix element for the $s$-channel becomes,
\begin{eqnarray}
\roverline{|\mathfrak{M}^s|^2}&=&\frac{{ A }^{ 2 }}{6}\bigg[16(\widetilde p_\parallel \cdot q_\parallel)
P^3q^0-16(\widetilde p_\parallel \cdot q_\parallel)q^3P^ 0
+8q_\parallel^{ 2 }P^0\widetilde p^3
-8q_\parallel^{ 2 }\widetilde p^0P^3
+8(p_\parallel \cdot q_\parallel)\widetilde P^3q^0\nonumber\\
&&-8(p_\parallel \cdot q_\parallel)\widetilde P^0q^3+4q^2_\parallel\widetilde P^0p^3
-4q_\parallel^{ 2 }\widetilde p^0P^3
+16(\widetilde p_\parallel \cdot q_\parallel)\widetilde P^0q^0
-8q_\parallel^{ 2 }\widetilde P^0\widetilde p^0
+8q_\parallel^{ 2 }\widetilde p^3\widetilde P^3\nonumber\\
&&-16(\widetilde p_\parallel \cdot q_\parallel)q^3\widetilde P^3+16(p_\parallel \cdot   q_\parallel)P^0q^0-16(p_\parallel \cdot   q_\parallel)P^3q^3 -8q_\parallel^{ 2 }p^0P^0 +
8q_\parallel^{ 2 }p^3P^3\bigg].
\label{s-matrix squared}
\end{eqnarray}	
 The above equation can be further written in Lorentz invariant form as
\begin{equation}
\roverline{|\mathfrak{M}^s|^2}=\frac{A^2}{6}\Big[32(\widetilde p_\parallel \cdot q_\parallel)(\widetilde P_\parallel \cdot q_\parallel) +24(p_\parallel \cdot q_\parallel)(P_\parallel \cdot q_\parallel) +8 q^2_\parallel(P_\parallel \cdot p_\parallel)\Big].
\end{equation}

\subsection{M-Matrix element squared for $u$-channel}
 The matrix element given by \eqref{u-matrix} can be 
 rewritten in a compact form 
\begin{equation}
\mathfrak{M}^s=A\overline { U }^{s'}
( P_{\slashed {Y}}){ \Gamma  }_{ 1 } 
{ U }^{s}( p_{\slashed {Y}}).
\end{equation}
Like in the $s$-channel \eqref{s-squared}, the same procedure have been followed 
to obtain the squared matrix element for the $u$-channel after averaging over the spin states  
\begin{align}
\roverline{|\mathfrak{M}^u|^2}=
\frac {{ B }^{ 2 } }{24  } 
{\rm Tr} \bigg[[{ \gamma  }^{ \mu }{ (1+
	{ \gamma  }^{ 0 }{ \gamma  }^{ 3 }
	{ \gamma  }^{ 5 })({ \gamma  }^{ 0 }
	{ q }_{ 0 }'-{ \gamma  }^{ 3 }{ q }_
	{ 3 }') }({ \gamma  }^{ \nu })(\slashed{ p }_
{ \parallel  }+\widetilde { 
	\slashed p_{ \parallel  } } 
{ \gamma  }^{ 5 })]\nonumber\\ 
\times[({ \gamma _{ \nu } })({ \gamma  }^{ 0 }
{ q }_{ 0 }'-{ \gamma  }^{ 3 }{ q }_{ 3 }')
{ (1-{ \gamma  }^{ 5 }{ \gamma  }^{ 3 }
	{ \gamma  }^{ 0})
	{ \gamma _{ \mu }(\slashed{ P }
		_{ \parallel  } }
	+\widetilde {\slashed P_{ \parallel  } } 
	{ \gamma  }^{ 5 })]}\bigg],
\end{align}	 
where $B$ is given by
\begin{align}
B=-{ e }^{ 2 }\bigg
[\frac {1 }{
	{ q }_{ \parallel  }
	^{ '2 }-m^2+i\epsilon  }\bigg].
\end{align}
The structure of $\roverline{|\mathfrak{M}^u|^2}$ is the same as 
$\roverline{|\mathfrak{M}^s|^2}$ which is given by \eqref{s-matrix squared}, 
except the fact that  $q$ is replaced by $q'$, thus $\roverline{|\mathfrak{M}^u|^2}$ takes the form
\begin{eqnarray}
\roverline{|\mathfrak{M}^u|^2}&=&\frac{B^2}{6}\bigg[16(\widetilde p_\parallel
\cdot q'_\parallel)
P^3q'^0-16(\widetilde p_\parallel.q'_\parallel)q'^3P^ 0
+8{q'_\parallel}^{ 2 }P^0\widetilde p^3
-8{q'_\parallel}^{ 2 }\widetilde p^0P^3
+8(p_\parallel.q'_\parallel)\widetilde P^3q'^0\nonumber\\&&-8(p_\parallel.q'_\parallel)\widetilde P^0q'^3+4{q'_\parallel}^{ 2 }\widetilde P^0p^3
-4{q'_\parallel}^{ 2 }\widetilde p^0P^3
+16(\widetilde p_\parallel.q'_\parallel)\widetilde P^0q'^0-8{q'_\parallel}^{ 2 }\widetilde P^0\widetilde p^0+8{q'_\parallel}^{ 2 }\widetilde p^3\widetilde P^3
\nonumber\\&&-16(\widetilde p_\parallel.q'_\parallel)q'^3\widetilde P^3+16(p_\parallel.
q'_\parallel)P^0q'^0 -16(p_\parallel.
q'_\parallel)P^3q'^3 
-8{q'_\parallel}^{ 2 }p^0P^0
+8{q'_\parallel}^{ 2 }p^3P^3\bigg].
\end{eqnarray}
 The above equation can be further written in Lorentz-invariant form as
\begin{equation}
\roverline{|\mathfrak{M}^u|^2}=\frac{B^2}{6}\Big[ 32(\widetilde p_\parallel \cdot q'_\parallel)(\widetilde P_\parallel \cdot q'_\parallel) + 24(p_\parallel \cdot q'_\parallel)(P_\parallel \cdot q'_\parallel) +8 q'^2_\parallel(P_\parallel \cdot p_\parallel)\Big].
\end{equation}

\subsection{M-Matrix element squared for interference term}
In Compton scattering, there are two possibilities of the interaction of electron with photon. 
Since photons are the identical particles, both $s$ and $u$-channel diagrams need to be taken into account,
where in $s$-channel, initially electron absorbs a photon and in $u$-channel, initially electron emits
a photon. So initially we are assuming that there is finite probability of 
crossing between these two processes, known as the
 interference term $( \roverline{\mathfrak{M}^s\mathfrak{M}^{u*}} +
\roverline{\mathfrak{M}^u\mathfrak{M}^{s*}}) $
in the total squared matrix element to make the total transition amplitude invariant.
We will first evaluate the $\roverline{\mathfrak{M}^s\mathfrak{M}^{u*}}$, which can be written by the eqs \eqref{s-matrix} and \eqref{u-matrix} as 
\begin{align}
\roverline{\mathfrak{M}^s\mathfrak{M}^{u*}}=\frac{AB}{6} \sum_{r,r'}\sum_{s,s'}
\left[ \overline { U }^{s'}( P_{\slashed {Y}}){ \Gamma  }
_{ 1 } { U }^{s}( p_{\slashed {Y}}) \right] 
\left[\overline { U }^{s}( p_{\slashed {Y}})\overline{\Gamma}_2  
{ U }^{s'}( P_{\slashed {Y}})\right],
\end{align}
where
\be
 { {\overline \Gamma  }_{ 1 } }
 &=&{ \varepsilon  }_{ \mu }^{r'*}(K)
 { \varepsilon  }^{r}_{ \nu }(k)
 { \gamma  }^{ \nu }({ \gamma  }^{ 0 }{ q }_{ 0 }
 -{ \gamma  }^{ 3 }{ q }_{ 3 })
 { (1-{ \gamma  }^{ 5 }{ \gamma  }
 	^{ 3 }{ \gamma  }^{ 0}) }{ \gamma  }^{ \mu },\\
 { {\overline \Gamma  }_{ 2 } } 
&=&{ \varepsilon  }_{ \nu }^{r'}(K)
 { \varepsilon  }^{r*}_{ \mu }(k)
 { \gamma  }^{ \nu }({ \gamma  }^{ 0 }
 { q' }_{ 0 }-{ \gamma  }^{ 3 }
 { q '}_{ 3 }){ (1-{ \gamma  }^
 	{ 5 }{ \gamma  }^{ 3 }
 	{ \gamma  }^{ 0}) }{ \gamma  }^{ \mu}.
\ee
The sum over the spin states for the electrons can be written as trace
\be
 \roverline{\mathfrak{M}^s\mathfrak{M}^{u*}}
&=& \frac{AB}{6}\sum_{r,r'}\bigg[\sum _{ s' } 
\overline { U }^{s'}( P_{\slashed {Y}}){ \Gamma  }_{ 1 }
\sum _{ s } \big[ { U }^{s}( p_{\slashed {Y}}) 
\overline { U }^{s}( p_{\slashed {Y}})\big] \overline\Gamma_2 
{ U }^{s'}( P_{\slashed {Y}})\bigg] \nonumber\\
& =& \frac{AB}{6}\sum_{r,r'}{\rm Tr} \bigg[ Q\sum 
_{  s'  }{ U^{s'}(P_\slashed {Y}) } \overline {
	U }^{s'} (P_\slashed {Y}) \bigg],\label{cross}
\ee
where 
\begin{align}
Q= { \Gamma  }_{ 1 }\sum _{s}
[U^{s}(p_\slashed {Y})\overline { U } ^{s}(p_\slashed {Y})]
{ {\overline  \Gamma  }_{2} }.
\end{align}
Now we sum over the spin states of photons, 
\begin{align}
\hspace*{-4cm}
\roverline{\mathfrak{M}^s\mathfrak{M}^{u*}}& = \frac{AB}{6}\sum _{ r,r'}
\varepsilon_{\mu}^{*}(K){\varepsilon}_{k}(K)
{ \varepsilon  }_{ \lambda }^{ \ast  }(k)
{ \varepsilon  }_{ \nu }(k){\rm Tr}\bigg[ \Big\{{ \gamma  }
^{ \mu }{ (1+{ \gamma  }^{ 0 }{ \gamma  }^{ 3 }
	{ \gamma  }^{ 5 })({ \gamma  }^{ 0 }{ q }
	_{ 0 }-{ \gamma  }^{ 3 }{ q }_{ 3 }) }
({ \gamma  }^{ \nu })(\slashed{ p }_{ \parallel  }+\widetilde {\slashed p_{ \parallel  } } 
{ \gamma  }^{ 5 })\Big\}\nonumber\\	&~~~
\times\Big\{({ \gamma  }^{ k })({ \gamma  }^{ 0 }
{ q }_{ 0 }'-{ \gamma  }^{ 3 }{ q }_{ 3 }')
{ (1-{ \gamma  }^{ 5 }{ \gamma  }^{ 3 }{ \gamma  }^{ 0}
	)( }{ \gamma  }^{ \lambda  })({\slashed P_
	{ \parallel  } }+\widetilde {\slashed P_
	{ \parallel  } } { \gamma  }^{ 5 })\Big\}\bigg]~,\label{crossav}
\end{align}
which can be further simplified, using the completeness condition of photons,
as 
\begin{align}
\roverline{\mathfrak{M}^s\mathfrak{M}^{u*}}&=\frac{AB}{24}{\rm Tr}\bigg[
 \Big\{{ \gamma  }^{ \mu }{ (1+
	{ \gamma  }^{ 0 }{ \gamma  }^{ 3 }
	{ \gamma  }^{ 5 }){\slashed q_{ \parallel  } 
	} }({ \gamma  }^{ \nu })(\slashed{ p }
_{ \parallel  }+\widetilde { \slashed p_
	{ \parallel  } } { \gamma  }^{ 5 })
\Big\}\nonumber \\
&~~~ \times\Big\{({ \gamma _{ \mu } })
{\slashed q'_{ \parallel  } }
{ (1-{ \gamma  }^{ 5 }{ \gamma  }^{ 3 }
	{ \gamma  }^{ 0})( }{ \gamma _{ \nu } })
({\slashed P_{ \parallel  }}+
\widetilde {\slashed P_{ \parallel  } } 
{ \gamma  }^{ 5 })\Big\}\bigg]\label{cross2}
\end{align}
and is split into 32 terms (\eqref{cross32} in Appendix A2). After evaluating 
the trace of each term\footnote{which 
	is calculated in Appendix A.2.1} , we finally obtain the first interference term in Compton scattering 
in a strong magnetic field
\be
\roverline{\mathfrak{M}^s\mathfrak{M}^{u*}}=0.
\label{cross4}
\ee
For calculating the other interference term, $\roverline{\mathfrak{M}^u\mathfrak{M}^{s*}}$, we follow the 
same procedure as for 
$\roverline{\mathfrak{M}^s\mathfrak{M}^{u*}}$ and obtain
\begin{align}
\roverline{\mathfrak{M}^u\mathfrak{M}^{s*}}&= \frac{AB}{24}~{\rm Tr} \bigg[  
\Big\{{ \gamma  }^{ \mu } { (1+{ \gamma  }^{ 0 }{ \gamma  }^{ 3 }
	{ \gamma  }^{ 5 }){\slashed q'_{ \parallel  } } }
({ \gamma  }^{ \nu })(\slashed{ p }_{ \parallel  }
+\widetilde { \slashed p_{ \parallel  } } 
{ \gamma  }^{ 5 })\Big\}\nonumber \\
&~~~ \times\Big\{({ \gamma _{ \mu } })
{\slashed q_{ \parallel  } }
{ (1-{ \gamma  }^{ 5 }{ \gamma  }^{ 3 }
	{ \gamma  }^{ 3 })( }{ \gamma _{ \nu } })
({\slashed P_{ \parallel  } }+\widetilde {\slashed P_{ \parallel  } }
{ \gamma  }^{ 5 })\Big\}\bigg].
\end{align}
 This can be further simplified similar to the first interference term. Thus the second interference term becomes
\be
\roverline{\mathfrak{M}^u\mathfrak{M}^{s*}}=0.\label{int1}
\ee
Thus the interference between the s- and u-channel in strong magnetic field 
limit vanishes.

\section{Crosssection}
Let us illustrate the usual procedure to compute the crosssection for the Compton scattering from
the above transition amplitude in the presence of strong magnetic field. In general, the crosssection is given by
 \begin{align}
 d\sigma =\frac { |S_{ fi }|^{ 2 } }{ TF } d\rho ~,\label{crosssection}
 \end{align}
where $F$ is the flux factor and $d\rho$ is the differential phase space.
In case of magnetic field, the differential phase factor for the electron
 gets modified due to the gauge dependence.
 We have used the Landau gauge ($A^\mu=(0,-BY,0,0)$), in which 
the Hamiltonian does not commute with the $Y$-component of 
the momentum ($p_Y$), so $p_Y$ will not 
 be a conserved quantity. So we will have the three dimensional 
 Dirac-delta function and the $d p_Y$ component of the momentum
 is missing in the phase space of the electron. Since the photon
is not affected by the magnetic field so the phase space factor of
the photon will remain the same as in vacuum. We will calculate the
crosssection in the lab frame, where the initial photon is along the 
direction of magnetic field and the electron is at rest initially. So we 
assign the initial $(p, k)$ and final $(P, K)$ four momentum of the 
electron and photon, respectively.
\be
p^{\mu}&=&(m,0,0,0),\quad \quad \quad {P}^{\mu} = { (E' },P_{ X }, P_{ Y }, 
P_{ Z })~, \nonumber\\
k^{ \mu  } &=& { (\omega  },0,0,{ \omega  }),\quad \quad \quad 
{ K }^{ \mu  } = { (\omega ' },K_{ X },K_{ Y },{ \omega ' }\cos\theta )~,
\label{labframe}
\ee
where, $\theta$ is the angle between the initial and the final direction of 
photon.
Hence the momentum appeared in the electron propagator in extreme relativistic
limit for the $s$- and 
$u$-channel are given by 
\be
{ q }^{ 0 } &=& { p }^{ 0 }+{ K }^{ 0 }=m+{ \omega  },
\quad \quad \quad { q }^{ 3 }={ \omega  }, \nonumber\\
{ q' }^{ 0 }&=& { p }^{ 0 }-{ K }^{ 0 }=m-\omega ',\quad  \quad 
 { q' }^{ 3 }=-{ \omega ' } \cos\theta, \label{propagatorlab}
\ee
respectively.

Now the differential phase space for the electron is modified by 
the mass-shell condition in the strong magnetic field, 
$ P_{ \parallel}^{ 2 } = { m }^{ 2 }$ \cite{Gusynin:1998nh,Rath:2017fdv},
  	   \be 
  	   { { \frac { { d }{ P }
  	   			^{ 0 }{ d }{ P }_{ X }
  	   			{ d }{ P }_{ Z } }{ { (2\pi ) }
  	   			^{ 2 } } \delta ({ P_{ \parallel  }
  	   		}^{ 2 } } }-{ m }^{ 2 })
  	   \Theta ({ { P }^{ 0 } }) 
  	 =\frac { { { d }{ P }_{ X } { d }{ P }_{ Z }d{ P}^{ 0 }
  	 	} }{ { (2\pi ) }^{ 2 } } \frac { \delta ({ { P }^{ 0 } }
  	 		-{ { E '} }) }{ 2{ E '} } ,\label{electron}
  	\ee
where $
E'=\sqrt { p_Z^{'2} +m^2}
$. However, the photon phase factor is simply
given by
  	 \begin{align}
  	 { { \frac { { d }^{ 4 }{ K } }
  	 		{ { (2\pi ) }^{ 3 } } \delta 
  	 		({ K }^{ 2 } } } )\Theta 
  	 ({ { K }^{ 0 } })
  	 &=  \frac { { { d }^{ 3 }{ K }d{ K }^{ 0 }
  	  	} }{ { (2\pi ) }^{ 3 } } \frac { \delta ({ { K }^{ 0 } }
  	  		-{ { \omega'  }}) }{ 2\omega ' }. \label{photon}
  	  \end{align}

 Now substituting the simplified photon and electron phase factors 
and using the property of Dirac-delta function, the crosssection for 
the Compton scattering is reduced into
\begin{align} 
\sigma =\dfrac{\pi^2}{2}
\int { \frac { { { d }{ P }_{ X}{ d }{ P}
_{ Z } } }{{ E '}  } \frac 
{ { d }{ K }_{ Y } }{ \omega' } 
{ \delta  }({ E '}+{ \omega  '}-{ E }-{ \omega  })
\frac{\roverline{|c\mathfrak{M}^s+d\mathfrak{M}^u|^2}}{F}}, 
\end{align}
where the flux factor for the process $1+2\longrightarrow 3+4$
is given by
\begin{equation}
F=\left| { v }_{ 1 }-{ v }
_{ 2 } \right| 2{ E }_{ 1 }2E_{ 2 },
\end{equation}
where $v_1$ and  $v_2$ are the velocities and $E_1$ and $E_2$ are the 
energies of the target and projectile, respectively. In the lab-frame, the 
flux factor becomes 
\begin{align}	
F &=4v_2m_1p_2=4m_1E_2.
\end{align}

In our case, where the target is the electron and the projectile 
is the photon, hence the flux factor becomes
\begin{align}	
F=4m\omega.
\end{align}
Thus the crosssection becomes
\begin{align} \sigma = \dfrac{\pi^2}{8}
\int { \frac { { { d }{ P }_{ X }{ d }
			{ P }_{ Z } } }{{ E '}  } 
	\frac { { d }{ K }_{ Y } }{ \omega' }
	{ \delta  }({ E '}+{ \omega  '}
	-{ E }-{ \omega  })
	\frac {  \roverline{|c\mathfrak{M}^s+d\mathfrak{M}^u|^2} }{ m\omega }  } ,
\end{align}
where,
\begin{align}
\roverline{|c\mathfrak{M}^s+d\mathfrak{M}^u|^2}=
{ c}^{ 2 }\roverline{|\mathfrak{M}^s|^2} +{ d }^{ 2 }\roverline{|\mathfrak{M}^u|^2} +cd~\roverline{\mathfrak{M}^s\mathfrak{M}^{u*}} +cd~\roverline{\mathfrak{M}^u\mathfrak{M}^{s*}}.
\end{align}
Since, the matrix element due to interference has no contribution as given in \eqref{cross4} and\eqref{int1}.
So finally 
\begin{align}
\roverline{|c\mathfrak{M}^s+d\mathfrak{M}^u|^2}=
{ c}^{ 2 }\roverline{|\mathfrak{M}^s|^2} +{ d }^{ 2 }\roverline{|\mathfrak{M}^u|^2}\label{squared}
\end{align}

Evaluating $c^2$ and $d^2$ in the lab frame 
\be
{ c }^{ 2 }&=&{ { e }^{ -\left
		( \frac { 3{ K^{ 2 }_{ Y } } }{ 4eB }  \right) 
		 } }\exp\left( \frac { -2{ q^{ 2 }_{ X } } }
		 { eB }  \right),\label{c} \\
 d^{ 2 } &=& { { e }^{ -\left( \frac { 3{ K^{ 2 }_{ Y } } }
{ 4eB }  \right)  } }\exp\left( \frac { 2q'^{ 2 }_{ X } }{ eB }  \right), \label{cd}
\ee
the $K_Y$-momentum integration results in
\be
\int \roverline{|c\mathfrak{M}^s+d\mathfrak{M}^u|^2}dK_Y &=&
\sqrt { \frac { 4eB }{ 3 }  } \left
        [ \exp\left( \frac { -2q^{ 2 }_{ X } }{ eB }
         \right) \right. \roverline{|\mathfrak{M}^s|^2}
+ \roverline{|\mathfrak{M}^u|^2}\exp
         \left( \frac { -2q'^{ 2 }_{ X } }{ eB }
          \right) \bigg]. \label{Kint}
\ee
 After doing the $P_{ X }$-momentum  integration, the crosssection 
comes to be
 	   \be
\sigma &=&\frac { eB\pi ^{ 2 } }{4 \sqrt { 3 }  } \int { \frac { {{ d }
 	   		{ P }_{ Z } } }{ { mE '}\omega' \omega} 
 	   { \delta  }({ E '}+{ \omega  '}-
 	   { E }-{ \omega  }) }
 	\left[  \exp\left( \frac { -2q^2_X }{ eB }  \right) 
 		\roverline{|\mathfrak{M}^s|^2} + \roverline{|\mathfrak{M}^u|^2}  
 	\exp\left( \frac { -2q'^2_X } { eB }  \right) \right].  
 	 \ee
where the remaining $P_Z$ integration will be done for the matrix elements 
in different channels as well as for the cross-terms one by one.
  	
\subsection{Crosssection due to the $s$-channel diagram }
 
The crosssection expression for the $s$-channel diagram is given by
 \begin{align} 
\sigma^s \com & 
 =\frac { eB\pi ^{ 2 } }{4 \sqrt { 3 }  } \int { \frac { { { d }{ P }_{ Z } } }
 	{ { E' }\omega' } { \delta  } ({ E '}+{ \omega ' }-{ E }
 	-{ \omega  }) } \frac { \exp\left ( \frac { - 2q^2_X }{ eB } 
 	\right) \roverline{|\mathfrak{M}^s|^2} }{ m\omega  }.
\label{sigmas}
  \end{align} 
Now we express $\roverline{|\mathfrak{M}^s|^2}$ in the
lab frame by using the eqs. \eqref{labframe}-\eqref{propagatorlab} as
 \be
\roverline{|\mathfrak{M}^s|^2}&=&\frac { 4A^{ 2 } }{ 3 } \left[ 4m(q^{ 0 })^{ 2 }+3m(q^{ 3 })^{ 2 }-7mq^{ 0 }q^{ 3 } \right]
.\label{squaredd}
 \ee
In the ultra-relativistic limit ($E'= { P}_{Z }$),
the real part of $ \roverline{|\mathfrak{M}^s|^2}$
becomes 
\begin{align}
\roverline{|\mathfrak{M}^s|^2}&=\frac { 4A^{ 2 } }{ 3 } [m^{ 2 }{ \omega  }]P_{ Z }\nonumber\\ &=e^{ 4 }\frac { { P }_{ Z } }{ 3\omega  } ,
\end{align} 
hence the crosssection due to the $s$-channel is 
 \begin{align}
 \sigma^{s} \com =e^4\frac { eB\pi ^{ 2 } } {4 \sqrt { 3 }  }
\int { \frac { { { d }{ P }_{ Z } } }{ { P }_ {Z}\omega ' } { \delta  }({ E' }
	+{ \omega'  }-{ E }-{ \omega  }) } \frac { \exp\left(
	\frac { -{ q_X }^{ 2 } }{ 2eB }  \right)
	\frac {{ P }_{ Z } }{3\omega }   }{ m\omega  }. 
\end{align}

Using the kinematic relations from the energy-momentum conservation 
in strong magnetic field limit ($p_\bot \simeq 0$) 
 \begin{align}
{ \omega'  }=\frac { {\omega  } }{ 1
+\frac { { \omega  }
(1-\cos\theta ) }{ m }  },\quad \quad
 { P }_{ Z }=\frac { (m{ \omega  }
 	+{ { \omega  } }^{ 2 })
 	(1-\cos\theta ) }{ m+{ \omega  }
 	(1-\cos\theta ) },\label{variables}
\end{align}
the above crosssection becomes 
\begin{align}
\sigma^{s} \com &=
e^4\frac { eB\pi ^{ 2 } }{12
	\sqrt { 3 } { m }\omega^{ 2 }  }
\exp\left( \frac { -{ 2q^{ 2 }_X } }{
	eB }  \right)  \int 
_{ 0 }^{ \pi  }\frac { (m+\omega )sin\theta  }{ m+\omega (1-\cos\theta ) }d\theta \nonumber\\&\times	 		 		 	 					
\delta\bigg[\sqrt{{\bigg[\frac{(m{\omega}+{{\omega}}^{2})(1-\cos\theta ) }
					{ m+{ \omega  } (1-\cos\theta ) 
					}\bigg ] }^{ 2 }+{ m }^{ 2 } }
-\frac { ({ { \omega  } }^{ 2 })(1-\cos\theta ) }{ m+{ \omega  } 
(1-\cos\theta ) } -m \bigg].	 \label{sigmass}	\end{align}

Using the property of Dirac-delta function 
	\begin{align}
	\delta \left[ f(x) \right] =\sum
	_{ i=1 }^{ n }{ \frac { \delta 
			(x-x_{ i }) }{ { f }^{ ' }(x_i) }  }, \label{delta}
	\end{align}
$\sigma^{s}$ becomes\footnote{which 
	is calculated in Appendix A.2},
	\begin{align} 
\sigma^{s} \com 
	=e^4
	\frac { eB\pi ^{ 2 } (m+\omega) }{ 
		12\sqrt { 3 } { m }\omega^{ 2 } } \exp
	\left( \frac { -{2 q^{ 2 } _X }
	}{ eB }  \right) \left[ \frac
	{ 1 }{ { { \omega  } }^{ 2 } } 
	-\frac { m+{ 2\omega  } }{ 9m
		{ { \omega  } }^{ 2 }+{ {
				2\omega  } }^{ 3 } }  
	\right],
  \end{align}
where the factor $\exp \left( \frac { -{2 q^{ 2 } _X }
	}{ eB }  \right)$ can be approximated to unity
in strong magnetic field limit. In the lab frame, considering
the direction of incoming photon along the direction of 
magnetic field (Z- direction), {\em i.e.} ${ q }_{ X }=p_{ X }+{ k }_{ X }
=m$, $ \sigma^{s}$ is further simplified into
\begin{align} 
\sigma^{s} \com =e^4
\frac { eB\pi ^{ 2 } (m+\omega) }{ 
	12\sqrt { 3 } { m }\omega^{ 2 } }\left[ \frac
{ 1 }{ { { \omega  } }^{ 2 } } 
-\frac { m+{ 2\omega  } }{ 9m
{ { \omega  } }^{ 2 }+{ {
2\omega  } }^{ 3 } } \right].
\end{align}
\subsection{Crosssection due to the $u$-channel diagram}

The crosssection due to the $u$-channel is given by
\begin{align}
\sigma^u \com & =\frac { eB\pi ^{ 4 } }{4 \sqrt { 3 }  } 
\int { \frac { { { d }{ P }_{ Z} } }{ { E' }
		\omega'} { \delta  }({ E '}
	+{ \omega ' }-{ E }-{ \omega  }
	) } \frac { \exp\left( \frac { -{2 q'^{ 2 }_X }
	}{ eB }  \right)\roverline{|\mathfrak{M}^u|^2}}{ m\omega  }.
\end{align}
Similar to the $s$-channel, $\roverline{|\mathfrak{M}^u|^2}$ 
has been translated into the lab frame by eqs. \eqref{labframe}-
\eqref{propagatorlab}
\begin{align}
\roverline{|\mathfrak{M}^u|^2}&=\frac { 4B^{ 2 } }{ 3 } \left[ 4m(q'^{ 0 })^{ 2 }+3m(q'^{ 3 })^{ 2 }-7mq'^{ 0 }q'^{ 3 } \right]\\
&=\frac { 4e^{ 4 } }{ 3 }\bigg [\frac { \left[ { m }^{ 3 }{ \omega  }^{ 2 }(4+3{ \cos ^{ 2 } \theta  }-7\cos  \theta )+{ m }^{ 3 }{ \omega  }(7\cos  \theta -8) (m+{ \omega  }(1-\cos  \theta )) \right] { \left[ m+{ \omega  }(1-\cos  \theta ) \right] ^{ 2 } } }{ { \left[ { m }^{ 2 }{ \omega  }^{ 2 }\left( 1-{ \cos   }^{ 2 }\theta  \right) -2{ m }^{ 2 }{ \omega \cos  \theta  \left( m+{ \omega  }(1-\cos  \theta ) \right)  } \right]  }^{ 2 } } \bigg]{ P }_{ Z }.
\end{align}
Substituting ${ { d } { P }_{ Z } }$, $\omega'$,
${ \delta  } ({ E '}+{ \omega  '} -{ E }-{ \omega  })$
from Eq.\eqref{variables}, the above matrix element squared gives the 
crosssection for the $u$-channel
\begin{align} 
\sigma^u & = e^4\frac { eB\pi ^{ 4 } } {3 \sqrt { 3 } { m }\omega  } 
	\exp\left( \frac { -2q^{ '2 }_X }
	{ eB }  \right) \times { \int 
		_{ 0 }^{ \pi  }\frac { (m+\omega )sin\theta  }{ m+\omega (1-\cos\theta ) }  } d\theta\nonumber\\\times&[\frac { \left[ { m }^{ 3 }{ \omega  }^{ 2 }(4+3{ \cos ^{ 2 } \theta  }-7\cos  \theta )+{ m }^{ 3 }{ \omega  }(7\cos  \theta -8)\times (m+{ \omega  }(1-\cos  \theta )) \right] \times { \left[ m+{ \omega  }(1-\cos  \theta ) \right] ^{ 2 } } }{ { \left[ { m }^{ 2 }{ \omega  }^{ 2 }\left( 1-{ \cos   }^{ 2 }\theta  \right) -2{ m }^{ 2 }{ \omega \cos  \theta \times \left( m+{ \omega  }(1-\cos  \theta ) \right)  } \right]  }^{ 2 } } ]{ P }_{ Z }\nonumber \\&\times	 		 		 	 					
\delta\bigg[ \sqrt { {\bigg [\frac 
		{	( m{ \omega  }+{ { \omega  } }^{ 2 }
			)(1-\cos\theta ) }
		{ m+{ \omega  }
			(1-\cos\theta ) 
		}\bigg ] }^{ 2 }+{ m }^{ 2 } }
-\frac { ({ { \omega  } }^{ 2 })(1-\cos\theta ) }{ m+{ \omega  }(1-\cos\theta ) } -m \bigg]	\label{sigmau} 		.
\end{align}	 		 					
	Using the same property of Dirac-delta function \eqref{delta},
the crosssection for the $u$-channel in the strong magnetic field
becomes
\begin{align}
\sigma^u =e^{ 4 }\frac { eB\pi ^{ 2 } }{ 12\sqrt { 3 }  }\exp\left( \frac { -2q^{ '2 }_X }
{ eB }  \right) \left[ \frac { 61{ m }^{ 2 }+78m{ \omega  }^{ 2 }+32{ \omega  }^{ 2 } }{ { m }^{ 2 }\omega [9{ m }{ \omega  }^{ 2 }+2{ \omega  }^{ 3 }] } 
-\frac { 1 }{ m{ \omega  }^{ 3 } }   \right].  
\end{align}
	In the lab frame, we consider the initial 
	direction of photon is along the direction of strong magnetic 
field, so, ${ q '}_{ X }$ (=$P_{ X }-{ k }_{ X }$) becomes
zero, therefore the $\sigma^u$ becomes
\begin{align}
\sigma^u =e^{ 4 }\frac { eB\pi ^{ 2 } }{ 12\sqrt { 3 }  } \left[ \frac { 61{ m }^{ 2 }+78m{ \omega  }^{ 2 }+32{ \omega  }^{ 2 } }{ { m }^{ 2 }\omega [9{ m }{ \omega  }^{ 2 }+2{ \omega  }^{ 3 }] } -
\frac { 1 }{ m{ \omega  }^{ 3 } }   \right].  
\end{align}
Thus the total crosssection for the Compton scattering (in units
of mb) is obtained as
\be
\sigma &=&\sigma ^{ s }+\sigma ^{ u }\nonumber\\&=& e^{ 4 }\frac { eB\pi ^{ 2 } }{ 12\sqrt { 3 }  } \left[ \frac { (m+\omega ) }{ { m }\omega ^{ 2 } } \left( \frac { 1 }{ { { \omega  } }^{ 2 } } -\frac { m+{ 2\omega  } }{ 9m{ { \omega  } }^{ 2 }+{ { 2\omega  } }^{ 3 } }  \right) +\frac { 61{ m }^{ 2 }+78m{ \omega  }^{ 2 }+32{ \omega  }^{ 2 } }{ { m }^{ 2 }\omega [9{ m }{ \omega  }^{ 2 }+2{ \omega  }^{ 3 }] } -\frac { 1 }{ m{ \omega  }^{ 3 } }  \right]. 
\ee
For the sake of comparison, we have also calculated the crosssection 
(in units of mb) for the Compton scattering in vacuum~\cite{Peskin:1995ev,Lahiri:2005sm} as
\be
\sigma_{\rm Vacuum} \com=\frac { { e }^{ 4 } }{ 8\pi m^{ 2 } }  \left[ \frac 
{ 1+r }{ { (1+2r) }^{ 2 } } 
+\frac { 2 }{ r^{ 2 } } 
-\frac { 2(1+r)-r^{ 2 } }
{ 2r^{ 3 } } ln(1+2r) \right], 
\ee
with a dimensionless variable, $ r=\frac{\omega}{m} $.

To see the effect of strong magnetic field on the Compton scattering, we have 
plotted the crosssection as a function of photon energy for the different strengths 
of magnetic fields in Figure 3, in addition to the crosssection in the 
vacuum only. We have found that the crosssection gets decreased in
the presence of strong magnetic filed. However, with the further increase
of strong magnetic field, the crosssection increases.

\begin{figure}[h]
	\centering
	\includegraphics[width=15cm,height=10cm]{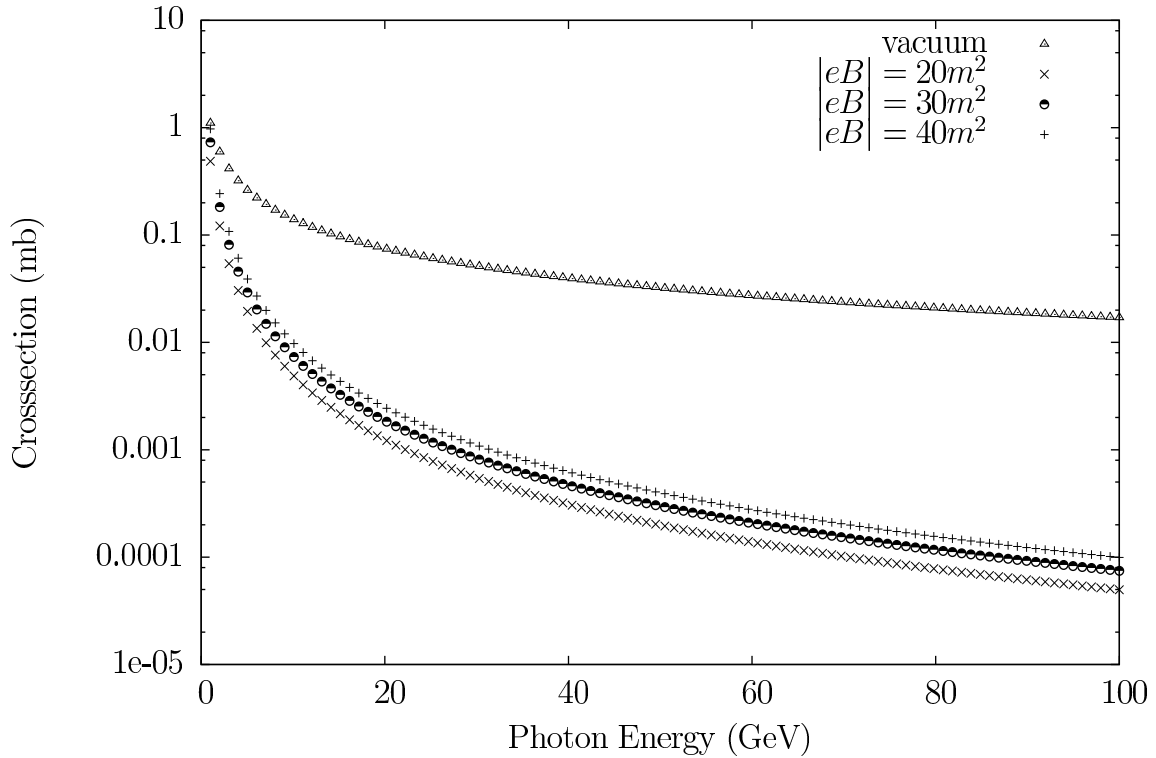}
	\caption{A plot between the crosssection and initial photon
		energy for different strengths of magnetic
		fields and vacuum (B=0) case.} 
\end{figure}
\section{Conclusions}
In neutron stars as well as in noncentral events of ultrarelativistic 
heavy ion collisions at Relativistic Heavy-ion Collider and Large Hadron 
Collider, a very strong magnetic field in the range $ 10^{16} - 10^{20}$ Gauss
is expected to be present. Also a very strong magnetic field 
($\sim 10^{23}$~Gauss) may have existed in the early universe.
Thus Compton scattering in strong magnetic field has huge importance
in astrophysics as well as in terrestrial laboratory, which 
therefore motivates us to revisit the Compton scattering in a strong magnetic 
field. For that purpose, using the Dirac spinor in strong magnetic field, we 
have first calculated the square of the S-matrix element and then
have summed (averaged) over the final (initial) states. Thereafter
we have illustrated the usual procedure to compute the 
crosssection in the lab-frame, where the photons are incident parallel to 
the magnetic field and the electron is at rest initially, by constructing
the Lorentz invariant phase space, incident flux factor, energy-momentum
conserving Dirac delta functions etc. in the strong magnetic field.
We have noticed that the Compton
scattering at the tree level gets suppressed due to the presence 
of strong magnetic field compared to the vacuum, when we increase the 
incident photon energy. But the crosssection increases linearly with the 
strength of magnetic field in strong magnetic field limit. In some
related works on Compton scattering in strong magnetic 
field \cite{ Gonthier:2014wja,Mushtukov:2015qul}, authors
have considered the resonant Compton scattering, spin-dependent 
influences etc. They have also considered the finite width of the 
Landau levels, so that electrons 
can excite to the higher Landau levels. But we consider the magnetic field 
to be strong enough so that no electron can excite to the higher landau 
levels, it  will remain only in its ground state.

\section {Acknowledgement}
We sincerely thank to Abhishek Tiwari, Mujeeb Hasan, Shubhalaxmi rath 
for their fruitful help in checking the manuscript. We are also thankful 
to Council of Scientific and Industrial Research (No: 03(1407)/17/EMR-II), 
Government of India for the financial support of this work.

	\appendix                                     
	\section{Matrix Element for Compton scattering:($\gamma   e^-
		\longrightarrow \gamma e^-$)\label{a.1}}
	This appendix contains detailed calculation of the squared matrix 
	element for the $s$-channel, the $u$-channel and the interference 
between them.
	\subsection{Matrix Element for $s$-channel process}
In this appendix  we will show the calculation of $\roverline{|\mathfrak{M}^s|^2}$. Then 
for the unpolarized crosssection, we need to average over the 
	quantum states of incoming particles and sum over the final states, 
	therefore we replace the above matrix element squared 
	\begin{equation}
	|\mathfrak{M}|^2 \rightarrow \frac{1}{(2s+1)(2r+1)}\sum_{all ~states}|\mathfrak{M}|^2
	\equiv \roverline{|\mathfrak{M}^s|^2}.\label{average}
	\end{equation} 
	We will now calculate the average square of the matrix element $\roverline{|\mathfrak{M}^s|^2}$ and then sum over the spin-states. Firstly we will do the spin-sum for the electron and
	 then after some simplification of $\roverline{|\mathfrak{M}^s|^2}$, we will do
	  the spin sum for the photon polarization vectors. Now the averaged squared matrix
	   element for $s$-channel process using(\refeq{matrixs})
	can be written as
	\begin{equation}
	\roverline{|\mathfrak{M}^s|^2}=\frac{A^{ 2 }}{6}
	\sum _{ r,r'} \sum _{ s,s' } \left[ \overline { U }^{s'}( P_{\slashed {Y}})
	{ \Gamma  }_
	{ 1 } { U }^{s}( p_{\slashed {Y}}) \right] 
	\left[ \overline { U 
	}^{s'}( P_{\slashed {Y}}){ \Gamma  }
	_{ 1 } { U }^{s}( p_{\slashed {Y}})
	\right] ^{ * }.\label{rs}
	\end{equation}
Since, $\left[ \overline { U }^{s'}( P_{\slashed {Y}}){ \Gamma  }_{ 1 } { U }^{s}( p_{\slashed {Y}}) \right] ^{ * }$ is a $1\times 1$  matrix, so it is just a complex number, so we can write complex conjugate
 equal to its  Hermitian conjugate.
	\begin{equation}
	\left[\overline { U }^{s'}( P_{\slashed {Y}})
	{ \Gamma  }_{ 1 } { U }^{s}( p_{\slashed {Y}})
	\right] ^{ * }=\left[\overline { U }^{s'}( P_{\slashed {Y}})
	{ \Gamma  }
	_{ 1 } { U }^{s}( p_{\slashed {Y}}) \right] ^{ \dagger  }
	=\quad \left[ { U^\dagger }^{s'}( P_{\slashed {Y}})
	{ { \gamma  }^{ 0 } }{ \Gamma  }
	_{ 1 }{ \gamma  }^{ 0 }
	{ \gamma  }^{ 0 } { U }^{s}( p_{\slashed {Y}})
	\right] ^{ \dagger  }.
	\end{equation}
	Using,
	\begin{equation}
	({ \gamma  }^{ 0 })^{ 2 }=1,
	\end{equation}
	the $1\times 1$ matrix becomes
	\begin{equation}
	\left[\overline { U }^{s'}( P_{\slashed {Y}}){ \Gamma  }
	_{ 1 } { U }^{s}( p_{\slashed {Y}}) \right]
	^{ * }=\left[ \overline  { U }^{s}( p_{\slashed {Y}})
	{ \Gamma  }_{ 1 }{ U }^{s'}( P_{\slashed {Y}}) \right].\label{conjugate}
	\end{equation}
 Using Eq. \eqref{conjugate}, $\roverline{|\mathfrak{M}^s|^2}$ becomes
	\begin{eqnarray}
	\roverline{|\mathfrak{M}^s|^2}=\frac{A^{ 2 }}{6}
	\sum _{ r,r'} \sum _{ s,s' }\left[\overline { U }^{s'}
	( P_{\slashed {Y}}){ \Gamma  }_{ 1 }
	{ U }^{s}( p_{\slashed {Y}}) \right]
	\left[ \overline  { U }^{s}( p_{\slashed {Y}})
	\overline { { \Gamma  }_{ 1 } } 
	{ U }^{s'}( P_{\slashed {Y}}) \right],
	\end{eqnarray}
	where 
	\begin{equation}
	{\overline  { \Gamma  }_{ 1 } } =
	{ { \gamma  }^{ 0 } }
	{ \Gamma^\dagger  }_{ 1 }
	{ \gamma  }^{ 0 }.
	\end{equation}
	Now, we need to do spin sum for both electrons and photons. First we will do 
	spin sum for electrons as follows 
	\begin{equation}
	\roverline{|\mathfrak{M}^s|^2}=\frac{A^{ 2 }}{6}\sum _{ r,r'}\bigg[\sum _{ s' } 
	\overline { U }^{s'}( P_{\slashed {Y}})
	{ \Gamma  }_{ 1 } 
	\sum _{ s }({ U }^{s}( p_{\slashed {Y}})
	\overline  { U }^{s}
	( p_{\slashed {Y}})) { \overline{
			\Gamma  }_{ 1 } } { U }^{s'}
	( P_{\slashed {Y}})\bigg].
	\end{equation}
	Simplifying $\roverline{|\mathfrak{M}^s|^2}$,
	\begin{equation}
	\roverline{|\mathfrak{M}^s|^2}=\frac{A^{ 2 }}{6}\sum _{ r,r'}\bigg[\sum _{ s' }
	\overline { U }^{s'}( P_{\slashed {Y}}) Q{ U }^{s'}
	( P_{\slashed {Y}})\bigg],
	\end{equation}
	where,	
	\begin{equation}
	Q={ \Gamma  }_{ 1 }\sum_{ s } 
	{ U }^{s}( p_{\slashed {Y}})\overline 
	{ U }^{s}( p_{\slashed {Y}})
	{ \overline{ \Gamma  }_{ 1 } }.\label{qq}
	\end{equation}\
	Since, $\roverline{|\mathfrak{M}^s|^2}$ is a number, we can multiply all the elements explicitly as
	\begin{equation}
	\roverline{|\mathfrak{M}^s|^2}=\frac{A^{ 2 }}{6}\sum _{ r,r'}\bigg[\sum _{ k,j }
	\sum _{ s'} \overline { U_k }^{s'}( P_{\slashed {Y}})
	Q_{ kj } { U_j }^{s'}( P_{\slashed {Y}})\bigg].
	\end{equation}
	Now, taking spin sum over final state of electron spinor
	\begin{eqnarray}
	\roverline{|\mathfrak{M}^s|^2}&=&
	\frac{A^{ 2 }}{6}\sum _{ r,r'}\sum _{ kj }Q_{ kj }
	\sum _{s' }
	\overline { U_k }^{s'}( P_{\slashed {Y}}) 
	{ U_j }^{s'}( P_{\slashed {Y}}),\nonumber\\
	&=&\frac{A^{ 2 }}{6}\sum _{ r,r'}\sum _{ k,j } \left[ \sum _{s' }
	{ Q{ U }^{s'}(P_{ { }{\slashed Y } }) } 
	\overline { U }^{s'}(P_{ {\slashed Y } })
	\right] _{ jk },\nonumber\\
	&=&\frac{A^{ 2 }}{6}\sum _{ r,r'}\sum _{ k } 
	\left[ \sum _{ s' }{ Q{ U }^{s'}(P_{ {\slashed Y } }) }
	\overline { U }^{s'}(P_{ {\slashed Y } }) \right] 
	_{ kk },\nonumber\\
	&=&\frac{A^{ 2 }}{6}\sum _{ r,r'}Tr\left[ Q\sum _{s' }{ { U }^{s'}(P_
		{ {\slashed Y } }) } \overline
	{ U }^{s'}(P_{ {\slashed Y } }) \right].
	\end{eqnarray}
	Now, we need to simplify the value of Q. To do this first simplify 
$ {\overline \Gamma}(1)$ 
	\begin{equation}
	{ {\overline \Gamma  }_{ 1 } } =
	{ { \gamma  }^{ 0 } }
	{ \Gamma  }_{ 1 }^{ \dagger  }
	{ \gamma  }^{ 0 }={ { \gamma  }^{ 0 } [ }{ \varepsilon  }
	_{ \mu }^{r' \ast  }(K)
	{ \gamma  }^{ \mu }
	{ (1+{ \gamma  }^{ 0 }
		{ \gamma  }^{ 3 }
		{ \gamma  }^{ 5 })
		({ \gamma  }^{ 0 }
		{ q }_{ 0 }-{ \gamma  }^{ 3 }
		{ q }_{ 3 })
	}){ \gamma  }^{ \nu }
	{ \varepsilon  }_{ \nu }^{r}(k)]
	^{ \dagger  } 
	{ \gamma  }^{ 0 }.
	\end{equation}
	Using the following identies of gamma matrices, 
	\begin{equation}
	({ \gamma  }^{ 0 })^{ 2 }=I,\quad 
	\quad \quad { \gamma  }^{ 0 }({ \gamma  }^{
		{ \nu }^{ \dagger  } }
	){ \gamma  }^{ 0 }
	={ \gamma  }^{ \nu },\quad  \quad { \gamma  }
	^{ \mu }{ \gamma  }^{ \nu }
	+{ \gamma  }^{ \nu }
	{ \gamma  }^{ \mu }=2{ g }^{ \mu \nu },
	\end{equation}
	we obtain,
	\begin{equation}
	{ {\overline \Gamma  }_{ 1 } }
	={ \varepsilon  }_{ \mu }^{ r'\ast  }(K){ \varepsilon  }_{ \nu }^{r}(k)
	{ \gamma  }^{ \nu }
	({ \gamma  }^{ 0 }{ q }_{ 0 }-{ \gamma  }
	^{ 3 }{ q }_{ 3 })
	{ (1-{ \gamma  }^{ 5 }
		{ \gamma  }^{ 3 }
		{ \gamma  }^{ 3 }) }
	{ \gamma  }^{ \mu }.\label{gammabar}
	\end{equation}
	Now, using the spin sum from \eqref{spinsum} and  substituting 
${ \Gamma  }_{ 1 }$  and $\overline{ \Gamma  }_{ 1 }$ from \eqref{gamma} 
and \eqref{gammabar}, respectively, into Q \eqref{qq} and then substituting
it in \eqref{ssquared}, we finally obtain \eqref{averaged M_s}. Taking the 
spin sum over the photon polarization index $r$ and $r'$ and using
	\begin{equation}
	\sum _{ r}^{ \quad  }
	{ { \varepsilon  }_{ \mu }^{r\ast  } }
	{ \varepsilon  }
	_{ \lambda  }^{r}=-{ g }_{ \mu\lambda  },
	\end{equation}
Eq. \eqref{averaged M_s} becomes
	\begin{equation}
	\roverline{|\mathfrak{M}^s|^2}
	=\frac{{ A }^{ 2 }}{24} Tr\bigg[[{ \gamma  }
	^{ \mu }{ (1+{ \gamma  }^{ 0 }{ \gamma  }^{ 3 }
		{ \gamma  }^{ 5 })
		{\slashed q_\parallel } }
	({ \gamma  }^{ \nu })(\slashed
	{ p }_{ \parallel  }
	+\widetilde {\slashed p_{ \parallel  } } 
	{ \gamma  }^{ 5 })]
	\nonumber\\ \times[({ \gamma _{ \nu } })
	{\slashed q_\parallel }{ (1-{ \gamma  }^{ 5 }
		{ \gamma  }^{ 3 }{ \gamma  }^{ 0})
		{ \gamma _{ \mu }(\slashed{ P }_{ \parallel 
			} }+\widetilde {\slashed
			P_{ \parallel  } }
		{ \gamma  }^{ 5 })}]\bigg].
	\end{equation}
Now, applying cyclic properties of traces and the following properties 
	\begin{equation}
	{ \gamma  }_{ \mu }  \slashed a{ \gamma  }
	^{ \mu }=-2\slashed a
	\quad and\quad { \gamma  }^{ 5 }
	{ \gamma  }^{ \nu }+{ \gamma  }^{ \nu }
	{ \gamma  }^{ 5 }=0 ,
	\end{equation}
and multiplying it carefully, we get
\be
\roverline{|\mathfrak{M}^s|^2}& =& \frac{{ A }^{ 2 }}{6}\Big[{ \widetilde p }
_{\parallel  \alpha  }{ \widetilde P }_{ \parallel \epsilon  }
{ q }_{\parallel \eta  }
{ q }_{\parallel \theta  }
Tr({ \gamma  }^{ \eta  }
{ \gamma  }^{ \alpha  }{ \gamma  }
^{ 5 }{ \gamma  }^{ \theta  }
{ \gamma  }^{ \epsilon  }{ \gamma  }^{ 5 })
-{ \widetilde p }_{\parallel  \alpha  }{ P }_{\parallel \mu }
{ q }_{ \parallel\eta  }
{ q }_{\parallel \theta  }
{\rm Tr} ({ \gamma  }^{ \eta  }
{ \gamma  }^{ \alpha  }{ \gamma  }^{ 5 
}{ \gamma  }^{ \theta  }
{ \gamma  }^{ \nu })\nonumber\\
& +& { \widetilde p }_{ \parallel \alpha  }{ \widetilde P }_
{ \parallel \nu }{ q }_{ \parallel\eta  }
{ q }_{ \parallel\theta  } {\rm Tr} ({ \gamma  }
^{ \eta  }{ \gamma  }^{ \alpha  }
{ \gamma  }^{ 5 }{ \gamma  }
^{ \theta  }{ \gamma  }^{ 3 }
{ \gamma  }^{ 0 }{ \gamma  }
^{ \nu })+{ \widetilde p }_{ \parallel \alpha  }{ P }
_{\parallel \rho  }{ q }
_{\parallel \eta  }{ q }_
{\parallel \theta  }Tr(
{ \gamma  }^{ \eta  }{ \gamma  
}^{ \alpha  }{ \gamma  }^{
5 }{ \gamma  }^{ \theta  }{
{ \gamma  }^{ 5 }\gamma  }^{ 3 }{ \gamma  }^{ 0 }
{ \gamma  }^{ \rho  })\nonumber \\ 
&-& { p }_{\parallel \beta  }
\widetilde P_{_\parallel \epsilon  }{ q }_
{ \parallel\eta  }{ q }
_{\parallel \theta  }
{\rm Tr} ({ \gamma  }^{ 5 }
{ \gamma  }^{ \eta  }{ \gamma  }
^{ \beta  }{ \gamma  }
^{ \theta  }{ \gamma  }^{ \epsilon  })+{
	p }_{\parallel \beta  }
P_{ \parallel \nu }
{ q }_{ \parallel\eta  }
{ q }_{ \parallel
	\theta  }(a)Tr({ \gamma  }^{ \eta  }
{ \gamma  }^{ \beta  }{ \gamma  }^{ \theta  }
{ \gamma  }^{ \nu })\nonumber \\ 
& - & p_{\parallel \beta  }\widetilde P_
{ _\parallel \nu }{ q }_{ \parallel\eta  }
{ q }_{\parallel \theta  }
{\rm Tr} ({ \gamma  }^{ \eta}
{ \gamma  }^{ \beta  }{ \gamma  }
^{ \theta  }{ \gamma  }^{ 3 }{ \gamma  }			    
^{ 0 }{ \gamma  }^{ \nu })-p_{\parallel 
	\beta  }P_{\parallel \rho  }{ q }_
{ \parallel\eta  }{ q }_{ \parallel\theta  }
(a)Tr({ \gamma  }^{ \eta  }{ \gamma  }^{ \beta  }
{ \gamma  }^{ \theta  }{
	{ \gamma  }^{ 5 }\gamma
}^{ 3 }{ \gamma  }^{ 0 }
{ \gamma  }^{ \rho  })
\nonumber \\  
&+&  {\widetilde P}_
{_\parallel \epsilon  }{ \widetilde p }_
{_\parallel \lambda  }{ q }_
{\parallel \eta  }{ q }
_{\parallel \theta  }
{\rm Tr} ({ \gamma  }^{ 5 }{
	{ \gamma  }^{ 0 }{ \gamma  }^{ 3 }\gamma  }^
{ \eta  }{ \gamma  }^{ \lambda  }{ \gamma  }^
{ \theta  }{ \gamma  }^{ \epsilon  })-{ \widetilde p }
_{_\parallel \lambda  }{ P }
_{\parallel \mu }{ q }_
{ \parallel\eta  }
{ q }_{\parallel			    	
	\theta  } {\rm Tr} ({ { \gamma  }^{ 0 }{
		\gamma  }^{ 3 }\gamma  }^{ \eta  }
{ \gamma  }^{ \lambda  }{ \gamma  }^{ \theta  }
{ \gamma  }^{ \mu })\nonumber \\  
&+&  { \widetilde p 
}_{ \lambda  }{ \widetilde P }_{_\parallel \nu }
{ q }_{ \parallel\eta  }
{ q }_{\parallel \theta  }
{\rm Tr} ({ { \gamma  }^{ 0 }
	{ \gamma  }^{ 3 }\gamma  }^{ \eta  }{ \gamma  }
^{ \lambda  }{ \gamma  }^{ \theta  }{
	{ \gamma  }^{ 3 }{ \gamma  }
	^{ 0 }\gamma  }^{ \nu })\nonumber\\
&+& { \widetilde{p} }_{_\parallel \lambda  
}{ P }_{\parallel \rho  }{ q }
_{\parallel \eta  }{ q }_{
	\parallel\theta  }  
{\rm Tr} ({ { \gamma  }^{ 0 }{ \gamma  }
	^{ 3 }\gamma  }^{ \eta  }{ \gamma  }^
{ \lambda  }{ \gamma  }^{ \theta  }
{ { \gamma  }^{ 5 }{ \gamma  }^{ 3 			    	 	  		
	}{ \gamma  }^{ 0 }\gamma  }^{ \rho  }) \nonumber\\
&-& { p }_{\parallel \delta  }{ \widetilde P}_
{ _\parallel\epsilon  }{ q }_{ \parallel\eta  }
{ q }_{ \parallel\theta  } {\rm Tr}
({ { \gamma  }^{ 0 }{ \gamma  }^{ 3 }
	{ \gamma  }^{ 5 }\gamma  }^{ \eta  }
{ \gamma  }^{ \delta  }{ \gamma  }
^{ \theta  }{ { \gamma  }^{ \epsilon  }
	{ \gamma  }^{ 5 } })\nonumber\\
&+&  { P}_{\parallel \nu }{ p }_{\parallel
	\delta  }{ q }_{\parallel \eta  }
{ q }_{\parallel \theta  }Tr
({ \gamma  }^{ 5 }{ { \gamma  }^{ 0 }
	{ \gamma  }^{ 3 }\gamma  }^{ \eta  }{
	\gamma  }^{ \delta  }{ \gamma  }^
{ \theta  }{ \gamma  }^{ \mu })-{ \widetilde p }
_{_\parallel \delta  }{ P}_{\parallel \nu }{ q }_{\parallel \eta  }{ q }_			    	 	  		 		 	
{ \parallel\theta  }Tr({ { \gamma  }
	^{ 0 }{ \gamma  }^{ 3 }{ \gamma  }
	^{ 5 }\gamma  }^{ \eta  }{ \gamma  
}^{ \delta  }{ \gamma  }^{ \theta  }
{ { \gamma  }^{ 3 }
	{ \gamma  }^{ 0 }\gamma  }^{ \nu })
\nonumber \\
&-  & { p }_{\parallel 
	\delta  }{ P }_{\parallel \rho  }{ q }_{ \parallel\eta  }{ q }_
{\parallel \theta  }Tr({ 
	{ \gamma  }^{ 0 }{ \gamma  }
	^{ 3 }{ \gamma  }^{ 5 }\gamma  }
^{ \eta  }{ \gamma  }^{ \delta  }
{ \gamma  }^{ \theta  }{
	{ { \gamma  }^{ 5 }
		\gamma  }^{ 3 }{ \gamma  }^{ 0 }
	\gamma  }^{ \rho  })\Big].	\label{sqav}		    	 	  		 		 	 \ee

 We obtain sixteen terms as shown below.
	
	\subsubsection{\underline{Calculation of all sixteen Terms }}
	\underline{Term-1}
	\begin{align}
	{ \widetilde p }_{\parallel  \alpha  }
	{ \widetilde P}
	_{\parallel \epsilon }{ q }
	_{ \parallel\eta  }{ q }
	_{ \parallel\theta  } {\rm Tr} ({ \gamma  }^
	{ \eta  }{ \gamma  }^
	{ \alpha  }
	{ \gamma  }^{ \theta  }
	{ \gamma  }^{ \epsilon})
	=8(\widetilde p_\parallel.q_{\parallel})
	(\widetilde P_\parallel.q_{\parallel})
	-4{ q_{\parallel}^{ 2 } }
	( \widetilde p_\parallel.\widetilde P_\parallel).\label{term1}
	\end{align}
	\underline{Term-2 }
	\begin{align}
	-{ \widetilde p }_{_\parallel \alpha  }{ P}
	_{\parallel \nu }
	{ q }_{ \parallel\eta  }
	{ q }_{\parallel
		\theta  } {\rm Tr} ({ \gamma  }
	^{ \eta  }
	{ \gamma  }^{ \alpha  }
	{ \gamma  }^
	{ 5 }{ \gamma  }^{ \theta  }
	{ \gamma  }^{ \nu })=0.
	\end{align}
	\underline{Term-3}
	\begin{align}
	{\widetilde p }_{_\parallel \alpha  }
	{ \widetilde P }_{ _\parallel \nu }{ q }_
	{ \parallel\eta  }{ q }_{\parallel \theta  }
	{ \rm Tr} ({ \gamma  }^
	{ \eta  }{ \gamma  }^
	{ \alpha  }
	{ \gamma  }^{ 5 }
	{ \gamma  }^{ \theta  }
	{ \gamma  }^{ 3 }{ \gamma  }^
	{ 0 }{ \gamma  }^{ \nu })=0.
	\end{align}
	\underline{Term-4}
	\begin{align}
	&{ \widetilde p }_{_\parallel \alpha  }{ P }_
	{ \parallel\rho  }
	{ q }_{\parallel \eta  }
	{ q }_{
		\parallel \theta  } 
	{\rm Tr} ({ \gamma 
	}^{ \eta  }{ \gamma  }^{ \alpha  }
	{ \gamma  }^{ 5 }
	{ \gamma  }^{ \theta  }
	{ { \gamma  }^{ 5 }\gamma  }^{ 3 }
	{ \gamma  }^{ 0 
	}{ \gamma  }^{ \rho  })\nonumber\\&=
	-8(\widetilde p_\parallel.q_\parallel)
	{ q }^{ 3 }{ P }^{ 0 }
	+8(\widetilde p_\parallel.q_\parallel){ q }^{ 0 }
	{ P }^{ 3 }-4{ q^{ 2 } _\parallel}
	{ \widetilde p }^{ 0 }{ P}^{ 3 }+4{ q_\parallel^{ 2 } }
	{ P}^{ 0 }{ \widetilde p }^{ 3 }.
	\end{align}
	\underline{Term-5}
	\begin{align}
	{ -p }_{ \parallel\beta  }\widetilde P_{_\parallel \epsilon  }
	{ q }_{\parallel \eta  }{  q }_
	{ \parallel\theta  } {\rm Tr} ({ \gamma  }^
	{ 5 }{ \gamma  }^{ \eta  }{ \gamma  }^
	{ \beta  }{ \gamma  }^{ \theta  }
	{ \gamma  }^{ \epsilon  })=0.
	\end{align}
	\underline{Term-6}
	\begin{gather}
	{ p }_{  \parallel\beta  }P_
	{ \parallel \nu }{ q }_{ \parallel \eta  }{ q }
	_{ \parallel \theta  } {\rm Tr} ({ \gamma  }
	^{ \eta  }{ \gamma  }^{ \beta  }{ \gamma  }
	^{ \theta  }{ \gamma  }^{ \nu })
	=8(q_ \parallel.{ p_ \parallel })(q_ \parallel.{
		P_ \parallel })-4{ q^2_ \parallel }
	{ (P_ \parallel }.{ p_ \parallel }).
	\end{gather}
	\underline{Term-7}
	\begin{align}
	&{ -p }_{ \parallel \beta  }\widetilde P_
	{_\parallel \nu }{ q }_{ \parallel \eta  }{
		q }_{  \parallel\theta  } {\rm Tr}
	({ \gamma  }^{ \eta  }{ \gamma  }
	^{ \beta  }{ \gamma  }^{ \theta  }
	{ \gamma  }^{ 3 }{ \gamma  }^{ 0 }
	{ \gamma  }^{ \nu })\nonumber
	\\&=8(p_ \parallel.q_ \parallel){ q }
	^{ 0 }{ \widetilde P}^{ 3 }-8(p_ \parallel.q_
	\parallel){ q }^{ 3 }{ \widetilde P }
	^{ 0 }+4{ q_ \parallel}^2{ \widetilde P }^{ 0 }
	{ p }^{ 3 }-4{ q^2_ \parallel }
	{ \widetilde P }^{ 0 }{ p }^{ 3 }.
	\end{align}
	\underline{Term-8}
	\begin{align}
	{ -p }_{  \parallel\beta  }P_
	{ \parallel \rho  }{ q }_{ \parallel
		\eta  }{q }_{  \parallel\theta  }
	{\rm Tr} ({ \gamma  }^{ \eta  }{ \gamma  }
	^{ \beta  }{ \gamma  }^{ \theta  }
	{ { \gamma  }^{ 5 }\gamma  }^{ 3 }
	{ \gamma  }^{ 0 }{ \gamma  }^{ \rho  })=0.
	\end{align}
	\underline{Term-9}
	\begin{align}
	{ \widetilde P }_{_\parallel \epsilon  }
	{ \widetilde p }_{_\parallel \lambda  } 
	{ q }_{ \parallel \eta  }{q }_
	{  \parallel\theta  } {\rm Tr} ({ \gamma  }
	^{ 5 }{ { \gamma  }^{ 0 }{ \gamma  }
		^{ 3 }\gamma  }^{ \eta  }{ \gamma  }^
	{ \lambda  }{ \gamma  }^{ \theta  }
	{ \gamma  }^{ \epsilon  })=0.
	\end{align}
	\underline{Term-10}
	\begin{align}
	&-{ \widetilde p }_{_\parallel \lambda  }{ P}_{ \parallel \nu }
	{ q }_{ \parallel \eta  }{q }
	_{ \parallel \theta  } {\rm Tr} ({ { \gamma  }
		^{ 0 }{ \gamma  }^{ 3 }\gamma  }^{ \eta  }
	{ \gamma  }^{ \lambda  }{ \gamma  }^{ \theta  }
	{ \gamma  }^{ \nu })\nonumber\\&
	=8(\widetilde p_\parallel.q_\parallel { q }^{ 0 }{ P }^{ 3 }-4
	q_\parallel^{ 2 }{ \widetilde p}^{ 0 }{ P }^{ 3 }
	+4q_\parallel^{ 2 }{ \widetilde p }^{ 3 }{ P }^
	{ 0 }-8(\widetilde p_\parallel.q_\parallel){ q }^{ 3 }{ P }^{ 0 }.
	\end{align}
	\underline{Term-11}
	\begin{align}
	&{ \widetilde p }_{_\parallel \lambda  }{ \widetilde P }_
	{ _\parallel \nu }{ q }_{ \parallel \eta  }{ q }
	_{ \parallel \theta  } {\rm Tr}
	({ { \gamma  }^{ 0 }{ \gamma  }^{ 3 }
		\gamma  }^{ \eta  }{ \gamma  }^{
		\lambda  }{ \gamma  }^{ \theta  }
	{ { \gamma  }^{ 3 }{ \gamma  }^{ 0 }
		\gamma  }^{ \nu })\nonumber\\&
	=16(\widetilde p_\parallel.q_\parallel){ q }
	^{ 0 }{ \widetilde P }^{ 0 }
	-8q_\parallel^{ 2 }{ \widetilde p }^{ 0 }{ \widetilde P }^
	{ 0 }-16(\widetilde p_\parallel.q_\parallel){ q }^{ 3 }{ \widetilde P }
	^{ 3 }+8q_\parallel^{ 2 }{ \widetilde p }^{ 3 }{ \widetilde P}^{ 3 }
	-8(\widetilde p_\parallel.q_\parallel)(\widetilde P_\parallel.q_\parallel)
	+4(q_\parallel)^{ 2 }(\widetilde p_\parallel.\widetilde P_\parallel).
	\end{align}	 
	\underline{Term-12}
	\begin{gather}
	{ \widetilde p }_{ _\parallel\lambda  }{ P }_{ \parallel \rho  }
	{ q }_{  \parallel\eta  }{  q }_{ \parallel \theta  }
	{\rm Tr} ({ { \gamma  }^{ 0 }{ \gamma  }^{ 3 }
		\gamma  }^{ \eta  }{ \gamma  }^{ \lambda  }
	{ \gamma  }^{ \theta  }{ { \gamma  }^{ 5 }
		{ \gamma  }^{ 3 }{ \gamma  }^{ 0 }
		\gamma  }^{ \rho  })=0.
	\end{gather}
	\underline{Term-13}
	\begin{align}
	{ p }_{ \parallel \delta  }{ \widetilde P }_{_\parallel \epsilon  }
	{ q }_{ \parallel \eta  }{q }_{  \parallel\theta  } 
	{\rm Tr} ({ { \gamma  }^{ 0 }{ \gamma  }^{ 3 }
		{ \gamma  }^{ 5 }\gamma  }^{ \eta  }
	{ \gamma  }^{ \delta  }{ \gamma  }^
	{ \theta  }{ { \gamma  }^{ \epsilon  }
		{ \gamma  }^{ 5 } })=0.
	\end{align}
	\underline{Term-14}
	\begin{align}
	{ P }_{  \parallel \nu }{ p }_{ \parallel \delta  }
	{ q }_{ \parallel \eta  }{ q }_{ \parallel \theta  }
	{\rm Tr} ({ \gamma  }^{ 5 }{ { \gamma  }
		^{ 0 }{ \gamma  }^{ 3 }\gamma  }
	^{ \eta  }{ \gamma  }^{ \delta  }
	{ \gamma  }^{ \theta  }{ \gamma  }^{ \mu })=0.
	\end{align}
	\underline{Term-15}
	\begin{align}
	-{ \widetilde p }_{ _\parallel\delta  }{ P }_{ \parallel \nu }{ q }_{  \parallel\eta  }
	{ q }_{  \parallel\theta  } {\rm Tr} ({ { \gamma  }
		^{ 0 }{ \gamma  }^{ 3 }{ \gamma  }^{ 5 }\
		\gamma  }^{ \eta  }{ \gamma  }^{ \delta  }
	{ \gamma  }^{ \theta  }{ { \gamma  }
		^{ 3 }{ \gamma  }^{ 0 }\gamma  
	}^{ \nu })=0.
	\end{align}
	\underline{Term-16}
	\begin{align}
	&-{ p }_{ \parallel \delta  }{ P }_{ \parallel \rho  }
	{ q }_{ \parallel \eta  }{ q }_{ \parallel \theta  } {\rm Tr} 
({ { \gamma  }^{ 0 }{ \gamma  }^{ 3
		}{ \gamma  }^{ 5 }\gamma  }^{
		\eta  }{ \gamma  }^{ \delta  }
	{ \gamma  }^{ \theta  }{ {
			{ \gamma  }^{ 5 }\gamma  }^{ 3 }
		{ \gamma  }^{ 0 }\gamma  }^{ \rho  })\nonumber 
	\\& =
	16(p_\parallel.q_\parallel){ q }^{ 0 }{ P }^{ 0 }
	-8q_\parallel^{ 2 }{ p }^{ 0 }{ P }^{ 0 } 
	-16(p_\parallel.q_\parallel){ q }^{ 3 }{ P }^
	{ 3 }+8q_\parallel^{ 2 }{ p }^{ 3 }
	{ P}^{ 3 }-8(p_\parallel.q_\parallel)(P_\parallel
	.q_\parallel)+4q_\parallel^{ 2 }(p_\parallel.P_\parallel).\label{term16}
	\end{align}
	Adding all the  sixteen terms from \eqref{term1}-\eqref{term16}, 
Eq. \eqref{sqav} becomes Eq. \eqref{s-matrix squared}.
	\subsection{Interference-Term of Matrix element}
 Now  we calculate the $\roverline{\mathfrak{M}^s\mathfrak{M}^{u*}}$, for this we write 
the matrix elements for $s$-channel and $u$-channel process individually as,
	\begin{eqnarray}
	\mathfrak{M}^s=A\left[  \overline { U }^{s'}( P_{\slashed {Y}})
	{ \Gamma  }_{ 1 } { U }^{s}( p_{\slashed {Y}}) \right] ,
	\\\nonumber \mathfrak{M}^u=B\left
	[  \overline { U }^{s'}( P_{\slashed {Y}}){ \Gamma  }_{ 2 }  { U }^{s}( p_{\slashed {Y}}) \right], 
	\end{eqnarray}
	so first interference term
	$\mathfrak{M}^s\mathfrak{M}^{u*}$
	can be written as
	\begin{align}
	\mathfrak{M}^s\mathfrak{M}^{u*}
	=AB\left[  \overline { U }^{s'}( P_{\slashed {Y}}){ \Gamma  }
	_{ 1 } { U }^{s}( p_{\slashed {Y}}) \right] { \left[  \overline { U }^{s'}( P_{\slashed {Y}}){ \Gamma  }_{ 2 }  { U }^{s}( p_{\slashed {Y}}) \right]  }^{ * }.
	\end{align}
	
	Since, $\left[ \overline { U } ^{s'}(P_\slashed Y){ \Gamma  }_{ 2 }U^{s}(p_\slashed Y) \right] ^{ * }$ 
	is a $1\times 1$  matrix, so it is just a complex number, so we can write complex conjugate 
	equal to its  Hermitian conjugate, so
	\begin{align}
	\left[  \overline { U }^{s'}( P_{\slashed {Y}}){ \Gamma  }_
	{ 2 }  { U }^{s}( p_{\slashed {Y}}) \right] ^{ * }=\left[  \overline { U }^{s'}( P_{\slashed {Y}}){ \Gamma  }_{ 2 }  { U }^{s}( p_{\slashed {Y}}) \right] 
	^{ \dagger  }=\quad \left[  { U^{ \dagger  } }^{s'}( P_{\slashed {Y}})
	{ { \gamma  }^{ 0 } }{ \Gamma  }_{ 2 }
	{ \gamma  }^{ 0 }{ \gamma  }^
	{ 0 } { U }^{s}( p_{\slashed {Y}}) \right] ^{ \dagger  }.
	\end{align}
	Then, first interference term $\mathfrak{M}^s\mathfrak{M}^{u*}$ becomes
	\begin{equation}
	\mathfrak{M}^s\mathfrak{M}^{u*}
	=AB\left[\overline { U }^{s'}( P_{\slashed {Y}}){
		\Gamma  }_{ 1 }{ U }^{s}( p_{\slashed {Y}}) \right]
	\left[ \overline { U }^{s}( p_{\slashed {Y}})
	{ { \Gamma  }_{ 2 } } \overline { U }^{s'}( P_{\slashed {Y}}) \right].
	\end{equation}
	
	Now we need to do spin sum for both electrons and photons. First we 
will do the spin sum for electrons giving spinors as indices $s$ and $s'$ for 
the initial and final state of electrons, respectively and then for 
initial (r) and final state (r') for photon polarization and taking 
	average over $\mathfrak{M}^s\mathfrak{M}^{u*}$ as we have 
done in Eq. \eqref{average}, we can write
	\begin{align}
	\roverline{\mathfrak{M}^s\mathfrak{M}^{u*}}=\frac{AB }{6}\sum_{r,r'}\sum_{s,s'}
	\left[ \overline { U }^{s'}( P_{\slashed {Y}}){ \Gamma  }
	_{ 1 } { U }^{s}( p_{\slashed {Y}}) \right] 
	\bigg[\overline { U }^{s}( p_{\slashed {Y}})\overline{\Gamma}_2  { U }^{s'}( P_{\slashed {Y}})\bigg],
	\end{align}
	where
	\begin{align}
	{ {\overline \Gamma  }_{ 1 } }
	={ \varepsilon  }_{ \mu }^{r'*}(K)
	{ \varepsilon  }^{r}_{ \nu }(k)
	{ \gamma  }^{ \nu }({ \gamma  }^{ 0 }{ q }_{ 0 }
	-{ \gamma  }^{ 3 }{ q }_{ 3 })
	{ (1-{ \gamma  }^{ 5 }{ \gamma  }
		^{ 3 }{ \gamma  }^{ 0}) }{ \gamma  }^{ \mu },
	\end{align}
	\begin{align}
	{ {\overline \Gamma  }_{ 2 } } 
	={ \varepsilon  }_{ \nu }^{r'}(K)
	{ \varepsilon  }^{r*}_{ \mu }(k)
	{ \gamma  }^{ \nu }({ \gamma  }^{ 0 }
	{ q' }_{ 0 }-{ \gamma  }^{ 3 }
	{ q '}_{ 3 }){ (1-{ \gamma  }^
		{ 5 }{ \gamma  }^{ 3 }
		{ \gamma  }^{ 0}) }{ \gamma  }^{ \mu}.
	\end{align}
	Now, First we will do the spin sum for initial state of electron
	\begin{align}
	\roverline{\mathfrak{M}^s\mathfrak{M}^{u*}}=\frac{AB }{6}\sum_{r,r'}\bigg[\sum _{ s' } 
	\overline { U }^{s'}( P_{\slashed {Y}}){ \Gamma  }_{ 1 }
	\sum _{ s } \big[ { U }^{s}( p_{\slashed {Y}}) 
	\overline { U }^{s}( p_{\slashed {Y}})\big] \overline\Gamma_2 { U }^{s'}( P_{\slashed {Y}})\bigg],
	\end{align}
	\begin{align*}
	=\frac{AB }{6}\sum_{r,r'}\bigg[\sum _{ s' } \overline { U }^{s'}( P_{\slashed {Y}}) Q  { U }^{s'}( P_{\slashed {Y}})\bigg],
	\end{align*}
	where,
	\begin{align}
	Q= { \Gamma  }_{ 1 }\sum _{s}
	[U^{s}(p_\slashed {Y})\overline { U } ^{s}(p_\slashed {Y})]
	{ {\overline  \Gamma  }_{2} }\label{qq'}.
	\end{align}
	Since, $\roverline{\mathfrak{M}^s\mathfrak{M}^{u*}}$  is a number we can multiply all the elements explicitly as
	\begin{align}
	\roverline{\mathfrak{M}^s\mathfrak{M}^{u*}}&=\frac{AB }{6}\sum_{r,r'}\bigg[\sum _{ i,j }\sum _
	{ s' } \overline { U } _
	{ i }^{s'}(P_\slashed {Y})  Q_{ ij }U_{ j }^{s'}(P_\slashed {Y}) \bigg],\nonumber\\
	&= \frac{AB }{6}\sum_{r,r'}\bigg[Q_{ ij }\sum _
	{  s' } \overline { U }
	_{ i }^{s'}(P_\slashed {Y})U_{ j }^{s'}(P_\slashed {Y})\bigg],\nonumber\\
	& = \frac{AB }{6}\sum_{r,r'}\bigg[\sum _{ i,j } \sum
	_{  s'  }Q U^{s'}(P_\slashed {Y})  \overline
	{ U } ^{s'}(P_\slashed {Y}) _{ ji }\bigg],\nonumber\\
	& =\frac{AB }{6}\sum_{r,r'} \sum _{ i }\bigg [\sum 
	_{ s'  }{ Q U^{s'}(P_\slashed {Y}) }\overline 
	{ U }^{s'} (P_\slashed {Y})\bigg ] _{ ii },\nonumber\\
	& =\frac{AB }{6}\sum_{r,r'} {\rm Tr} \bigg[ Q\sum 
	_{  s'  }{ U^{s'}(P_\slashed {Y}) } \overline {
		U }^{s'} (P_\slashed {Y}) \bigg].
	\end{align}
	Now, using the spin sum from \eqref{spinsum} and  substituting 
	$Q$ from \eqref{qq'} in equation \eqref{cross}, 
$ \roverline{\mathfrak{M}^s\mathfrak{M}^{u*}}$
yields to \eqref{crossav}.
	
	Now, taking spin sum over all the polarization vectors using
	\begin{align}
	\sum_{r}
	{ { \varepsilon  }_{ \mu}^{r \ast  } }
	{ \varepsilon  }_{ \nu  }^{r}=-{ g }_{ \mu\nu  },
	\end{align}
$\roverline{\mathfrak{M}^s\mathfrak{M}^{u*}}$ in \eqref{crossav} becomes 
\eqref{cross2}.
Now, simplifying $ \roverline{\mathfrak{M}^s\mathfrak{M}^{u*}}$ and using the properties of gamma matrices, it becomes
		\begin{align}
		\roverline{\mathfrak{M}^s\mathfrak{M}^{u*}}&=C Tr\Bigg[
		\big[{ \gamma^\mu  } 
		{\slashed q_\parallel   }\gamma^\nu{\slashed p_\parallel  }
		\gamma_\mu +\gamma^\mu \gamma^0\gamma^3\gamma^5
		{\slashed q_\parallel }\gamma^\nu
		{\slashed p_\parallel  }\gamma_\mu+
		\gamma^\mu \gamma^0\gamma^3\gamma^5
		{\slashed q_\parallel  }\gamma^\nu \widetilde { \slashed p_
			{ \parallel  } }
		\gamma^5\gamma_\mu+{ \gamma^\mu  } {\slashed q_\parallel  }
		\gamma^\nu\widetilde { \slashed p_
			{ \parallel  } }\gamma^5 
		\gamma_\mu\big]\nonumber\\
		&~~~~\times\big[    
		\slashed{q'}_\parallel\gamma_\nu
		\slashed{P}_\parallel +\slashed{q'}_\parallel
		\gamma_\nu\widetilde { \slashed P_
			{ \parallel  } }\gamma^5-
		\slashed{q'}_\parallel\gamma^5\gamma^3
		\gamma^0\gamma_\nu\slashed{P}_\parallel -
		\slashed{q'}_\parallel\gamma^5\gamma^3
		\gamma^0\gamma_\nu\widetilde { \slashed P_
			{ \parallel  } }\gamma^5\big]
		\Bigg].\label{cross3}
		\end{align}
		Simplifying above equation as follows,
		\[ \underbrace{{ \gamma^\mu  } 
			{\slashed q_\parallel } \gamma^\nu{\slashed p_\parallel }
			\gamma_\mu}_{\text{E}} +
		\underbrace{\gamma^\mu
			\gamma^0\gamma^3\gamma^5
			{\slashed q_\parallel }\gamma^\nu
			{\slashed p_\parallel }\gamma_\mu}
		_{\text{F}}+
		\underbrace{\gamma^\mu 
			\gamma^0\gamma^3\gamma^5 
			{\slashed q_\parallel }\gamma^\nu \widetilde { \slashed p_
				{ \parallel  } }
			\gamma^5\gamma_\mu}_{\text{G}}+
		\underbrace{{ \gamma^\mu  } {\slashed q_\parallel }
			\gamma^\nu\widetilde { \slashed p_
				{ \parallel  } }
			\gamma^5 \gamma_\mu}_{\text{H}}\] .
		\begin{align*}
		E=-2 \slashed p_\parallel 
		\gamma^\nu\slashed q_\parallel,\\
		F=-2\gamma^5\gamma^3
		\slashed q_\parallel\gamma^\nu
		\slashed p_\parallel\gamma^0+
		2\gamma^5\gamma^0\slashed q_\parallel
		\gamma^\nu\slashed  p_\parallel\gamma^3+
		2\gamma^5\gamma^0\gamma^3\slashed p_\parallel
		\gamma^\nu\slashed q,\\
		G=-2\gamma^3\slashed q\gamma^\nu
		\widetilde { \slashed p_
			{ \parallel  } }\gamma^0+2\gamma^0\slashed q_\parallel
		\gamma^\nu\widetilde { \slashed p_
			{ \parallel  } }\gamma^3+
		2\gamma^0\gamma^3\widetilde { \slashed p_
			{ \parallel  } }\gamma^\nu\slashed q_\parallel,\\
		H=-2\gamma^5\widetilde { \slashed p_
			{ \parallel  } }
		\gamma^\nu\slashed q_\parallel,
		\end{align*}
		and substituting the value of E,F,G and H in Eq. 
\eqref{cross3} and multipling among themselves, we get 
		\be
		\roverline{\mathfrak{M}^s\mathfrak{M}^{u*}}&= &C {\rm Tr} 
		\bigg[-2 \slashed p_\parallel \gamma^\nu 
		\slashed q_\parallel\slashed q'_\parallel\gamma_\nu\slashed P_\parallel -
		2\gamma^5\gamma^3\slashed q_\parallel
		\gamma^\nu\slashed p_\parallel\gamma^0\slashed q'
		\gamma_\nu\slashed P_\parallel \nonumber \\
		&+& 2\gamma^5\gamma^0
		\slashed q\gamma^\nu\slashed p_\parallel
		\gamma^3\slashed q'_\parallel\gamma_\nu\slashed P_\parallel
		+2\gamma^5\gamma^0\gamma^3 \slashed p_\parallel
		\gamma^\nu \slashed q_\parallel\slashed q'\gamma_\nu
		\slashed P_\parallel \nonumber \\
		&-& 2\gamma^3\slashed q_\parallel\gamma^\nu
		\widetilde { \slashed p_
			{ \parallel  } }\gamma^0\slashed q'_\parallel
		\gamma_\nu\slashed P_\parallel +2\gamma^0
		\slashed q\gamma^\nu\widetilde { \slashed p_
			{ \parallel  } }
		\gamma^3\slashed q'\gamma_\nu\slashed P_\parallel
		\nonumber \\
		&+& 2\gamma^0\gamma^3\widetilde { \slashed p_
			{ \parallel  } }
		\gamma^\nu\slashed q_\parallel\slashed q'_\parallel
		\gamma_\nu\slashed P_\parallel -2\gamma^5
		\widetilde { \slashed p_
			{ \parallel  } }\gamma^\nu\slashed q_\parallel
		\slashed q'_\parallel\gamma_\nu\slashed P_\parallel \nonumber\\
		&-& 2 \slashed p_\parallel\gamma^\nu \slashed
		q_\parallel\slashed q'_\parallel\gamma_\nu\widetilde { \slashed P_{ \parallel  } } \gamma^5-
		2\gamma^5\gamma^3\slashed q_\parallel
		\gamma^\nu\slashed p\gamma^0
		\slashed q'_\parallel\gamma_\nu\widetilde { \slashed P_
			{ \parallel  } }
		\gamma^5\nonumber \\
		&+& 2\gamma^5\gamma^0\slashed q_\parallel\gamma^
		\nu\slashed p_\parallel\gamma^3\slashed q'_\parallel\gamma_\nu
		\widetilde { \slashed P_
			{ \parallel  } } \gamma^5+2\gamma^5\gamma^0\gamma^3
		\slashed p_\parallel\gamma^\nu \slashed q_\parallel\slashed q'_\parallel
		\gamma_\nu\widetilde { \slashed P_
			{ \parallel  } } \gamma^5\nonumber \\
		&-& 2\gamma^3\slashed q_\parallel\widetilde { \slashed p_
			{ \parallel  } }\gamma^0
		\slashed q'\gamma_\nu\widetilde { \slashed P_
			{ \parallel  } } \gamma^5
		+2\gamma^0\slashed q_\parallel\gamma^\nu\widetilde { \slashed p_
			{ \parallel  } }
		\gamma^3\slashed q'\gamma_\nu\widetilde { \slashed P_
			{ \parallel  } }
		\gamma^5\nonumber \\
		&+& 2\gamma^0\gamma^3\widetilde { \slashed p_
			{ \parallel  } }\gamma^\nu
		\slashed q_\parallel\slashed q'_\parallel\gamma_\nu\widetilde { \slashed P_
			{ \parallel  } }
		\gamma^5-2\gamma^5\widetilde { \slashed p_
			{ \parallel  } }\gamma^\nu\slashed q_\parallel
		\slashed q'_\parallel\gamma_\nu\widetilde { \slashed P_
			{ \parallel  } } \gamma^5 \nonumber\\
		&-& 2 \slashed p\gamma^\nu
		\slashed q\slashed q'
		\gamma^5\gamma^3\gamma^0
		\gamma_\nu\slashed P-
		2\gamma^5\gamma^3\slashed q
		\gamma^\nu\slashed p\gamma^0\slashed q'
		\gamma^5\gamma^3\gamma^0\gamma_\nu
		\slashed P\nonumber \\
		&+& 2\gamma^5\gamma^0\slashed q
		\gamma^\nu\slashed p\gamma^3\slashed q'
		\gamma^5\gamma^3\gamma^0\gamma_\nu\slashed P
		+2\gamma^5\gamma^0\gamma^3 
		\slashed p\gamma^\nu \slashed q
		\slashed q'\gamma^5\gamma^3\gamma^0
		\gamma_\nu\slashed P\nonumber \\
		&-& 2\gamma^3\slashed q\gamma^\nu\widetilde { \slashed p_
			{ \parallel  } }
		\gamma^0\slashed q'\gamma^5\gamma^3\gamma^0
		\gamma_\nu\slashed P+2\gamma^0\slashed q
		\gamma^\nu\widetilde { \slashed p_
			{ \parallel  } }\gamma^3\slashed q'
		\gamma^5\gamma^3\gamma^0\gamma_\nu\slashed P
		\nonumber \\
		&+& 2\gamma^0\gamma^3\widetilde { \slashed p_
			{ \parallel  } }
		\gamma^\nu\slashed q
		\slashed q'\gamma^5\gamma^3\gamma^0\gamma_\nu\slashed P
		-2\gamma^5\widetilde { \slashed p_
			{ \parallel  } }\gamma^\nu
		\slashed q\slashed q'
		\gamma^5\gamma^3\gamma^0\gamma_\nu \slashed P \nonumber\\
		&-& 2 \slashed p_\parallel\gamma^\nu \slashed q_\parallel
		\slashed q'_\parallel\gamma^5\gamma^3\gamma^0
		\gamma_\nu\widetilde { \slashed P_
			{ \parallel  } }\gamma^5-
		2\gamma^5\gamma^3\slashed q_\parallel
		\gamma^\nu\widetilde { \slashed p_
			{ \parallel  } }\gamma^0\slashed q'
		\gamma^3\gamma^0\gamma_\nu
		\widetilde { \slashed P_
			{ \parallel  } }\nonumber \\
		&+& 2\gamma^5\gamma^0\slashed q
		\gamma^\nu\slashed p_\parallel\gamma^3\slashed q'
		\gamma^3\gamma^0\gamma_\nu\widetilde { \slashed P_
			{ \parallel  } }
		+2\gamma^5\gamma^0\gamma^3 \slashed p_\parallel
		\gamma^\nu \slashed q\slashed q'_\parallel
		\gamma^3\gamma^0\gamma_\nu
		\widetilde { \slashed P_
			{ \parallel  } }\nonumber \\
		&-& 2\gamma^3\slashed q_\parallel\gamma^\nu
		\widetilde { \slashed p_
			{ \parallel  } }\gamma^0\slashed q'_\parallel
		\gamma^3\gamma^0\gamma_\nu\widetilde { \slashed P_
			{ \parallel  } }
		+2\gamma^0\slashed q_\parallel\gamma^\nu\widetilde { \slashed p_
			{ \parallel  } }
		\gamma^3\slashed q'_\parallel\gamma^3
		\gamma^0\gamma_\nu\widetilde { \slashed P_
			{ \parallel  } }\nonumber \\
		&+& 2\gamma^0\gamma^3\widetilde { \slashed p_
			{ \parallel  } }\gamma^\nu
		\slashed q_\parallel\slashed q'_\parallel\gamma^3\gamma^0
		\gamma_\nu\widetilde { \slashed P_
			{ \parallel  } }-2\gamma^5\widetilde { \slashed p_
			{ \parallel  } }
		\gamma^\nu\slashed q_\parallel\slashed q'_\parallel
		\gamma^3\gamma^0\gamma_\nu\widetilde { \slashed P_
			{ \parallel  } }\bigg]~.\label{cross32}
		\ee
 We obtain thirty-two terms as shown below.

		\subsubsection{\underline{Calculation of Thirty-two Terms }}
		\begin{flushleft}
			\textbf{\underline{Term-1 }}\vspace{-1em}
		\end{flushleft}
		
		\begin{align}
		-2 {\rm Tr} (\slashed p_\parallel\gamma^\nu \slashed q_\parallel\slashed q'_\parallel
		\gamma_\nu\slashed P_\parallel)=-32( q_\parallel.q'_\parallel)(p_\parallel.P_\parallel). \label{crossterm1}
		\end{align}
		
		\begin{flushleft}
			\textbf{\underline{Term- 2}}\vspace{-1em}
		\end{flushleft}
		\begin{align}
		2 {\rm Tr} (\gamma^5\gamma^3\slashed q_\parallel
		\gamma^\nu\slashed p_\parallel\gamma^0
		\slashed q'_\parallel\gamma_\nu\slashed P_\parallel)=0.
		\end{align}

		\begin{flushleft}
			\textbf{\underline{Term-3}}\vspace{-2em}
		\end{flushleft}
		\begin{align}
		2 {\rm Tr} (\gamma^5\gamma^0\slashed q_\parallel
		\gamma^\nu\slashed p_\parallel\gamma^3
		\slashed q'_\parallel\gamma_\nu\slashed P_\parallel)=0.
		\end{align}

		\begin{flushleft}
			\textbf{\underline{Term-4}}\vspace{-2em}
		\end{flushleft}
		\begin{align}
		2 {\rm Tr} (\gamma^5\gamma^0\gamma^3
		\slashed p_\parallel\gamma^\nu \slashed q_\parallel
		\slashed q'_\parallel\gamma_\nu\slashed P_\parallel)=0.
		\end{align}
		
		\begin{flushleft}
			\textbf{\underline{Term-5}}\vspace{-2em}
		\end{flushleft}
	
		\begin{eqnarray}
		-2 {\rm Tr} (\gamma^3\slashed q_\parallel\gamma^\nu
		\widetilde { \slashed p_
			{ \parallel  } }\gamma^0\slashed q'_\parallel
		\gamma_\nu\slashed P_\parallel)&=&16(\widetilde {  p_
			{ \parallel  } }
		.P_\parallel)(\widetilde q_\parallel \cdot q'_\parallel)-16(q_\parallel \cdot q'_\parallel)(P_\parallel \cdot P_\parallel)\nonumber\\
		&&-16(q'_\parallel.\widetilde {  p_
			{ \parallel  } }
		)(q^3P^0 + q^0P^3) + 16(q'_\parallel.P_\parallel)(q^3\widetilde {  p }^0 + q^0 \widetilde p^3)\nonumber\\
		&&+16(q_\parallel.\widetilde { p_
			{ \parallel  } })q'^3P^0-
		16(q_\parallel.P_\parallel)q'^3\widetilde {  p }^0\nonumber\\ 
		&&-16(q_\parallel.P_\parallel)q'^0\widetilde {  p }^3+16(q_\parallel.\widetilde {  p_
			{ \parallel  } })q'^0P^3
		\end{eqnarray}

		\begin{flushleft}
			\textbf{\underline{Term-6 }}\vspace{-2em}
		\end{flushleft}
		\begin{align}
		2 {\rm Tr} (\gamma^0\slashed q_\parallel\gamma^\nu
		\widetilde { \slashed p_
			{ \parallel  } }\gamma^3\slashed q'_\parallel
		\gamma_\nu\slashed P_\parallel)
		&=-16( \widetilde { p_
			{ \parallel  } }.P_\parallel)q^0q'^3+
		16( q'_\parallel.\widetilde {  p_
			{ \parallel  } })q^0P^3-16(q'_\parallel.P_\parallel)q^0\widetilde p^3
		+16(\widetilde {  p_
			{ \parallel  } }.P_\parallel)q^ 3q'^0\nonumber\\
		&~~~-16( q_\parallel.p_\parallel)q'^0P^3+16(q_\parallel.P_\parallel) q'^0\widetilde {  p }^3+16(q_\parallel.q'_\parallel)P^3\widetilde {  p }^0
		-16(q'_\parallel.P_\parallel) q^3\widetilde {  p }^0\nonumber\\
		&~~~+16( q_\parallel.P_\parallel)q'^3\widetilde {  p }^0
		-16(q_\parallel.q'_\parallel)\widetilde {  p }^3P^0+16( q'_\parallel.p_\parallel)q^0P^0
		-16( q_\parallel.\widetilde {  p_
			{ \parallel  } }) q'^3P^0.
		\end{align}

		\begin{flushleft}
			\textbf{\underline{Term-7 }}\vspace{-2em}
		\end{flushleft}
		\begin{eqnarray}
		2 {\rm Tr} (\gamma^0\gamma^3\widetilde { \slashed p_
			{ \parallel  } }\gamma^\nu
		\slashed q_\parallel\slashed q'_\parallel
		\gamma_\nu\slashed P_\parallel)
		&=&-32( q_\parallel. q'_\parallel)P^3\widetilde {  p }^0
		+32(q_\parallel.q'_\parallel)\widetilde {  p }^3P^0\\
		&=&32(p_\parallel \cdot P_\parallel) (q_\parallel \cdot 
q'_\parallel).
		\end{eqnarray}
		\begin{flushleft}
			\textbf{\underline{Term-8}}\vspace{-2em}
		\end{flushleft}
		\begin{align}
		-2 {\rm Tr} (\gamma^5\widetilde { \slashed p_
			{ \parallel  } }\gamma^\nu\slashed q_\parallel
		\slashed q'_\parallel\gamma_\nu\slashed P_\parallel)=0.
		\end{align}
		\begin{flushleft}
			\textbf{\underline{Term-9}}\vspace{-2em}
		\end{flushleft}
		\begin{align}
		-2 {\rm Tr} (\gamma^5\widetilde { \slashed p_
			{ \parallel  } }\gamma^\nu
		\slashed q_\parallel\slashed q'_\parallel\gamma_\nu\slashed P_\parallel)=0.
		\end{align}
		\begin{flushleft}
			\textbf{\underline{Term-10 }}\vspace{-1em}
		\end{flushleft}
		\begin{align}
		-2 {\rm Tr} (\gamma^5\gamma^3\slashed q_\parallel
		\gamma^\nu\slashed p\gamma^0
		\slashed q'_\parallel\gamma_\nu\widetilde { \slashed P_
			{ \parallel  } }
		\gamma^5)
		)&= 16(p_\parallel.\widetilde { P_
			{ \parallel  } })q^3q'^0-
		16(q'_\parallel.p_\parallel)q^3\widetilde {  P }^0+16(q'_\parallel.\widetilde {  P_
			{ \parallel  } })q^3p^0
		-16(p_\parallel.\widetilde { P_
			{ \parallel  } })q^0q'^3\nonumber\\
		&~~~+16(q_\parallel.p_\parallel)q'^3\widetilde {  P }^0
		-16(q_\parallel.\widetilde {  P_
			{ \parallel  } })q'^3p^0-16(q_\parallel.q'_\parallel)\widetilde {  P }^0p^3
		+16(q'_\parallel.\widetilde {  P_
			{ \parallel  } })p^3q^0\nonumber\\ 
		&~~~-16(q_\parallel.\widetilde { P_
			{ \parallel  } })q'^0p^3
		+16(q_\parallel.q'_\parallel)\widetilde {  P }^3p^0-16(q'_\parallel.p_\parallel)q^0\widetilde {  P }^3
		+16(q'_\parallel.p_\parallel)q'^0\widetilde {  P }^3.
		\end{align}
		\begin{flushleft}
			\textbf{\underline{Term-11 }}\vspace{-1em}
		\end{flushleft}
		\begin{align}
		2 {\rm Tr} (\gamma^5\gamma^0\slashed q_\parallel
		\gamma^\nu\slashed p_\parallel\gamma^3
		\slashed q'_\parallel\gamma_\nu\widetilde { \slashed P_
			{ \parallel  } } \gamma^5)
		)&=-16( p_\parallel.\widetilde {  P_
			{ \parallel  } })q^0q'^3
		+16( q'_\parallel.p_\parallel)q^0(a)\widetilde {  P }^3-16(q'_\parallel.\widetilde { P_
			{ \parallel  } })q^0p^3+
		16(p_\parallel.\widetilde { P_
			{ \parallel  } })q^3q'^0\nonumber\\
		&~~~-16( q_\parallel.\widetilde { P_
			{ \parallel  } })q'^0k^3
		+16(q_\parallel.\widetilde { P_
			{ \parallel  } }) q'^0p^3+16(q_\parallel.q'_\parallel)\widetilde {  P }^3p^0
		-16(q'_\parallel.\widetilde {  P_
			{ \parallel  } }) q^3p^0\nonumber\\
		&~~~+16( q_\parallel.\widetilde { P_
			{ \parallel  } })q'^3p^0
		-16(q_\parallel.q'_\parallel)p^3\widetilde {  P }^0+16( q'_\parallel.p_\parallel)q^0\widetilde {  P}^0
		-16( q_\parallel.p_\parallel) q'^3\widetilde {  P }^0.
		\end{align}
		\begin{flushleft}
			\textbf{\underline{Term-12 }}\vspace{-1em}
		\end{flushleft}
		\begin{eqnarray}
		2 {\rm Tr} ( \gamma^5\gamma^0\gamma^3\slashed
		p_\parallel\gamma^\nu\slashed q_\parallel(a)\slashed q'
		\gamma_\nu\widetilde { \slashed P_
			{ \parallel  } } \gamma^5)
		&=&-32( q_\parallel.q'_\parallel)\widetilde { P }^3p^0+3
		2(q_\parallel.q'_\parallel)p^3\widetilde {P }^0\nonumber\\
		&=&-32(p_\parallel \cdot P_\parallel) (q_\parallel \cdot 
q'_\parallel).
		\end{eqnarray}

		\begin{flushleft}
			\textbf{\underline{Term-13}}\vspace{-1em}
		\end{flushleft}
		\begin{align}
		-2 {\rm Tr} ( \gamma^5\gamma^0\gamma^3
		\slashed p_\parallel\gamma^\nu\slashed q_\parallel
		\slashed q'\gamma_\nu\widetilde {  P_
			{ \parallel  } } \gamma^5)=0.
		\end{align}
		
		\begin{flushleft}
			\textbf{\underline{Term-14}}\vspace{-1em}
		\end{flushleft}
		\begin{align}
		2 {\rm Tr} (\gamma^0\slashed q\gamma^\nu
		\widetilde { \slashed p_
			{ \parallel  } }\gamma^3\slashed q'_\parallel
		\gamma_\nu\widetilde { \slashed P_
			{ \parallel  } } \gamma^5)=0.
		\end{align}
		\begin{flushleft}
			\textbf{\underline{Term-15}}\vspace{-1em}
		\end{flushleft}
		\begin{align}
		2 {\rm Tr} (\gamma^0\gamma^3\widetilde { \slashed p_
			{ \parallel  } }
		\gamma^\nu\slashed q_\parallel\slashed q'_\parallel\gamma_\nu\widetilde { \slashed P_
			{ \parallel  } } \gamma^5)=0.
		\end{align}
		\begin{flushleft}
			\textbf{\underline{Term-16 }}\vspace{-1em}
		\end{flushleft}
		\begin{eqnarray}
		-2 {\rm Tr} (\gamma^5\widetilde { \slashed p_
			{ \parallel  } }\gamma^\nu
		\slashed q_\parallel\slashed q'_\parallel\gamma_\nu
		\widetilde { \slashed P_
			{ \parallel  } } \gamma^5)&=&-32( q_\parallel.q'_\parallel)(\widetilde {  p_
			{ \parallel  } }.\widetilde {  P_
			{ \parallel  } })\nonumber\\
		&=& 32(p_\parallel \cdot P_\parallel) (q_\parallel \cdot 
q'_\parallel).
		\end{eqnarray}

		\begin{flushleft}
			\textbf{\underline{Term-17}}\vspace{-1em}
		\end{flushleft}
		\begin{align}
		-2 {\rm Tr} (\slashed p_\parallel\gamma^\nu 
		\slashed q_\parallel\slashed q'_\parallel
		\gamma^5\gamma^3\gamma^0
		\gamma_\nu\slashed P_\parallel)=0.
		\end{align}

		\begin{flushleft}
			\textbf{\underline{Term-18}}\vspace{-1em}
		\end{flushleft}
		\begin{align}
		-2 {\rm Tr} (\gamma^5\gamma^3\slashed q_\parallel
		\gamma^\nu\slashed p_\parallel\gamma^0\slashed q'
		\gamma^5\gamma^3\gamma^0\gamma_\nu
		\slashed P_\parallel)&=64p^3q'^0q^3P^0
		-64q'^0q^3P^3p^0-32(p_\parallel.P_\parallel)q'^0q^0
		+32(q_\parallel.p_\parallel)q'^0P^0\nonumber\\
		&~~~-32(q_\parallel.P_\parallel)q'^0p^0
		+32(p.P)q'^3q^3-32(q_\parallel.P_\parallel)q^3p^3
		+32(q'_\parallel.p_\parallel)q^3P^3\nonumber\\
		&~~~+16(q_\parallel.q'_\parallel)(p_\parallel.P_\parallel
		-16(p_\parallel.q_\parallel)(P_\parallel.q'_\parallel)+16(P_\parallel.q_\parallel)(p_\parallel.q'_\parallel).
		\end{align}

		\begin{flushleft}
			\textbf{\underline{Term-19}}\vspace{-1em}
		\end{flushleft}
		\begin{align}
		2 {\rm Tr} (\gamma^5\gamma^0\slashed q_\parallel
		\gamma^\nu\slashed p_\parallel\gamma^3
		\slashed q'_\parallel\gamma^5\gamma^3
		\gamma^0\gamma_\nu\slashed P_\parallel)
		&=64q'^3q^0P^3p^0
		-64q'^3q^0P^0p^3 +32q'^3q^3(p_\parallel.P_\parallel)
		-32(q_\parallel.p_\parallel)q'^3P^3\nonumber\\
		&~~~+32(q_\parallel.P_\parallel)q'^3p^3-
		32(p_\parallel.P_\parallel)q'^0q^0+32(q'_\parallel.P_\parallel)p^0q^0-
		32(q'_\parallel.p_\parallel)q^0P^0\nonumber\\
		&~~~+16(q_\parallel.q'_\parallel)(p_\parallel.P_\parallel
		-16(p_\parallel.q_\parallel)(P_\parallel.q'_\parallel)+16(P_\parallel.q_\parallel)(p_\parallel.q'_\parallel).
		\end{align}

		\begin{flushleft}
			\textbf{\underline{Term-20}}\vspace{-1em}
		\end{flushleft}
		\begin{align}
		2 {\rm Tr} (\gamma^5\gamma^0\gamma^3
		\slashed p_\parallel\gamma^\nu \slashed q_\parallel
		\slashed q'_\parallel\gamma^5\gamma^3\gamma^0
		\gamma_\nu\slashed P_\parallel)&=
		-64p^3P^0q'^0q^3
		+32(P_\parallel.q'_\parallel)p^3q^3+128q'^3P^0p^0q^3-
		32(q'_\parallel.P_\parallel)p^0q^0\nonumber\\
		&~~~ +32(q'_\parallel.p_\parallel)q^0P^0
		-32(q_\parallel.p_\parallel)q'^0P^0-32(P_\parallel.q_\parallel)(p_\parallel.q'_\parallel)
		+32(p_\parallel.q_\parallel)(P_\parallel.q'_\parallel)\nonumber\\
		&~~~+32(q_\parallel.q'_\parallel)(p_\parallel.P_\parallel)
		+128p^0q'^3q^0P^3 +64P^3p^0q^3q'^0P^3
		+32(P_\parallel.q_\parallel)p^0q'^0\nonumber\\
		&~~~ +64p^3P^0q'^3q^0
		-64P^3p^0q'^3q^0-32(P_\parallel.q_\parallel)q'^3p^3
		+32(p_\parallel.q_\parallel)P^3q'^3\nonumber\\
		&~~~-32(p_\parallel.q'_\parallel)P^{ 3 }q^3.
		\end{align}

		\begin{flushleft}
			\textbf{\underline{Term-21}}\vspace{-1em}
		\end{flushleft}
		\begin{align}
		-2 {\rm Tr} (\gamma^3\slashed q_\parallel
		\gamma^\nu\widetilde { \slashed p_
			{ \parallel  } } \gamma^0
		\slashed q'_\parallel\gamma^5\gamma^3
		\gamma^0\gamma_\nu\slashed P_\parallel)=0.
		\end{align}

		\begin{flushleft}
			\textbf{\underline{Term-22}}\vspace{-1em}
		\end{flushleft}
		\begin{align}
		2 {\rm Tr} (\gamma^0\slashed q_\parallel
		\gamma^\nu\widetilde { \slashed p_
			{ \parallel  } }
		\gamma^3\slashed q'_\parallel
		\gamma^5\gamma^3\gamma^0
		\gamma_\nu\slashed P_\parallel)=0.
		\end{align}
		
		\begin{flushleft}
			\textbf{\underline{Term-23}}\vspace{-1em}
		\end{flushleft}
		\begin{align}
		2 {\rm Tr} (\gamma^0\gamma^3\widetilde { \slashed p_
			{ \parallel  } }\gamma^\nu
		\slashed q_\parallel\slashed q'_\parallel\gamma^5\gamma^3
		\gamma^0\gamma_\nu\slashed P_\parallel)=0.
		\end{align}
		
		\begin{flushleft}
			\textbf{\underline{Term-24}}\vspace{-1em}
		\end{flushleft}
		\begin{eqnarray}
		-2 {\rm Tr} (\gamma^5\widetilde { \slashed p_
			{ \parallel  } }\gamma^\nu
		\slashed q_\parallel\slashed q'_\parallel\gamma^5
		\gamma^3\gamma^0\gamma_\nu\slashed P_\parallel)&=&-32(\widetilde { p_
			{ \parallel  } }.P_\parallel)q'^3q^0+32(\widetilde {  p_
			{ \parallel  } }.P_\parallel)q^3q'^0\nonumber\\
		&=&32(\widetilde p_\parallel \cdot P_\parallel) (\widetilde q_\parallel \cdot q'_\parallel)
		\end{eqnarray}

		\begin{flushleft}
			\textbf{\underline{Term-25}}\vspace{-1em}
		\end{flushleft}
		\begin{eqnarray}
		-2 {\rm Tr} (\slashed p_\parallel\gamma^\nu \slashed q_\parallel
		\slashed q'_\parallel\gamma^5\gamma^3
		\gamma^0\gamma_\nu\widetilde { \slashed P_
			{ \parallel  } }
		\gamma^5) &=&-32(p_\parallel.\widetilde {  P_
			{ \parallel  } })q'^3
		q^0+32(p.\widetilde {  P_
			{ \parallel  } })q^3q'^0\nonumber\\
		&=&32(\widetilde P_\parallel \cdot p_\parallel) (\widetilde q_\parallel \cdot q'_\parallel)
		\end{eqnarray}

		\begin{flushleft}
			\textbf{\underline{Term-26}}\vspace{-1em}
		\end{flushleft}
		\begin{align}
		-2 {\rm Tr} (\gamma^5\gamma^3\slashed q_\parallel
		\gamma^\nu\slashed p_\parallel\gamma^0
		\slashed q'_\parallel\gamma^3\gamma^0
		\gamma_\nu\widetilde { \slashed P_
			{ \parallel  } }) =0.
		\end{align}

		\begin{flushleft}
			\textbf{\underline{Term-27}}\vspace{-1em}
		\end{flushleft}
		\begin{align}
		2 {\rm Tr} (\gamma^5\gamma^0\slashed q_\parallel
		\gamma^\nu\slashed p_\parallel\gamma^3
		\slashed q'_\parallel\gamma^3\gamma^0
		\gamma_\nu\widetilde { \slashed P_
			{ \parallel  } })=0.
		\end{align}

		\begin{flushleft}
			\textbf{\underline{Term-28}}\vspace{-1em}
		\end{flushleft}
		\begin{align}
		2 {\rm Tr} (\gamma^5\gamma^0\gamma^3
		\slashed p_\parallel\gamma^\nu 
		\slashed q_\parallel\slashed q'_\parallel
		\gamma^3\gamma^0\gamma_\nu\widetilde { \slashed P_
			{ \parallel  } })=0.
		\end{align}

		\begin{flushleft}
			\textbf{\underline{Term-29}}\vspace{-1em}
		\end{flushleft}
		\begin{align}
		-2 {\rm Tr} (\gamma^3\slashed q_\parallel\gamma^\nu
		\widetilde { \slashed p_
			{ \parallel  } }\gamma^0\slashed q'_\parallel
		\gamma^3\gamma^0\gamma_\nu\widetilde { \slashed P_
			{ \parallel  } })
		&=64\widetilde p^3q'^0q^3\widetilde P^0-
		64q'^0q^3\widetilde P^3\widetilde p^0-32(\widetilde {p_
			{ \parallel  } }.\widetilde { P_
			{ \parallel  } })q'^0q^0+
		32(q_\parallel.\widetilde {  P_
			{ \parallel  } })q'^0k^0\nonumber\\
		&~~~-32(q_\parallel.\widetilde {  P_
			{ \parallel  } })q'^0\widetilde p^0+
		32(\widetilde {  p_
			{ \parallel  } }.\widetilde {  P_
			{ \parallel  } })q'^3q^3-32(q_\parallel(a).\widetilde {  P_
			{ \parallel  } })q^3\widetilde p^3+
		32(q'.\widetilde {  p_
			{ \parallel  } })q^3\widetilde P^3\nonumber\\ 
		&~~~+16(q_\parallel.q'_\parallel)(\widetilde {  p_
			{ \parallel  } }.\widetilde {  P_
			{ \parallel  } })-
		16(\widetilde { p_
			{ \parallel  } }.q_\parallel)(\widetilde {  P_
			{ \parallel  } }.q'_\parallel)+16(\widetilde {  P_
			{ \parallel  } }.q_\parallel)(\widetilde {  p_
			{ \parallel  } }.q'_\parallel).
		\end{align}

		\begin{flushleft}
			\textbf{\underline{Term-30}}\vspace{-1em}
		\end{flushleft}
		\begin{align}
		2 {\rm Tr} (\gamma^5\gamma^0\slashed q_\parallel
		\gamma^\nu\widetilde { \slashed p_
			{ \parallel  } }\gamma^3
		\slashed q'_\parallel\gamma^5\gamma^3
		\gamma^0\gamma_\nu\widetilde { \slashed P_
			{ \parallel  } })
		&=64q'^3q^0\widetilde P^3\widetilde p^0
		-64q'^3q^0\widetilde P^0\widetilde p^3 +32q'^3q^3(\widetilde {  p_
			{ \parallel  } }.\widetilde {  P_
			{ \parallel  } })
		-32(q_\parallel.\widetilde { p_
			{ \parallel  } })q'^3\widetilde P^3\nonumber\\
		&~~~+32(q_\parallel.\widetilde {  P_
			{ \parallel  } })q'^3\widetilde p^3
		-32(\widetilde {  p_
			{ \parallel  } }.\widetilde {  P_
			{ \parallel  } })q'^0q^0+32(q'_\parallel.\widetilde {  P_
			{ \parallel  } })\widetilde p^0q^0
		-32(q'_\parallel.\widetilde { p_
			{ \parallel  } })q^0\widetilde P^0\nonumber\\
		&~~~+16(q_\parallel.q'_\parallel)(\widetilde { p_
			{ \parallel  } }.\widetilde {  P_
			{ \parallel  } }
		-16(\widetilde { p_
			{ \parallel  } }.q_\parallel)(\widetilde { P_
			{ \parallel  } }.q'_\parallel)+16(\widetilde {  P_
			{ \parallel  } }.q_\parallel)(\widetilde { p_
			{ \parallel  } }.q'_\parallel).
		\end{align}

		\begin{flushleft}
			\textbf{\underline{Term-31}}\vspace{-1em}
		\end{flushleft}
		\begin{align}
		2 {\rm Tr} (\gamma^5\gamma^0\gamma^3 
		\widetilde { \slashed p_
			{ \parallel  } }\gamma^\nu \slashed q_\parallel
		\slashed q' \gamma^5\gamma^3\gamma^0
		\gamma_\nu\widetilde { \slashed P_
			{ \parallel  } })&=-
		64\widetilde p^3\widetilde P^0q'^0q^3
		+32(\widetilde { P_
			{ \parallel  } }.q'_\parallel)\widetilde p^3q^3+128q'^3\widetilde P^0\widetilde p^0q^3
		-32(q'_\parallel.\widetilde {  P_
			{ \parallel  } })\widetilde p^0q^0\nonumber\\
		&~~~ +32(q'.\widetilde {  p_
			{ \parallel  } })q^0\widetilde P^0
		-32(q_\parallel.\widetilde {  p_
			{ \parallel  } })q'^0\widetilde P^0-32(\widetilde {  P_
			{ \parallel  } }.q_\parallel)(\widetilde {  p_
			{ \parallel  } }.q'_\parallel)
		+32(\widetilde {  p_
			{ \parallel  } }.q_\parallel)(\widetilde { P_
			{ \parallel  } }.q'_\parallel)\nonumber\\
		&~~~+32(q_\parallel.q'_\parallel)(\widetilde p.\widetilde P)
		+128\widetilde p^0q'^3q^0\widetilde P^3 +64\widetilde P^3\widetilde p^0q^3q'^0
		+32(\widetilde { P_
			{ \parallel  } }.q_\parallel)\widetilde p^0q'^0\nonumber\\
		&~~~ +64\widetilde p^3\widetilde P^0q'^3q^0
		-64\widetilde P^3\widetilde p^0q'^3q^0-32(\widetilde { P_
			{ \parallel  } }.q_\parallel)q'^3\widetilde p^3
		+32(\widetilde { p_
			{ \parallel  } }.q_\parallel)\widetilde P^3q'^3\nonumber\\
		&~~~-32(\widetilde { p_
			{ \parallel  } }.q'_\parallel)\widetilde P^3q^3.
		\end{align}
		
		\begin{flushleft}
			\textbf{\underline{Term-32}}\vspace{-1em}
		\end{flushleft}
		\begin{align}
		-2 {\rm Tr} (\widetilde { \slashed p_
			{ \parallel  } }\gamma^\nu \slashed q_\parallel
		\slashed q'_\parallel\gamma^5\gamma^3\gamma^0
		\gamma_\nu\widetilde { \slashed P_
			{ \parallel  } })=0 .\label{crossterm32}
		\end{align}
	Adding all the  thirty two terms from 
\eqref{crossterm1}-\eqref{crossterm32}, \eqref{cross32} becomes \eqref{cross4}.
	
\section{Crosssection}

In this appendix section, we will show the detailed calculation for the crosssection. So to calculate
 crosssection first we need to see  differential phase factor. In magnetic field, differential phase
  factor is given by
		\begin{align} d\rho ={ { \frac
				{ { d }{ P }^{ 0 }{ d }
					{ P }_{ X }{ d }{ P}_{ Z } }
				{ { (2\pi ) }^{ 2 } } \delta
				({ P^{ 2 } }- } }{ m }
		^{ 2 })\Theta { ({ P }^{ 0 }
			){ { \frac { { d }^{ 4 }{ K } }{ { (2\pi ) }^{ 3 } } 
					\delta ({ K^{ 2 } }
				} })
				\Theta { ({ K }^{ 0 }) } } .\label{phase}
			\end{align}
Now substituting the S-matrix element squared from 
\eqref{totalmatrix} and the differential phase factor from \eqref{phase}  
into  \eqref{crosssection}, the crosssection for the Compton scattering 
in strong magnetic field becomes
			\begin{align}
			\sigma\com =&\int { { \frac { { d }{ P }
						^{ 0 }{ d }{ P }_{ X }{ d }{ P }
						_{ Z} }{ { (2\pi ) }^{ 2 } } 
					\delta ({ P^{ 2 } }- } } { m }
			^{ 2 })\Theta { ({ P }^{ 0 })
				{ { { { \frac { { d }^{ 4 }{ K }
								}{ { (2\pi ) }^{ 3 } } \delta ({ K^{ 2 }
							} } })\Theta { ({ K }
							^{ 0 }) } } } }\nonumber \\ &~~~~~ \times{ { { 
						\left( 2\pi  \right)  }^{ 6 }\pi 
				}\delta_\slashed Y  }^{ 3 }({ P }+
			{ K }-p_{  }-k)\roverline{|{ c \mathfrak{M}^s  }
				+{ d \mathfrak{M}^u 
				}|^{ 2 }}.
			\end{align}
To solve the crosssection expression, we will first simplify the photon phase
factor and then electron phase factor. Using the property of the Dirac-delta 
function
				\begin{align}
				\delta ({{ K^0}}^2-{ \omega'  }
				^{ ^{ 2 } })\theta ({ { K }^
					{ 0 }})=\frac { \left[
					\delta ({ { K }^{ 0 } 
					}+{ { \omega ' } })+
					\delta ({ { K }^{ 0 } 
					}-{ { \omega'  } }) 
					\right]  }{ 2{ \omega ' } } \theta 
({ { K }^{ 0 } }),\label{diracdelta}
				\end{align}
the photon phase factor in Eq. \eqref{photon} can be simplified as
				\begin{align}
				{ { \frac { { d }^{ 4 }{ K } }
						{ { (2\pi ) }^{ 3 } } \delta 
						({ K }^{ 2 } } } )\theta 
				({ { K }^{ 0 } })&=\frac 
				{ { { d }^{ 3 }{ K }_{ f }d{ K }
						^{ 0 } } }{ { (2\pi ) 
					}^{ 3 } } \frac { \left[ \delta
					({ { K }^{ 0 } }+
					{ { \omega'  } })+
					\delta ({ { K }^{ 0 } }
					-{ { \omega  '}}) \right]  }
				{ 2{ \omega  '} } 
				\theta ({ { K }^{ 0 } }),\nonumber\\ 
				&=  \frac {
					{ { d }^{ 3 }{ K }d{ K }^{ 0 }
					} }{ { (2\pi ) }^{ 3 } } 
					\frac { \delta ({ { K }^{ 0 } }
						-{ { \omega'  }}) }{ 2\omega ' } .\label{photon1}
					\end{align}
Using the mass-shell condition in strong magnetic field:
$ P_{ \parallel}^{ 2 } ={ m }^{ 2 }$, the differential phase factor for the 
electron in \eqref{electron} can be simplified as
					\begin{align} 
					{ { \frac { { d }{ P }
								^{ 0 }{ d }{ P }_{ X }
								{ d }{ P }_{ Z } }{ { (2\pi ) }
								^{ 2 } } \delta ({ P_{ \parallel  }
							}^{ 2 } } }-{ m }^{ 2 })
					\theta ({ { P }^{ 0 } }) & =
					{ \frac { { { d }{ P }_{ X }{ d }
								{ P }_{ Z }d{ P }^{ 0 }
							} }{ { (2\pi ) }^{ 2 } }  }
						\delta ({ { P^{ 0 } } }^{ 2 }-
						{ p^{ 2 }_Z }-{ E '}
						^{ 2 }+{ p^{ 2 }_Z })
						\theta ({ { P }^{ 0 } }),\nonumber\\&
						={ \frac { { d }{ P }
								_{ X }{ d }{ P}_{ Z}{ d{ P }
									^{ 0 } } }{ { (2\pi ) }^{ 2 } }
						}\delta ({ { P^{ 0 } } }^{ 2 }-{ E '}^{ 2 })\theta ({ { P }^{ 0 } }).
						\end{align}
Now using equation \eqref{diracdelta}, the electron phase factor is modified 
into
						\begin{align} 
						{ { \frac { { d }{ P }
									^{ 0 }{ d }{ P }_{ X }
									{ d }{ P }_{ Z } }{ { (2\pi ) 
									}^{ 2 } } \delta ({ P_{ \parallel  } }
								^{ 2 } } }-{ m }^{ 2 })\theta 
						({ { P }^{ 0 } }) & ={ \frac { { { d }
									{ P }_{ X }{ d }{ P }_{ Z }d{ P }^{ 0 }
								} }{ { (2\pi ) }^{ 2 } }  }\frac
							{ \left[ \delta ({ { P }^{ 0 } }+{ E' }
								)+\delta ({ { P }^{ 0 } }-{ { E' } })
								\right]  }{ 2{ E' } }
							\theta ({ { P }^{ 0 } }), \nonumber\\
							&=\frac { { { d }{ P }_{ X }
									{ d }{ P }_{ Z }d{ P}^{ 0 }
								} }{ { (2\pi ) }^{ 2 } }
								\frac { \delta ({ { P }^{ 0 } }
									-{ { E '} }) }{ 2{ E '} } .\label{electron1}
								\end{align}
Now, substituting the simplified phase factors for photon and and electron
from \eqref{photon1} and  \eqref{electron1},  respectively, the crosssection 
for Compton scattering expression becomes
								
								\begin{align}
								\sigma\com =\int \frac
								{ { { d }{ P }_{ X }{ d }{ P }
										_{ Z }d{ P }^{ 0 }} }
								{ { (2\pi ) }^{ 2 } } \frac { \delta
									({ { P }^{ 0 } }-{ { E' } }) 
								}{ 2{ E '} } \frac { { { d }^{ 3 }
									{ K }d{ K }^{ 0 } } }{ 
								{ (2\pi ) }^{ 3 } } \frac { \delta ({
									{ K }^{ 0 } }-{ { \omega ' }
								}) }{ 2\omega' }  
							\frac { { { { \left( 2\pi  \right)  }
										^{ 6 }\pi }\delta^{ 3 }
									_\slashed Y  }({ P }
								+{K }-p-k)\roverline{|{ c \mathfrak{M}^s  }
									+{ d \mathfrak{M}^u 
									}|^{ 2 } }}{ F }.\end{align}
							
Using the property of the Dirac-delta function: $\int { f(x)\delta
(x-{ x }_{ 0 })dx= } f(x_{ 0 })$, the Dirac-Delta functions ($\delta ({ { K }
^{ 0 } }-{ { \omega'  } })$ and $\delta ({ { P }^{ 0 } }-{{ E'  } })$) 
will be eliminated. Then substituting $ { K}^{ 0 }=\omega'$ and 
${ P }^{ 0 }=E'$ in the remaining function, the differential phase factor
of photon ${ d }{ K }_{ X }$ ${ d }{ K }_{ Z }$ will be eliminated by the 
energy-momentum conserving Dirac-delta function, hence the crosssection 
finally reduces to 
							\begin{align} \sigma =\dfrac{\pi^2}{2}
							\int { \frac { { { d }{ P }_{ X}{ d }{ P}
										_{ Z } } }{{ E '}  } \frac 
								{ { d }{ K }_{ Y } }{ \omega' } 
								{ \delta  }({ E '}+{ \omega  '}-{ E }-{ \omega  })
								\frac {\roverline{|{ c \mathfrak{M}^s  }
										+{ d \mathfrak{M}^u 
										}|^{ 2 }}}{ F }  } .\label{cross55} \end{align}
							\textbf{\underline{Flux factor}}:
For the process $$1+2\longrightarrow 3+4$$, the flux factor is given by
							\begin{equation}
							F=\left| { v }_{ 1 }-{ v }
							_{ 2 } \right| 2{ E }_{ 1 }2E_{ 2 },
							\end{equation}
where, $v_1$ and  $v_2$ are the velocities of the target and projectile,
respectively and $E_1$ and $E_2$
are their corresponding energy. Assuming the lab frame where the initial momentum of
target will be zero, then flux factor in lab-frame becomes,
							\begin{align}	
							F &=4v_2\sqrt { { p }_{ 1 }^{ 2 }+
								{ m }_{ 1 }^{ 2 } }\sqrt { { p }
								_{ 2 }^{ 2 }+{ m }_{ 2 }^{ 2 } }
							=4v_2m_1p_2=4m_1E_2.
							\end{align}
For our case target is electron and projectile is photon, hence flux factor becomes
							\begin{align}	
							F=4m\omega.\label{flux}
							\end{align}
Now, substituting the expression of flux factor from \eqref{flux}, 
the crosssection in \eqref{cross55} becomes
							
							\begin{align} \sigma\com = \dfrac{\pi^2}{8}
							\int { \frac { { { d }{ P }_{ X }{ d }
										{ P }_{ Z } } }{{ E '}  } 
								\frac { { d }{ K }_{ Y } }{ \omega' }
								{ \delta  }({ E '}+{ \omega  '}
								-{ E }-{ \omega  })
								\frac { \roverline{|{ c \mathfrak{M}^s  }
										+{ d \mathfrak{M}^u 
										}|^{ 2 }} }{ m\omega }  } .
							\end{align}
							
Now using \eqref{c}-\eqref{cd}, \eqref{squared}, \eqref{Kint} and 
$dP_X$ integration
								$$\int _{ 0 }^{ \sqrt { eB }  }
								{ dP_{ X } } 
=\sqrt { eB },$$
the crosssection gets simplified into 
								\begin{align}  \sigma &=\frac { eB\pi ^{ 2 }
								}{4 \sqrt { 3 }  } \int { \frac { {{ d }
										{ P }_{ Z } } }{ { E '}\omega'} 
								{ \delta  }({ E '}+{ \omega  '}-
								{ E }-{ \omega  }) }
							\times \frac { \left(  \exp\left( \frac { -2q^2_X }{ eB }  \right) 
								\roverline{| \mathfrak{M}^s 
									|  ^{ 2 }} 
								+\roverline{| \mathfrak{M}^u 
									|  ^{ 2 }}  
								\exp\left( \frac { -2q'^2_X }
								{ eB }  \right) \right)  }{m\omega} .
							\end{align}
Now, we will integrate over $dP_Z$ for all the matrix elements due to different channels and cross-terms one by one.\\
							
							\underline{\textbf{FOR $s$-CHANNEL DIAGRAM}}:
							
The crosssection expression for the $s$-channel diagram is given by
					\begin{align} 
					\sigma^s \com & 
					=\frac { eB\pi ^{ 2 } }{4 \sqrt { 3 }  } \int { \frac { { { d }{ P }_{ Z } } }
						{ { E' }\omega' } { \delta  } ({ E '}+{ \omega ' }-{ E }
						-{ \omega  }) } \frac { \exp\left ( \frac { - 2q^2_X }{ eB } 
						\right) \roverline{|\mathfrak{M}^s|^2} }{ m\omega  }.
					\end{align}
We will get eq.\eqref{sigmass} from steps done in eqs.\eqref{squaredd}-
\eqref{variables}. Now our next task is to  find
the roots and integrate over the energy conserving Dirac-delta function, 
firstly we find the roots of the Dirac-delta function as follows
						\begin{align}
						{ \delta  }({ E' }+
						{ \omega ' }-{ E }
						-{ \omega  })&={ \delta  }
						(\sqrt { { P^{ 2 }_{ Z } }+
							{ m }^{ 2 } } { +\omega'  }
						-{ p }_{ Z }-{ \omega  }),
						\nonumber
						\\&={ \delta  }(\sqrt { { \bigg[\frac
								{ (m{ \omega  }+{ { \omega  }
									}^{ 2 })(1-\cos\theta ) 
								}{ m+{ \omega  }(1-\cos\theta ) }
								\bigg] }^{ 2 }+{ m }^{ 2 } } -\frac
						{ ({ { \omega  } }^{ 2 })
							(1-\cos\theta ) }{ m+
							{ \omega  }(1-\cos\theta ) }- m).
						\end{align}
Now finding the roots of the equation inside the Dirac-delta function
						\begin{align}
						{ \left[ \frac { (m{ \omega  }+{ { \omega  } }
								^{ 2 })(1-\cos\theta ) }{ m+
								{ \omega  }(1-\cos\theta ) }
							\right]  }^{ 2 }+{ m }^{ 2 }={ \left
							[ \frac { ({ { \omega  } }^{ 2 })
								(1-\cos\theta ) }{ m+{ \omega  }
								(1-\cos\theta ) } +m \right]  }
						^{ 2 },\nonumber\\
						\end{align}
simplifying this equation and finally, we get a quadratic equation,
						\begin{align}
						[2{ { \omega  } }^{ 2 }
						{ m }^{ 2 }-{ { \omega  }
						}^{ 2 }{ m }^{ 2 }(1-\cos\theta )]
						(1-\cos\theta )=0,
						\end{align}
whose roots are given by 
						\begin{align}
						\cos\theta =1 ,\quad\quad 
\cos\theta =-1.
						\end{align}
							
Now, putting $x=\cos\theta$ in eq.\eqref{sigmass}, then the crosssection 
for the $s$-channel becomes
									\begin{align} 
									&\sigma^s \com\nonumber\\ &=e^4
									\frac { eB\pi ^{ 2 } (m+\omega) }{ 
										12\sqrt { 3 } { m }\omega^{ 2 } } \exp
									\left( \frac { -{2 q^{ 2 }_X } }
									{ eB }  \right) \int _{ -1 }^{ 1 }
									{ \frac { dx }{ m+{ \omega  }(1-x) }
										\delta \left[ \sqrt { \left[ \frac {
												(m{ \omega  }+{ { \omega  }
												}^{ 2 })(1-x) }{ m+{ \omega  }(1-x) }  \right] ^{ 2 }+{ m }^{ 2 } } 
										-\frac { ({ { \omega  } }^{ 2 })(1-x) }
										{ m+{ \omega  }(1-x) } -m \right]  }. \end{align}		
Since, we have two roots $x=1$ and $x=-1$, so applying the property of 
the Dirac-delta function 
									\begin{align}
									\delta \left[ f(x) \right] =\sum
									_{ i=1 }^{ n }{ \frac { \delta 
											(x-x_{ i }) }{ { f }^{ ' }(x_i) }} ,
									\end{align}
$\sigma^s $ becomes
									\begin{align} 
									\sigma^s \com
									=e^4
									\frac { eB\pi ^{ 2 } (m+\omega) }{ 
										12\sqrt { 3 } { m }\omega^{ 2 } }
									\exp\left( \frac { -{2 q^{ 2 }_X } }
									{ eB }  \right) \int _{ -1 }^{ 1 }
									{ \frac { dx }{ m+{ \omega  }(1-x) }
									} \left[ \frac { \delta (x-1) }{ {
										|f }^{ ' }(x){ | }_{ x=1 } } +
								\frac { \delta (x+1) }{ { |f }^
									{ ' }(x){ | }_{ x=-1 } }  \right] , 
								\end{align}	
								where,
								\begin{align}
								f^{ ' }\left( x \right)
								=\Bigg[\frac { \frac { [4{ 
											{ \omega  } }^
										{ 2 }{ m }^{ 2 }(1-x)(-1)+
										{ { \omega  } }^{ 4 }(2)
										(-1)(1-x)-2m{ { \omega  } }^{
											3 }(-2)(1-x)+2{ m }^{ 3 }{ \omega  }
										(-1)][m+{ \omega  }(1-x)] }
									{ 2\sqrt { { { 2\omega  } }^{ 2 }
											{ m }^{ 2 }{ (1-x) }^{ 2 }+
											{ { \omega  } }^{ 4 }
											{ (1-x) }^{ 2 }+2m{ { \omega  } }^{ 3 }{ (1-x) }^{ 2 }
											+{ m }^{ 4 }+2{ m }^{ 3 }{ \omega  }(1-x) }  }  }{ { (m+{ \omega  }
										(1-x)) }^{ 2 } }\nonumber\\+\frac
								{ { \omega  }\sqrt { { { 2\omega  } }
										^{ 2 }{ m }^{ 2 }{ (1-x) }^{ 2 }+{ { \omega  }
										}^{ 4 }{ (1-x) }^{ 2 }+2m{ { \omega  }
									}^{ 3 }{ (1-x) }^{ 2 }+{ m }^{ 4 }
									+2{ m }^{ 3 }{ \omega  }(1-x) }  }
							{ { (m+{ \omega  }
									(1-x)) }^{ 2 } }
							\nonumber\\
							-{ { \omega  } }^{ 2 }
							\left[ \frac { (-1)(m+{ \omega  }(1-x))+{ \omega  }_{ i }(1-x) }
							{ { (m+{ \omega  }(1-x)) }^{ 2 } } 
							\right] \Bigg] . 
							\end{align}	
							We obtain,
							\begin{eqnarray}
							|f^{ ' }\left( x \right) |_{x=1}=\frac { { { \omega  } }^{ 2 } }{ m },\quad\quad\quad\quad\quad\quad\quad\quad\nonumber\\\\|f^{ ' }\left( x \right) |_{x=-1}=\left| \frac { -9m{ { \omega  } }^{ 2 }-{ { 2\omega  } }^{ 3 } }{ { (m+{ 2\omega  }) }^{ 2 } }  \right| =\frac { 9m{ { \omega  } }^{ 2 }+{ { 2\omega  } }^{ 3 } }{ { (m+{ 2\omega  }) }^{ 2 } }.\nonumber\label{fx}
							\end{eqnarray} 
Substituting the values of $|f^{ ' }\left( x \right) |_{x=1}$  and 
$|f^{ ' }\left( x \right) |_{x=-1}$ and
 integrating using the property of Dirac-delta function
							\begin{align}
							\int _{ -a }^{ a }{ f(x)\delta (x-a)dx }
							=f(a)\left[ 2\Theta (a)-1 \right] \label{property},
							\end{align}
							\begin{align} \sigma^s \com
							=e^4
							\frac { eB\pi ^{ 2 } (m+\omega) }{ 
								12\sqrt { 3 } { m }\omega^{ 2 } } \exp
							\left( \frac { -{2 q^{ 2 } _X }
							}{ eB }  \right) \left[ \frac
							{ 1 }{ { { \omega  } }^{ 2 } } 
							-\frac { m+{ 2\omega  } }{ 9m
								{ { \omega  } }^{ 2 }+{ {
										2\omega  } }^{ 3 } }  
							\right]. \end{align}
The factor $\left( \frac { -{2 q^{ 2 } _X }}{ eB }  \right)$ can be eliminated by using strong magnetic
 field limit. In the lab frame, we consider the initial direction of photon 
along the magnetic field (Z) direction, so 
							\begin{equation}
							{ q }_{ X }=p_{ X }+{ k }_{ X }=m.\label{kx}
							\end{equation}
							Also we have $eB\gg { m }^{ 2 }$, so finally $\sigma^s \com$ becomes
							\begin{align} \sigma^s \com
							=e^4
							\frac { eB\pi ^{ 2 } (m+\omega) }{ 
								12\sqrt { 3 } { m }\omega^{ 2 } }\left[ \frac
							{ 1 }{ { { \omega  } }^{ 2 } } 
							-\frac { m+{ 2\omega  } }{ 9m
								{ { \omega  } }^{ 2 }+{ {
										2\omega  } }^{ 3 } }  
							\right].\end{align}
							\underline{\textbf{FOR $u$-CHANNEL DIAGRAM :}}
							
Putting  $x=\cos\theta$ in \eqref{sigmau}, then $\sigma^u$ reduces to
				\begin{eqnarray}
				\sigma^u\com &=&e^4\frac 
				{ eB\pi ^{ 2 }(m+\omega ) }
				{3\sqrt { 3 } { m }\omega  }
				\exp\left( \frac { -{ 2q'^{ 2 }_X }
				}{ eB }  \right)\nonumber \\
				& \times& \int \bigg[\frac { \left[ { m }^{ 3 }{ \omega  }^{ 2 }(4+3{ x^{ 2 } }-7x)+{ m }^{ 3 }{ \omega  }(7x-8)\times (m+{ \omega  }(1-x)) \right] \times { \left[ m+{ \omega  }(1-x) \right] ^{ 2 } } }{ { \left[ { m }^{ 2 }{ \omega  }^{ 2 }\left( 1-x^{ 2 } \right) -2{ m }^{ 2 }{ \omega x\times \left( m+{ \omega  }(1-x) \right)  } \right]  }^{ 2 } } \bigg]{ P }_{ Z }\nonumber\\& \times& \delta \left[ \sqrt { \left[ \frac { (m{ \omega  }+{ { 
\omega  } }^{ 2 })(1-x) }{ m+{ \omega  }(1-x) } \right] ^{ 2 }+{ m }^{ 2 } } 
					-\frac { ({ { \omega  } }^{ 2 })
						(1-x) }{ m+{ \omega  }(1-x) } 
-m \right] \nonumber\\
					&=&e^4\frac 
					{ eB\pi ^{ 2 }(m+\omega ) }
					{3\sqrt { 3 } { m }\omega  } \exp\left
					( \frac { -{2 q^{' 2 }_X  }}{ eB }
					\right) \nonumber\\ &\times &
\int \bigg[\frac { \left[ { m }^{ 3 }{ \omega  }^{ 2 }(4+3{ x^{ 2 } }-7x)+{ m }^{ 3 }{ \omega  }(7x-8)\times (m+{ \omega  }(1-x)) \right] \times { \left[ m+{ \omega  }(1-x) \right] ^{ 2 } } }{ { \left[ { m }^{ 2 }{ \omega  }^{ 2 }\left( 1-x^{ 2 } \right) -2{ m }^{ 2 }{ \omega x\times \left( m+{ \omega  }(1-x) \right)  } \right]  }^{ 2 } } \bigg]{ P }_{ Z }\nonumber\\ 
				&\times &  \left
				[ \frac { \delta (x-1) }{ { 
						|f }^{ ' }(x){ | }_{ x=1 } }
				+\frac { \delta (x+1) }{ { |f }^{ ' }
					(x){ | }_{ x=-1 } }  \right]. 		   			
				\end{eqnarray}
Substituting values of $|f^{ ' }\left( x \right) |_{x=1}$  and $|f^{ ' }
\left( x \right) |_{x=-1}$  from \eqref{fx} and integrating using property of Dirac-delta function as given 
in \eqref{property}, finally the crosssection for the $u$-channel $\sigma^u$  becomes
			\begin{align}
			\sigma^u\com =e^{ 4 }\frac { eB\pi ^{ 2 } }{ 12\sqrt { 3 }  }\exp\left( \frac { -2q^{ '2 }_X }
			{ eB }  \right) \left[ \frac { 61{ m }^{ 2 }+78m{ \omega  }^{ 2 }+32{ \omega  }^{ 2 } }{ { m }^{ 2 }\omega [9{ m }{ \omega  }^{ 2 }+2{ \omega  }^{ 3 }] } -\frac { 1 }{ m{ \omega  }^{ 3 } }   \right]  
			\end{align}
The factor $\exp ( \frac { -{2 q^{ 2 } _X }}{ eB }  )$ becomes
unity in strong magnetic field limit. In lab frame, we consider the initial 
direction of photon along the magnetic field (Z-direction), so 
${ q '}_{ X }=P_{ X }-{ k }_{ X }=P_{ X }$.
Also we have $P_\bot=0$, so that we can write
				
				\begin{equation}
				{ q '_{ X }}=0. \label{qdash}
				\end{equation} 
				Finally $	\sigma^u$ becomes
			\begin{align}
			\sigma^u\com =e^{ 4 }\frac { eB\pi ^{ 2 } }{ 12\sqrt { 3 }  } \left[ \frac { 61{ m }^{ 2 }+78m{ \omega  }^{ 2 }+32{ \omega  }^{ 2 } }{ { m }^{ 2 }\omega [9{ m }{ \omega  }^{ 2 }+2{ \omega  }^{ 3 }] } -\frac { 1 }{ m{ \omega  }^{ 3 } }   \right]  
			\end{align}
				
\end{document}